\documentclass[12pt]{article}
\usepackage{amsmath}
\usepackage{graphicx}
\usepackage{enumerate}
\usepackage{natbib}
\usepackage{url} 
\usepackage{authblk}
\usepackage{scalerel}
\usepackage{xr}
\usepackage{subcaption}

\newcommand{\blind}{1}

\usepackage{geometry}
\usepackage{amsmath}
\allowdisplaybreaks[4]
\usepackage{float}
\usepackage{mathrsfs}
\usepackage{amsfonts}   
\usepackage{xcolor}
\usepackage{times}
\usepackage{bm}
\usepackage{colonequals}
\usepackage{graphicx}
\usepackage{caption}
\usepackage{enumerate}
\usepackage{cmll}
\usepackage{geometry}
\usepackage{amsthm}
\usepackage{threeparttable}
\usepackage{booktabs}
\usepackage{multirow}
\usepackage{makecell}
\usepackage{natbib}
\usepackage{graphicx,float,booktabs,mathtools}
\usepackage{easyReview}

\usepackage[plain,ruled]{algorithm2e}

\newcommand{\pr}{\text{pr}}

\usepackage{color}

\newtheorem{theorem}{\bf{Theorem}}

\newtheorem{remark}{\bf{Remark}}
\newtheorem{lemma}{\bf{Lemma}}
\newtheorem{definition}{\bf{Definition}}

\newtheorem{example}{\bf{Example}}
\newtheorem{proposition}{\bf{Proposition}}
\newtheorem{condition}{\bf{Condition}}

\newtheorem{assumption}{\bf{Assumption}}

\newcommand{\ind}{\perp \!\!\! \perp}

\usepackage{titlesec}

\titleformat{\section}
{\normalfont\fontsize{12.5}{12.5}\bfseries}{\thesection}{1em}{}
\titleformat{\subsection}
{\normalfont\fontsize{12}{12}\bfseries}{\thesubsection}{1em}{}

\titlespacing*\section{0pt}{10pt}{3pt}
\titlespacing*\subsection{0pt}{10pt}{1pt}
\titlespacing*\subsubsection{0pt}{1pt}{1pt}

\makeatletter
\g@addto@macro\normalsize{%
	\setlength\abovedisplayskip{3.5pt}
	\setlength\belowdisplayskip{3.5pt}
	\setlength\abovedisplayshortskip{3.5pt}
	\setlength\belowdisplayshortskip{3.5pt}
}
\makeatother

\makeatletter
\def\thm@space@setup{\thm@preskip=1pt
	\thm@postskip=1pt}
\makeatother

\newcommand{\sI}{\scaleto{\mathrm{I}\mathstrut}{7pt}}
\newcommand{\sE}{\scaleto{\mathrm{E}\mathstrut}{7pt}}

\def\T{{ \mathrm{\scriptscriptstyle T} }}

\newcommand\blfootnote[1]{%
	\begingroup
	\renewcommand\thefootnote{}\footnote{#1}%
	\addtocounter{footnote}{-1}%
	\endgroup
}

\begin{document}
\pagenumbering{gobble}

\def\spacingset#1{\renewcommand{\baselinestretch}%
	{#1}\small\normalsize} \spacingset{1}

\newcommand{\wei}[1]{{\color{magenta}{[\textbf{Wei:} #1]}}}
\newcommand{\MARKER}[1]{}

\if1\blind
{
	\title{\bf Semiparametric Efficient Fusion of Individual Data and Summary Statistics
	}
\author[a${\dag}$]{Wenjie Hu}
\author[b${\dag}$]{Ruoyu Wang}
\author[c$*$]{Wei Li}
\author[a$*$]{Wang Miao}

\affil[a]{Department of Probability and Statistics, Peking University, Beijing, China}
\affil[b]{Department of Biostatistics, Harvard T.H. Chan School of Public Health, Boston, MA, USA}
\affil[c]{Center for Applied Statistics and School of Statistics, Renmin University of China, Beijing, China}
\date{}
\maketitle
\blfootnote{$\dag$ Equal contribution.}
\blfootnote{$*$ Co-corresponding authors. Emails: weilistat@ruc.edu.cn; mwfy@pku.edu.cn}
} 
\fi

\if0\blind
{
\bigskip
\bigskip
\bigskip
\begin{center}
{\bf \Large Semiparametric Efficient Fusion of Individual Data and Summary Statistics}
\end{center}
\medskip
} \fi

\bigskip

\begin{abstract}
Suppose we have individual data from an internal study and various summary statistics from relevant external studies. 
External summary statistics have the potential to improve  statistical inference for the internal population;
however, it may lead to efficiency loss or bias if not used properly.
We study the fusion of individual data and summary statistics in a semiparametric framework to investigate the efficient use of external summary statistics.
Under a weak transportability assumption, we establish the semiparametric efficiency bound for estimating a general functional of the internal data distribution, which is no larger than that using only internal data and underpins the potential efficiency gain of integrating individual data and summary statistics. 
We propose a data-fused efficient estimator that achieves this efficiency bound. 
In addition, an adaptive fusion estimator is proposed to eliminate the bias of the original data-fused estimator when the transportability assumption fails. We establish the asymptotic oracle property of the adaptive fusion estimator.
Simulations and application to a Helicobacter pylori infection dataset demonstrate the promising  performance of the proposed method.
\end{abstract}

{\it Keywords:} Causal inference; Data fusion; Integrative data analysis; Semiparametric efficiency bound.

\newpage
\spacingset{1.7} 
\pagenumbering{arabic}
\section{Introduction}
Suppose individual data from an internal study is available to investigate a particular scientific purpose.
It is appealing to fuse external datasets from different sources with the internal data to improve statistical inference.
Methods for data fusion with external individual data have grown in popularity in recent years \citep[e.g.,][]{yang2020combining,li2023improving,li2023efficient,sun2018semiparametric,chen2021minimax}. 
However, sometimes it is  impossible to access the individual data due to ethics and privacy concerns
and one can only have certain summary statistics from external studies.
Meta-analysis has been widely applied  to integrate summary statistics  on a common parameter from multiple studies  \citep[e.g.,][]{singh2005combining,lin2010relative,li2020causal},
but it becomes challenging if participating studies are analyzed with different statistical models and are concerned with different parameters.
In order to assimilate various types of summary statistics from multiple sources, previous authors have developed a suite of methods which essentially view external summary statistics as certain constraints on the internal data distribution.
For example, \citet[Section 3.2 Example 3]{bickel1993efficient}  established the semiparametric theory for using external summary statistics as moment equation constraints;
\cite{qin2000miscellanea}  proposed an empirical likelihood method and \cite{chatterjee2016constrained} proposed a constrained maximum likelihood approach, which leverage summary statistics obtained from a large external dataset to improve estimation efficiency for a parametric model of the internal dataset.
In situations where the uncertainty of external summary statistics is negligible, these methods can achieve higher efficiency than the maximum likelihood estimator based solely on the internal individual data.
Nonetheless, \cite{zhang2020generalized} cautioned  that if  the uncertainty of external summary statistics is not negligible, 
the efficiency gain of the constrained maximum likelihood estimator is not guaranteed, and paradoxically, it can be even less efficient than the maximum likelihood estimator based on internal individual data solely. 

The external summary statistics provide additional information, and when incorporated appropriately, they are expected to improve the efficiency of statistical inference. The efficiency paradox arises mainly because the use of the external summary statistics is not optimized. Therefore, it is of both practical and theoretical interest to investigate the efficient fusion of individual data and summary statistics. Semiparametric theory is a widely used framework to study the optimality of estimators, providing guidance for constructing efficient estimators under minimal assumptions on the data distribution. However, the classic semiparametric theory, which is designed for independent and identically distributed (i.i.d.) individual data \citep{bickel1993efficient, li2023efficient, li2023data},
is not applicable to the current setting involving both individual data and summary statistics. The literature on semiparametric theory for non-i.i.d. data is relatively sparse \citep{strasser1989tangent,mcneney2000application}, and, to our knowledge, none of these works address the incorporation of summary statistics. \citet{zhang2020generalized} developed a generalized data integration method (GIM)  under a parametric conditional density model, which avoids the efficiency paradox and is optimal in the sense that it achieves the smallest asymptotic variance over a class of maximum log-pseudolikelihood estimators. Recently, \cite{chen2024integrating} extended the method of \cite{zhang2020generalized} to account for prior probability shifts between internal and external populations, thereby improving robustness against population heterogeneity.  However, the general semiparametric theory for fusing individual data and summary statistics is not available yet. In particular, it is not clear how to efficiently estimate a general functional in a semi/non-parametric model when given both individual data and summary statistics.

In addition to potential efficiency loss, 
 integrating external summary statistics can also introduce estimation bias  if the summary statistics are not transportable.
This bias frequently occurs in the presence of population heterogeneity, biased sampling, or model misspecification.
It has been noted in  conventional meta-analysis, and various robust methods have been proposed to address this problem \citep{singh2005combining,shen2020fusion,wang2023robust}, ensuring consistent and asymptotically normal estimation despite untransportable summary statistics. 
For the fusion of individual data, \citet{chen2021minimax}, \citet{kallus2018removing}, \citet{yang2023elastic}, \citet{li2023improving}, \citet{yang2020improved} and \citet{gao2025improving} considered combining randomized trial data and observational data to remove unmeasured confounding bias.
However, these methods require the availability of individual data from multiple data sources.
For integration of   individual data and summary statistics, 
\citet{zhai2022data, huang2023simultaneous}   proposed penalized constrained maximum likelihood methods that extend the empirical likelihood framework of \citet{qin2000miscellanea}, \citet{chatterjee2016constrained} and \cite{zhang2020generalized} to accommodating possibly untransportable external summary statistics.
However, these methods focus on parameters in a parametric model and may not be suitable for general semi/non-parametric inference problems, such as estimation of the average treatment effect estimation in causal inference or the outcome mean in missing data analysis.

In this paper,  we develop a semiparametric framework for the integration of internal individual data and external summary statistics. In contrast to previous works that primarily focus on parametric models,  our approach accommodates general semiparametric and nonparametric models. 
We derive the semiparametric efficiency bound for inference on a general functional of the internal data distribution in the presence of external summary statistics, which is shown to be no larger than that obtained using only the internal data. 
We construct a data-fused efficient estimator that achieves the efficiency bound under data fusion. By adopting the efficient estimator which incorporates the external summary statistics in an optimal way, the efficiency paradox is resolved. The data-fused efficient estimator has a closed form. This makes it computationally efficient when all the efficient influence functions in the proposed estimator are available and easy to compute, although obtaining the efficient influence function may be challenging in complex problems. Moreover, the data-fused efficient influence function inherits the Neyman orthogonality property from the internal-data influence functions it incorporates. This property enables the use of flexible machine learning methods—such as neural networks—for estimating nuisance parameters in the data-fused efficient estimator. 
To address potential issues with untransportable summary statistics, we further propose an adaptive fusion estimator by
leveraging some carefully designed weighting matrices.  
The adaptive fusion estimator is continuous with respect to the observed data, which is a desirable property that improves numerical stability \citep{fan2001variable}. In addition, it is easy to compute due to its closed-form expression, provided that the relevant efficient influence functions involved are available and easy to compute.
The adaptive fusion estimator is consistent and asymptotically normal even if some external summary statistics are untransportable. It is also asymptotically equivalent to the oracle estimator that uses only transportable summary statistics. The asymptotic results hinge on the fact that one can consistently assess whether the summary statistics are transportable in large sample size. However, distinguishing between transportable and untransportable summary statistics in finite samples can be challenging, particularly when the internal and external populations are similar but not identical. This problem may lead to undercoverage of confidence intervals based on the adaptive fusion estimator's asymptotic distribution. We thus propose a re-bootstrap procedure to mitigate the undercoverage issue in finite samples. This procedure maintains robust coverage rates even under the challenging setting where the heterogeneity between populations is comparable to the magnitude of the estimation error. We discuss theoretical results for the scenario where the transportability of certain summary statistics holds asymptotically at certain rates in Supplementary Material. 
We evaluate the performance of the proposed estimation methods via simulations and apply them to test the causal effect of a combined therapy on Helicobacter pylori infection.

The rest of this paper is organized as follows. In Section \ref{sec: classic theory}, we briefly review the classic semiparametric theory and discuss the gap between the classic theory and the problem under consideration. In Section \ref{sec:efficiency}, we establish the semiparametric efficiency theory for fusing individual data and summary statistics. In Section \ref{sec:adaptive fusion}, we propose an adaptive fusion estimator to integrate the summary statistics in the presence of untransportable components and a re-bootstrap procedure to make inference. We report extensive simulation results and a real data analysis in Section \ref{sec: numerical}, followed by discussions. Additional simulations, discussion on the scenario when transportability holds asymptotically and all proofs are provided in Supplementary Material.

{\bf Notation.} For any integer $j$, index set $\mathcal{I}$, vector $v$ and matrix $V$, let $v_{j}$ be the $j$-th component of $v$, $V_{jj}$ the $j$-th diagonal element of $V$, $v_{\mathcal{I}}$ the vector consisting of the components of $v$ in $\mathcal{I}$ and $V_{\mathcal{I}}$ the matrix consisting of elements with indices in $\mathcal{I}\times\mathcal{I}$. For any positive sequences $\{a_{1n}\}$ and $\{a_{2n}\}$, $a_{1n} \asymp a_{2n}$ means $C^{-1} a_{1n} \leq a_{2n} \leq C a_{1n}$ for some $C > 1$.

\section{The Classic Semiparametric Theory}\label{sec: classic theory}
Suppose we have $n$ i.i.d. individual-level observations $Z_1, \ldots, Z_n$ of $Z$ from the internal distribution/study  $P_0\in \mathcal{P}_0$ and a $q$-dimensional vector of summary statistics $\tilde{\beta}=(\tilde\beta_1,\ldots,\tilde\beta_q)^\T$ based on individual observations  $(W_1,\ldots, W_m)$ from the external distribution/study  $P_1\in \mathcal{P}_1$, where $\mathcal{P}_0$ and $\mathcal{P}_1$ denote collections of models for internal data and external data, respectively. 
The external sample size $m$ is known,  the external individual data are unavailable, and $\tilde \beta$ is assumed to be an estimator of some functional $\beta(P_1)$ of $P_{1}$.
The parameter of interest is a $p$-dimensional   functional of the internal data distribution,   $\tau = \tau(P_0)$, which may differ from $\beta(P_1)$.
Throughout the paper, we let $E(\cdot)$ denote the expectation with respect to    $P_0$ and $\widehat{E}(\cdot)$  the empirical mean in the internal data, unless otherwise specified. 
We briefly review the classical semiparametric theory when only the internal i.i.d. individual data are used for estimating $\tau$.
Most reasonable estimators in statistical inference problems are regular and asymptotically linear (RAL).
Following \citet[][Chapter 3]{tsiatis2006semiparametric},    RAL estimators in a parametric model indexed by a finite-dimensional parameter $\theta$, say $\{P_0(Z;\theta);\theta\in \mathbb R^k\}$,  are described as follows.
\begin{definition}\label{classical estimator}
An estimator $T_n=T_n(Z_1,\ldots, Z_n)$ of $\tau$ is regular  if $n^{1/2}\{T_n(Z_1^{(n)},\ldots, Z_n^{(n)})- \tau(\theta_n)\}$ has a limiting distribution that does not depend on the local data generating process where for each $n$ the data $\{Z_1^{(n)},\ldots, Z_n^{(n)}\}$ are i.i.d. distributed according to $P_0(Z;\theta_n)$ with $n^{1/2}(\theta_n-\theta)$ converging to a constant; 
an estimator $T_n$   is asymptotically linear if $T_n = \tau + n^{-1}\sum_{i=1}^n \phi(Z_i) + o_P(n^{-1/2})$ for some vector function $\phi$ with $E\{\phi(Z)\}=0$
and $E\{\phi(Z)\phi^\T(Z)\}$ finite and nonsingular;
and an estimator $T_n$  is RAL if it is both regular and asymptotically linear.
\end{definition}
The function $\phi$ is called an influence function for $\tau$, 
describing the influence of each observation on the estimation of $\tau$.
Regularity is often desirable, which rules out pathological estimators such as the superefficient estimator of Hodges and estimators that invoke more information than is contained in the model.
Moreover, it can be shown that the most efficient regular estimator is asymptotically linear  \citep{hajek1970characterization}; 
hence, it is reasonable to restrict attention to RAL estimators.
For a differentiable parameter $\tau(\theta)$   in the parametric model $P_0(Z;\theta)$, letting $S_\theta$ denote the score for $\theta$,  
then the      Cramer--Rao bound,  $V_\theta=\{\partial \tau(\theta)/\partial \theta\} \{E(S_\theta S^\T_\theta)\}^{-1} \{\partial \tau(\theta)/\partial \theta^\T\}$, characterizes the smallest possible asymptotic variance  for  RAL estimators of $\tau$.
However,  lack of flexibility and thus potential misspecification of parametric models  incur  untransportable inferences, and in many situations, one is   only  interested in 
a finite-dimensional parameter rather than the full data distribution. 
This leads to the adoption of semiparametric or nonparametric models that admit infinite-dimensional parameters embodying less restrictive assumptions beyond the parameter of interest.
\cite{bickel1993efficient} described the efficiency theory for inference in semiparametric and nonparametric models. 
One can view a semiparametric model as the collection of many parametric submodels that satisfy the semiparametric assumptions and contain the true data generating process but impose no additional restrictions.  
A (pathwise) differentiable functional $\tau$ on a semiparametric model needs to be differentiable  on all   parametric submodels and satisfy
$\partial \tau(\theta)/\partial \theta = E\{\phi S_\theta\}$ for some squared integrable function $\phi$ and    score function  $S_\theta$ of an arbitrary parametric submodel.
An estimator is said to be regular on a semiparametric model if it is regular on all parametric submodels.
A key concept in the semiparametric theory is the semiparametric efficiency bound, which is the supremum of the Cramer-Rao bounds for all parametric submodels.
The semiparametric efficiency bound is the lower bound for the asymptotic variance of any RAL estimator.
The influence function attaining the semiparametric efficiency bound is called the efficient influence function,
and the corresponding estimator is the efficient estimator.

In the rest of this paper, we let $\phi_{\mathrm{e}}$ denote the efficient influence function,  $E(\phi_{\mathrm{e}}\phi_{\mathrm{e}}^\T)$ the efficiency bound, and $\hat{\tau}_{\mathrm{e}}^{\sI}$   an efficient estimator for $\tau$  based on the internal data in the class of semiparametric or nonparametric models under consideration.
We illustrate these concepts with an influential causal inference example; see \cite{bang2005doubly} and \citet{hahn1998role} for details.

\begin{example}\label{causal}
Suppose we have   internal individual   data on $Z=(D,X,Y)$ from an observational study $P_0$ about the effect of a binary treatment $D$ on the outcome $Y$ with
covariates $X$. 
Let $Y_d$ denote the potential outcome if the treatment were set to $D=d$ for $d=0,1$ and $\tau = E(Y_1 - Y_0)$ the average treatment effect. 
Let $p(X) = \pr(D=1\mid X)$ be the treatment propensity score and $\mu_d(X) = E(Y\mid D=d, X)$ ($d = 0, 1$)  the outcome regression function.
Under the ignorability assumption  ($Y_d\ind D\mid X$ and  $0<p(X)<1$), 
$\tau$ is identified from the observed data  with $\tau =E\{\mu_1(X) - \mu_0(X)\}$.
The efficient influence function for $\tau$ in the nonparametric model that imposes no other restrictions than the ignorability is   
\[
\phi_{\mathrm{e}}(Z;\tau)  = \frac{D}{p(X)}\{Y - \mu_1(X)\} - \frac{1-D}{1-p(X)}\{Y - \mu_0(X)\} + \mu_1(X) - \mu_0(X) - \tau.\]
The efficiency bound for $\tau$ is $E(\phi_{\mathrm{e}}^2)$.
The efficient estimator  $\hat{\tau}_{\mathrm{e}}^{\sI}$ can be obtained by  firstly estimating   $\{p(X), \mu_d(X)\}$ and then solving  $\widehat{E}\{\phi_{\mathrm{e}}(Z;\tau)\}=0$ with these nuisance estimators plugged in. 
For instance,  one can specify and fit parametric working  models $p(X;\hat\zeta)$ and $\mu_d(X;\hat\psi_t)$, then
\[
\begin{aligned}
\hat{\tau}_{\mathrm{e}}^{\sI} =& \frac{1}{n}\sum_{i=1}^n \biggl [\frac{D_i}{p(X_i; \hat{\zeta})}\{Y_i - \mu_1(X_i; \hat{\psi}_1)\}  - \frac{1-D_i}{1-p(X_i;\hat{\zeta})}\{Y_i - \mu_0(X_i;\hat{\psi}_0)\}\\
&+ \mu_1(X_i;\hat{\psi}_1) - \mu_0(X_i;\hat{\psi}_0) \biggl].
\end{aligned}
\]
\end{example} 

We aim to combine both the internal individual data and the external summary statistics $\tilde \beta$ to improve the estimation of $\tau$. Suppose $\beta = \beta(P_{0})$ is well defined in the internal data, and let $\eta_{\mathrm{e}}$ be the efficient influence function for $\beta$ based on internal individual data, which depends on $\beta$ in general. For simplicity, we omit the dependency in the notation $\eta_{\mathrm{e}}$ when evaluated at the true value $\beta$, and use $\eta_{\mathrm{e}}(\beta^{\dag})$ to denote the corresponding influence function evaluated at some $\beta^{\dag}$ that may not be equal to $\beta$.
Applying the classical semiparametric theory, 
\citet[Section 3.2 Example 3]{bickel1993efficient}  established a well-known result that  the efficient  influence function for $\tau$ is  
$\phi_{\mathrm{e}} - E\left( \phi_{\mathrm{e}} \eta_{\mathrm{e}}^\T \right)\left\{E (\eta_{\mathrm{e}}\eta_{\mathrm{e}}^\T)  \right\}^{-1}\eta_{\mathrm{e}}$ when $\beta$ is known.   
This influence function   motivates the   estimator $T_{n, \mathrm{e}}(\beta) = \hat{\tau}_{\mathrm{e}}^{\sI} - \widehat{\Sigma}_{\phi\eta}\widehat{\Sigma}_{\eta\eta}^{-1}\widehat{E}(\eta_{\mathrm{e}})$ with   $\widehat{\Sigma}_{\phi\eta}$ and $\widehat{\Sigma}_{\eta\eta}$ being consistent estimators for $E(\phi_{\mathrm{e}}\eta_{\mathrm{e}}^\T)$ and $E(\eta_{\mathrm{e}}\eta_{\mathrm{e}}^\T)$ based on the internal data. 
The estimator $T_{n, \mathrm{e}}(\beta)$ is efficient  when $\beta$ is known and has  asymptotic variance  no larger than $\hat{\tau}_{\mathrm{e}}^{\sI}$.
However, the estimator $T_{n, \mathrm{e}}(\beta)$ is infeasible if $\beta$ is unknown. In practice, the true value $\beta$ is typically unknown but is instead estimated by some external summary statistics $\tilde{\beta}$.
Thus, a primitive data-fused estimator of $\tau$ is the plug-in estimator $T_{n, \mathrm{e}}(\tilde{\beta}) = \hat{\tau}_{\mathrm{e}}^{\sI} - \widehat{\Sigma}_{\phi\eta}\widehat{\Sigma}_{\eta\eta}^{-1}\widehat{E}\{\eta_{\mathrm{e}}(\tilde{\beta})\}$. 
The estimator $T_{n, \mathrm{e}}(\tilde{\beta})$ is expected to still deliver better efficiency than $\hat{\tau}_{\mathrm{e}}^{\sI}$.
Nonetheless, somewhat paradoxically, the estimator $T_{n, \mathrm{e}}(\tilde{\beta})$ that incorporates external information may be less efficient than the internal data-based estimator $\hat{\tau}_{\mathrm{e}}^{\sI}$, which we will show later. The phenomenon that incorporating external information may lead to efficiency loss has been previously pointed out and resolved under a parametric conditional density model by \cite{zhang2020generalized}.

Roughly speaking, the above efficiency paradox arises because the estimator $T_{n, \mathrm{e}}(\tilde{\beta})$ ignores the uncertainty of $\tilde{\beta}$ and thus fails to incorporate the external summary statistics in an efficient way. 
This motivates us to investigate the efficiency of the data-fused estimators and derive the most efficient one in a semiparametric framework where the parameter of interest is a generic functional of 
internal data distribution.
However, the classical semiparametric theory for i.i.d. individual data does not apply to the setting here, which involves the external summary statistics $\tilde{\beta}$ that are distributed differently from the internal data.
Although there exists a sparse literature on the  semiparametric theory for non-i.i.d data \citep{strasser1989tangent,mcneney2000application}, 
their theory is not applicable here.
In the next section, we extend the classical semiparametric theory to 
the data fusion setting where both individual data and summary statistics are available.

\section{Semiparametric   Theory for   Fusion of     Individual data   and   Summary Statistics} \label{sec:efficiency}

\subsection{Assumptions and Data-fused RAL Estimators}

In order to make use of the external summary statistic, we make the following assumptions.
{ 
\begin{assumption}\label{beta}
The external summary statistic $\tilde{\beta}$ is a RAL estimator of a $q$-dimensional parameter/functional $\beta(P_{1})$    of the external data distribution $P_1$;
$m^{1/2}\{\tilde{\beta} - \beta(P_1)\} \rightarrow N(0, \Sigma_1)$;
a consistent covariance  estimator  $\widehat{\Sigma}_1$ for $\Sigma_1$ is also available;
and $m/n\rightarrow \rho \in (0, +\infty)$.
\end{assumption}
}
\begin{assumption}[Transportability]\label{well defined}
 $\beta(P_0) = \beta(P_1)$.  
\end{assumption}
Assumption \ref{beta}  is met with standard estimation methods under mild regularity conditions and has been widely adopted in meta-analysis   \citep{singh2005combining,xie2011confidence,kundu2019generalized,zhang2020generalized}.
Note that the functional $\beta(\cdot)$ is not necessarily the same as $\tau(\cdot)$, the parameter of interest.
Assumption \ref{well defined}  requires that the parameter $\beta$ is well-defined  in both the internal and external data and has the same   value.
This transportablity assumption establishes the  connection between the internal data distribution $P_0$ and external data distribution $P_1$,
which is essential for efficiency improvement with external summary statistics. 
Analogous assumptions such as mean/distribution exchangeability have been used in previous work  \citep[e.g.,][]{dahabreh2019generalizing,li2023improving}. 
{ For example, \cite{gao2025improving} assumed  that the conditional mean of the (potential) outcome is the same in the internal and external studies, at least for some covariate values. Their assumption is neither implied by nor implies our Assumption \ref{well defined}. Both are plausible under different scenarios, and we focus on Assumption \ref{well defined} because it is particularly well-suited for data fusion when only summary statistics are available.}
In Section 4, we discuss scenarios where Assumption 2 fails and bias may arise.  Furthermore, we consider the scenario where the transportablity of a subset of the summary statistics holds only asymptotically at a suitable rate in Supplementary Material.
In the rest of the paper, we denote $\beta = \beta(P_0)$, 
and we assume that    $\tau$  is pathwise differentiable on $\mathcal{P}_{0}$ at $P_0$ and $\beta$ is pathwise differentiable on $\mathcal{P}_{0}$ and $\mathcal P_1$ at $P_0$ and $P_1$, respectively.

We focus on the estimation of $\tau$ in the semiparametric model 
\[
\mathcal P_{\rm trans}=\{P_0\times P_1 \in \mathcal P_0\times \mathcal P_1: P_0, P_1 \text{  satisfy Assumption~\ref{well defined}}\}.
\] 
We consider the following class of estimators that incorporate both internal individual data and the external summary statistics.
\begin{definition}[Data-fused RAL estimator]\label{RAL estimator}
Let $T_n(Z_1,\ldots,Z_n,\tilde{\beta})$ denote a data-fused estimator  of $\tau$ and we write $T_n(\tilde{\beta})$ for shorthand.
\begin{itemize}
\item[(i)]   $T_n(\tilde{\beta})$   is regular if for every parametric submodel $P_0(Z;\theta)\times P_1(W;\theta) \in P_{\rm trans}$,
the quantity $n^{1/2}\big\{T_n(Z_1^{(n)}, \ldots, Z_n^{(n)}, \tilde{\beta}^{(m)}) - \tau(P_0(Z;\theta_n))\big\}$ has a limiting distribution that does not depend on the local data generating process, 
where the data $\{Z_1^{(n)},\ldots, Z_n^{(n)}\}$ is an i.i.d. sample from $P_0(Z;\theta_n)$, 
and $\tilde{\beta}^{(m)}$ is obtained from an i.i.d. sample $\{W_1^{(m)},\ldots, W_m^{(m)}\}$ from $P_1(W;\theta_n)$,  with $m/n\rightarrow \rho\in (0, \infty)$ and $n^{1/2}(\theta_n-\theta)$ converging to a constant.

\item[(ii)]  $T_n(\tilde{\beta})$   is asymptotically linear  if 
$T_n(\tilde{\beta}) = \tau +n^{-1}\sum_{i=1}^n  \psi(Z_i) + \gamma(\tilde{\beta}) + o_P(n^{-1/2})$
with  $E\{\psi(Z)\}=0$, $E\{\psi(Z) \psi^\T (Z)\}$ finite and nonsingular,  $\gamma(\tilde{\beta})$   continuously differentiable in $\tilde{\beta}$ and $\gamma(\beta) = 0$.
\item[(iii)] $T_n(\tilde{\beta})$  is regular and asymptotically linear (RAL) if it satisfies both (i) and (ii).
\end{itemize}

\end{definition}

Analogous to the classical semiparametric theory, 
Definition~\ref{RAL estimator}  (i)  characterizes the regularity with respect to both the internal data distribution and external data distribution.
This class of data-fused regular estimators contains all the regular estimators based only on the internal individual data. 
Following the spirit of classical asymptotic linearization,
Definition~\ref{RAL estimator} (ii) treats $\tilde{\beta}$ as an additional  sample  to the internal data and uses $\psi(Z_i)$ as well as $\gamma(\tilde \beta)$  to depict      the influence  of $(Z_1,\ldots, Z_n,\tilde{\beta})$ on the estimation of $\tau$.
The restrictions on  $\gamma(\tilde{\beta})$ ensure  that $T_n(\tilde{\beta})$ satisfying Definition~\ref{RAL estimator} (ii)  is consistent and asymptotically normal.
The class of data-fused RAL estimators in Definition~\ref{RAL estimator}  (iii) includes all the RAL estimators that use only the internal data.

\begin{proposition} \label{characterization}
Under  Assumptions \ref{beta}, \ref{well defined} and a regularity condition (Condition \ref{cond: regular beta-tilde}) in Supplementary Material,
a data-fused RAL estimator $T_n(\tilde{\beta})$ has the following representation,
\begin{equation}\label{complex}
T_n(\tilde{\beta}) = \tau + \frac{1}{n}\sum_{i=1}^n \{\phi(Z_i) - \xi \eta_{\mathrm{e}}(Z_i)\} + \xi (\tilde{\beta} - \beta)+o_P(n^{-1/2}),
\end{equation}
and its asymptotic variance   is
\[
E(\phi\phi^\T) +\xi E(\eta_{\mathrm{e}}\eta_{\mathrm{e}}^\T)\xi^\T - 2\xi E(\eta_{\mathrm{e}}\phi^\T) +\rho^{-1} \xi\Sigma_1\xi^\T,
\]
where $\phi$ is an influence function for $\tau$  based only on the internal data, $\xi = \xi(P_0)$ is a $p\times q$ matrix, and the forms of $\phi$ and $\xi$ depend on the estimator $T_n(\tilde{\beta})$.
\end{proposition}
Condition \ref{cond: regular beta-tilde} in Supplementary Material is a regularity condition concerning the continuity and boundedness of $\tilde{\beta}$'s density, invoked primarily for technical purposes.
Proposition \ref{characterization} reveals how the estimation of $\beta$ in the external data affects the efficiency of a data-fused RAL estimator $T_n(\tilde{\beta})$.
If the estimation $\tilde \beta$ in the external study is very precise, or if the true value $\beta$ is known,
then $T_n(\tilde{\beta})$ reduces to   $T_n(\beta)$, which is  a RAL estimator  with influence function $\psi = \phi- \xi\eta_{\mathrm{e}}$. 
Recall that $T_{n, \mathrm{e}}(\beta)$ is the efficient estimator when $\beta$ is known and  $T_{n, \mathrm{e}}(\tilde \beta)$ is the plug-in estimator obtained by replacing $\beta$ by its estimate $\tilde \beta$ in the definition of $T_{n, \mathrm{e}}(\beta)$.
Then, $T_{n, \mathrm{e}}(\tilde{\beta}) = \tau + \widehat{E}(\phi_{\mathrm{e}} - A\eta_{\mathrm{e}}) + A(\tilde{\beta} - \beta) + o_P(n^{-1/2}) $ with $A=E\left( \phi_{\mathrm{e}} \eta_{\mathrm{e}}^\T \right)\left\{E (\eta_{\mathrm{e}}\eta_{\mathrm{e}}^\T)  \right\}^{-1}$ and its asymptotic variance is
$E(\phi_{\mathrm{e}}\phi_{\mathrm{e}}^\T) +  A\{\Sigma_1/\rho- E(\eta_{\mathrm{e}}\eta_{\mathrm{e}}^\T)\}A^\T$.
This result has three meaningful implications: First, when the  external data sample size  is much larger than  the internal data, i.e., $\rho \rightarrow + \infty$, 
the asymptotic variance of the plug-in estimator $T_{n, \mathrm{e}}(\tilde \beta)$  approximates  the semiparametric efficiency bound   when $\beta$ is known;
this has been noted by \citet{qin2000miscellanea} and \citet{chatterjee2016constrained} for the estimation of parameters in a parametric model for the internal data; here we extend this result to semiparametric models where the parameter of interest is a functional of data distribution.
Second,  when the external data sample size is much smaller than the internal data, i.e., $\rho\rightarrow 0$, the asymptotic variance of $T_{n, \mathrm{e}}(\tilde \beta)$  diverges,  suggesting that it has a slower convergence rate than $n^{-1/2}$. 
In this case,   the large variability in external summary statistics will severely damage the estimation efficiency of the plug-in estimator. 
Third,  for sufficiently small $\rho$ such that $\Sigma_1/\rho - E(\eta_{\mathrm{e}}\eta_{\mathrm{e}}^\T)>0$ (positive definite),  the asymptotic variance of $T_{n, \mathrm{e}}(\tilde \beta)$  is larger than that of the efficient estimator using only internal data.  
This explains why   the efficiency paradox arises:
the external summary statistics are used as the true values while their uncertainty is not negligible.  A concrete example where the efficiency paradox occurs is provided in Section \ref{Supp examples} in   Supplementary Material.

\subsection{The Efficiency Bound}\label{sec:Efficiency bound}

To assess how external summary statistics can improve the efficiency for estimating $\tau$, we establish the semiparametric efficiency bound for the data-fused RAL estimators.
Theorem \ref{convolution} characterizes their asymptotic distribution.

\begin{theorem}[Convolution theorem]\label{convolution}
Under Assumptions \ref{beta}, \ref{well defined} and  regularity Condition \ref{cond: regular beta-tilde} in Supplementary Material,   for any data-fused RAL estimator  $T_n(\tilde{\beta})$ we have
\[
n^{1/2}\left[\begin{matrix}
T_n(\tilde{\beta}) - \tau - n^{-1}\sum_{i=1}^n \left\{\phi_{\mathrm{e}}(Z_{i})- M\eta_{\mathrm{e}}(Z_{i})\right\} - M(\tilde{\beta} - \beta)\\
 n^{-1}\sum_{i=1}^n \left\{\phi_{\mathrm{e}}(Z_{i})- M\eta_{\mathrm{e}}(Z_{i})\right\} + M(\tilde{\beta} - \beta)
\end{matrix}
\right]\rightarrow \begin{pmatrix}
\Delta_0\\
S_0
\end{pmatrix},
\]
where  $M = E(\phi_{\mathrm{e}}\eta_{\mathrm{e}}^\T ) \left\{ \Sigma_1/\rho + E\left(\eta_{\mathrm{e}}\eta_{\mathrm{e}}^\T \right) \right\}^{-1}$, $\Delta_0$ and $S_0$ are independent, and $S_0 \sim N(0, B)$ with $B = E(\phi_{\mathrm{e}}\phi_{\mathrm{e}}^\T ) - E(\phi_{\mathrm{e}}\eta_{\mathrm{e}}^\T) \left\{ \Sigma_1/\rho + E(\eta_{\mathrm{e}}\eta_{\mathrm{e}}^\T ) \right\}^{-1}E(\phi_{\mathrm{e}}\eta_{\mathrm{e}}^\T )^\T$. 
\end{theorem}
Hereafter, we refer to the following 
as the data-fused efficient influence function,
\[\phi_{\mathrm{e}}- M\eta_{\mathrm{e}} + M(\tilde{\beta} - \beta).\]
In classical semiparametric theory, convolution theorem is a key venue for establishing the asymptotic bound for the limiting distribution of RAL estimators.
Here we extend it to the data fusion setting with both individual data and summary statistics.
The proof follows from \cite{bickel1993efficient}, 
the innovation here is that we incorporate the external summary statistics, which is different from the i.i.d case considered in classical semiparametric theory.
Theorem \ref{convolution} asserts that the asymptotic distribution of any data-fused RAL estimator $T_n(\tilde{\beta})$ can be decomposed into two independent parts $\Delta_0$  and $S_0$, where $S_0$ follows a normal distribution. 
We have $n^{1/2}\{T_n(\tilde{\beta}) - \tau\} \to \Delta_0 + S_0$ 
and $\mathrm{var}(\Delta_0 + S_0) = \mathrm{var}(\Delta_0) + \mathrm{var}(S_0) \geq \mathrm{var}(S_0) = B$, which is a lower bound for the asymptotic variance of any data-fused RAL estimator $T_n(\tilde{\beta})$.

\begin{theorem}\label{efficiency theorem}
Under Assumptions \ref{beta}, \ref{well defined} and  regularity Condition \ref{cond: regular beta-tilde} in Supplementary Material, the efficiency bound for data-fused RAL estimators given in Definition \ref{RAL estimator} is 
\begin{equation}\label{efficiency bound}
\begin{aligned}
B = E(\phi_{\mathrm{e}}\phi_{\mathrm{e}}^\T) - E(\phi_{\mathrm{e}}\eta_{\mathrm{e}}^\T) \left\{\Sigma_1/\rho + E(\eta_{\mathrm{e}}\eta_{\mathrm{e}}^\T) \right\}^{-1}E(\phi_{\mathrm{e}}\eta_{\mathrm{e}}^\T)^\T.
\end{aligned}
\end{equation}
\end{theorem}

Note that when only internal data are available, the efficiency bound is $E(\phi_{\mathrm{e}}\phi_{\mathrm{e}}^\T)$.
Theorem \ref{efficiency theorem}  suggests that the efficiency bound does not increase with the inclusion of external summary statistics.
However, $B$ is larger than  $E(\phi_{\mathrm{e}}\phi_{\mathrm{e}}^\T) - E(\phi_{\mathrm{e}}\eta_{\mathrm{e}}^\T) \left\{ E(\eta_{\mathrm{e}}\eta_{\mathrm{e}}^\T) \right\}^{-1}E(\phi_{\mathrm{e}}\eta_{\mathrm{e}}^\T)^\T$, which is the efficiency bound  when $\beta$ is known. This  indicates that $\tilde{\beta}$  provides no more information than the true   value of  $\beta$ for estimating $\tau$,
and $B$ reduces to the latter as $\rho\rightarrow +\infty$.
The efficiency bound $B$ also  depends on the efficiency of   $\tilde \beta$,  captured by $\Sigma_1$; specifically,
$B$ increases as  $\Sigma_1$ increases.  
{
To benchmark the efficiency bound with summary statistics, 
the following result describes the efficiency bound under Assumption \ref{well defined} when the individual data in external data are also available. 
	\begin{theorem}\label{individual external}
	Suppose  we have  individual data $(Z_1,\ldots,Z_n)$, $(W_1,\ldots, W_m)$ for  both the internal and external studies   and Assumption 2  holds, and $m/n\rightarrow \rho \in (0,+\infty)$. Then the semiparametric efficiency bound for $\tau$ is
	\begin{equation}\label{bound ind}
		E(\phi_{\mathrm{e}}\phi_{\mathrm{e}}^\T) - E(\phi_{\mathrm{e}}\eta_{\mathrm{e}}^\T) \left\{\Sigma_{1,{\rm e}}/\rho + E(\eta_{\mathrm{e}}\eta_{\mathrm{e}}^\T) \right\}^{-1}E(\phi_{\mathrm{e}}\eta_{\mathrm{e}}^\T)^\T,
	\end{equation}
	where $\Sigma_{1,{\rm e}}$ is the efficiency bound for $\beta$ in the external data. 
	\end{theorem}
    The only difference between efficiency bounds \eqref{bound ind} and \eqref{efficiency bound} is that $\Sigma_{1}$ in \eqref{efficiency bound} is replaced by $\Sigma_{1, {\rm e}}$ in \eqref{bound ind}. Note that $\Sigma_{1, {\rm e}} \leq \Sigma_{1}$ as $\Sigma_{1,{\rm e}}$ is the efficiency bound for $\beta$ in the external dataset and $\Sigma_{1}$ is the asymptotic variance of some RAL estimator of $\beta$. 
    The strict inequality   holds when $\tilde{\beta}$ is an inefficient estimator of $\beta$ based on the external   data. 
    Thus,  the bound with external individual data   is smaller or at least equal to that with external summary statistics. 
    Therefore, in practice it is best to report and to integrate an efficient  summary statistic, 
    which enjoys both the   ease in data collection, computation and 
    the  full use of the external study without efficiency loss.
    Otherwise,   integrating an inefficient summary statistic can lead to efficiency loss,  compared to  the situation when  external individual data are available.
    Although, the potential efficiency loss is less pronounced when the external study has a much larger sample size than the internal study, i.e., $\rho$ is large.}
Theorem \ref{efficiency theorem} also shows that external summary statistics bring no efficiency gain if $E(\phi_{\mathrm{e}}\eta_{\mathrm{e}}^\T) = 0$,
in which case,  knowing $\beta$ does not benefit   the   estimation of $\tau$.
This happens if $P_0$ factorizes as $P_0(Z)=f_1(Z)f_2(Z)$ and $\tau(P_0) = \tau(f_1), \beta(P_0) = \beta(f_2)$, i.e., $\tau$ and $\beta$ are functionals of variationally independent components of the internal data distribution.  

Applying Theorem \ref{efficiency theorem} to the estimation of the generalized linear model, we have the following proposition that provides a formal justification of the result conjectured by \citet{zhang2020generalized}. 

\begin{proposition}\label{glm}
Suppose $P_0=P_1$ and $E(Y\mid X_1,X_2) = g^{-1}(X_1^\T\tau  + X_2^\T \zeta)$ with $g$ being  the canonical link function.
Suppose in the external study    $g^{-1}(X_2^\T\beta)$ is used as  a working model for $E(Y\mid X_2)$ and $\beta$ is estimated by solving 
 estimating equation $\widehat{E}_1[X_2\{Y - g^{-1}(X_2^\T\beta) \} ] = 0$.
Here $\widehat{E}_1$ means the empirical mean in the external study.
Then the resultant estimator $\tilde \beta$ does not bring efficiency gain for estimating $\tau$.\end{proposition}

{
The proposed methods can be extended to integrate summary statistics from multiple external studies. 
Suppose there are  $S$  independent external studies. The $s$-th $(1\leq s\leq S)$ study has $m_s$ observations from population $P_s$ and provides summary statistics $\tilde{\beta}_{s}$ on a functional $\beta_{s}(P_s)$. Define $\tilde{\beta}_{[S]} = (\tilde{\beta}_{1}^{\T}, \dots, \tilde{\beta}_{S}^{\T})^{\T}$. Then, we have the following result.
\begin{theorem}\label{multi-source}
	Suppose $m_s/n\rightarrow \rho_s\in(0,\infty)$, $m_s^{1/2}\{\tilde{\beta}_{s} - \beta_{s}(P_s)\} \rightarrow N(0, \Sigma_s)$ for $s = 1,\dots, S$,  
	$(\beta_{1}^\T(P_1),\ldots, \beta_{S}^\T(P_S))^\T = (\beta_{1}^\T(P_0), \ldots, \beta_{S}^\T(P_0))^\T \eqqcolon \beta_{[S]}(P_0)$, and  regularity Condition \ref{cond: regular beta-tilde multiple} in Supplementary Material holds.
	Denote the efficient influence function for $\beta_{[S]}(P_0)$ by $\eta_{\mathrm{e}, [S]} = (\eta_{\mathrm{e},1}^\T, \ldots, \eta_{\mathrm{e},S}^\T)^\T$.
	Then  the data-fused efficiency bound for $\tau$ is  
	\[
	E(\phi_{\mathrm{e}}\phi_{\mathrm{e}}^\T) - E(\phi_{\mathrm{e}}\eta_{\mathrm{e}, [S]}^\T) \left\{ \Sigma_{[S]} + E(\eta_{\mathrm{e}, [S]}\eta_{\mathrm{e}, [S]}^\T ) \right\}^{-1}E(\phi_{\mathrm{e}}\eta_{\mathrm{e}, [S]}^\T )^\T,
	\]
	where $\Sigma_{[S]} = \mathrm{diag}(\Sigma_1/\rho_1, \ldots, \Sigma_S/\rho_S)$. 
\end{theorem}

The proof of Theorem~\ref{multi-source}  is analogous to that of  Theorem~\ref{efficiency theorem} and is omitted.
{
\begin{remark}\label{remark:multi-source}
	 Suppose the summary statistics $\tilde{\beta}_{S + 1}$ is available from the $(S+1)$-th external study with sample size $m_{S+1}$, $m_{S+1}/n\rightarrow \rho_{S+1}\in(0,\infty)$, $m_{S+1}^{1/2}\{\tilde{\beta}_{S+1} - \beta_{S+1}(P_{S+1})\} \rightarrow N(0, \Sigma_{S+1})$, and 
	$\beta_{S+1}(P_{S+1}) = \beta_{S+1}(P_{0})$. 
	Let $\eta_{\mathrm{e}, S+1}$ be the efficient influence function for $\beta_{S+1}(P_0)$ based on the internal data. In  Supplementary Material, we show that, compared to the efficiency bound in Theorem \ref{multi-source}, an additional summary statistic  $\tilde{\beta}_{\rm S+1}$   leads to efficiency gain of
	\begin{equation}\label{efficiency bound reduction}
		E(\phi_{\rm e}e_{S+1}^{\T})\{E(e_{S+1}e_{S+1}^{\T}) + \Sigma_{S+1}/\rho_{S+1}\}^{-1}E(\phi_{\rm e}e_{S+1}^{\T})^{\T},
	\end{equation}
	where $e_{S+1} = \eta_{{\rm e}, S+1} - E(\eta_{{\rm e}, S+1}\eta_{\mathrm{e}, [S]}^\T) \left\{ \Sigma_{[S]} + E(\eta_{\mathrm{e}, [S]}\eta_{\mathrm{e}, [S]}^\T ) \right\}^{-1}(\eta_{{\rm e}, [S]} + \epsilon_{[S]})$ is the residual of projecting $\eta_{{\rm e}, S+1}$ onto the linear space spanned by $\eta_{{\rm e},[S]} + \epsilon_{[S]}$ and $\epsilon_{[S]}$ is an independent random vector following $N\left(0, \Sigma_{[S]}\right)$.  
    From the perspective of study design, an estimate or approximation of \eqref{efficiency bound reduction} is useful for choosing the summary statistics for data fusion. 
\end{remark}
}
}

\subsection{An Efficient Data-fused Estimator}\label{sec:efficient estimation}

Let $\hat{\beta}_{\mathrm{e}}^{\sI}$ denote  the efficient estimator of $\beta$    and $\hat{\tau}_{\mathrm{e}}^{\sI}$ the efficient estimator of   $\tau$  based only  on internal individual data. Recall that $\widehat{\Sigma}_{\phi\eta}, \widehat{\Sigma}_{\eta\eta}$ are consistent estimators of $E(\phi_{\mathrm{e}}\eta_{\mathrm{e}}^\T), E(\eta_{\mathrm{e}}\eta_{\mathrm{e}}^\T)$ based on the internal data, respectively,  and $\widehat{\Sigma}_1$ is a consistent estimator of $\Sigma_1$.
Motivated by Theorem~\ref{convolution}, we propose the following data-fused estimator: 
\begin{equation}
\label{efficient}
\begin{aligned}
\hat{\tau}_{\mathrm{e}} &= \hat{\tau}_{\mathrm{e}}^{\sI} - \widehat{\Sigma}_{\phi\eta} \left(\widehat{\Sigma}_1/\rho + \widehat{\Sigma}_{\eta\eta}\right)^{-1}(\hat{\beta}_{\mathrm{e}}^{\sI} - \tilde{\beta}).
\end{aligned}
\end{equation}

\begin{theorem}\label{efficient estimator}
Under Assumptions \ref{beta}, \ref{well defined}, and regularity Condition \ref
{cond: regular internal} in Supplementary Material,  we have $n^{1/2}(\hat{\tau}_{\mathrm{e}} - \tau)$ is asymptotically normal with asymptotic variance equal to the semiparametric efficiency bound $B$    in  \eqref{efficiency bound}.
\end{theorem}

Theorem \ref{efficient estimator} shows that $\hat{\tau}_{\mathrm{e}}$   attains  the efficiency bound for estimating a general functional in semiparametric or nonparametric models
when both internal individual data and external summary statistics are available.
This generalizes previous   results \citep{zhang2020generalized} on parameter estimation in parametric models.
We refer to $\hat{\tau}_{\mathrm{e}}$  as the data-fused efficient estimator.
This estimator is at least as efficient as any RAL estimator using only internal data, and thus, resolves the efficiency paradox.
{ The estimator $\hat{\tau}_{\mathrm{e}}$  has a  closed form, which facilitates computation in many settings where the efficient influence functions $\phi_{\mathrm{e}}$ and $\eta_{\mathrm{e}}$ are readily available and easy to compute. In particular, the efficient influence functions are easy to handle in the examples presented in the main text and Supplementary Material, as well as in many other important applications \citep{tsiatis2006semiparametric}.
In addition, under a generic parametric model $f(y\mid x; \tau)$ for the conditional density of outcome $Y$ given covariates $X$ that includes many practically important models such as generalized linear models, and the models considered by \cite{chatterjee2016constrained}, \cite{zhang2020generalized}, \cite{zhang2022integrative}, \cite{zhai2022data}, and \cite{zhai2024integrating}, the efficient influence functions can be derived provided that $\tilde{\beta}$ is the solution of some external data-based estimating equation; see Section \ref{app: cond density} in Supplementary Material  for more details. 
However, we caution that the calculation of efficient influence functions   may be 
difficult in complicated semiparametric models such as 
those with missing covariates or complex censoring. Additional theoretical or computational efforts are required to apply the proposed method in these scenarios.
}

{
\begin{remark}
	The data-fused efficient influence function described after Theorem \ref{convolution} may involve complex nuisance functions, e.g., $\mu_{1}(X)$, $\mu_{0}(X)$, and $p(X)$ in Example \ref{causal}.
	Fortunately, it is Neyman orthogonal \citep{chernozhukov2018debiased}, provided that the efficient influence functions $\phi_{\rm e}$ and $\eta_{e}$ in the internal study are Neyman orthogonal---a condition met by many estimands \citep{chernozhukov2018debiased, kennedy2024semiparametric}. This enables the use of flexible machine learning methods such as neural networks to estimate nuisance functions without compromising the validity of inference, provided these estimators converge at appropriate rates (e.g., faster than $n^{-1/4}$) and  cross-fitting is employed. See Supplementary Material \ref{supp simulations} for simulation results when neural networks are used to estimate nuisance functions.
\end{remark}
}

Denote the asymptotic covariance of $(\hat{\tau}_{\mathrm{e}}^{\sI}, \hat{\beta}_{\mathrm{e}}^{\sI})$ by
\begin{equation*}
\Sigma = \left\{\begin{matrix}
E(\phi_{\mathrm{e}}\phi_{\mathrm{e}}^\T) & E(\phi_{\mathrm{e}}\eta_{\mathrm{e}}^\T)\\ 
E(\eta_{\mathrm{e}}\phi_{\mathrm{e}}^\T) & E(\eta_{\mathrm{e}}\eta_{\mathrm{e}}^\T)
\end{matrix}\right\}.
\end{equation*}
The estimator   $\hat{\tau}_{\mathrm{e}}$ in \eqref{efficient} and its asymptotic variance are determined once  we  obtain   $(\hat{\tau}_{\mathrm{e}}^{\sI}, \hat{\beta}_{\mathrm{e}}^{\sI}, \widehat{\Sigma})$ and  $(\tilde \beta, \widehat{\Sigma}_1)$, where $\widehat{\Sigma}$ is a consistent estimator of $\Sigma$. Specifically, let $\widehat{\Sigma}_{\phi\phi}$ be a consistent estimator of $E(\phi_{\mathrm{e}}\phi_{\mathrm{e}}^\T)$. The asymptotic variance of $\hat{\tau}_{\mathrm{e}}$ can be estimated by $\widehat{\Sigma}_{\phi\phi} -  \widehat{\Sigma}_{\phi\eta} \left(\widehat{\Sigma}_1/\rho + \widehat{\Sigma}_{\eta\eta}\right)^{-1}\widehat{\Sigma}_{\phi\phi}^{\T}$.
The estimators $(\hat{\tau}_{\mathrm{e}}^{\sI}$, $\hat{\beta}_{\mathrm{e}}^{\sI},\widehat{\Sigma})$ can be obtained with internal  individual data,
$\tilde \beta$ is available from external data, and  $\widehat{\Sigma}_1$  is routinely reported as summary statistics together with $\tilde \beta$. 
One can also consistently estimate $\Sigma_1$ using internal data if  $P_1$ is a marginal distribution of $P_0$ and the method for estimating $\tilde{\beta}$ is known. 
Otherwise,  one can use a  positive-definite working matrix $\Omega$ for  $\Sigma_1$ without compromising consistency of  $\hat{\tau}_{\mathrm{e}}$,
but in this case, there is no guarantee of efficiency gain from external summary statistics. 
Additional discussion on the choice of the working covariance matrix and the efficiency is provided in the proof of Proposition~\ref{CD estimator} in Supplementary Material.

If $\beta$ is the same functional as $\tau$, 
then $\hat{\tau}_\mathrm{e}$ reduces to 
\[
\frac{\hat{\tau}_{\mathrm{e}}^{\sI}/\widehat{\mathrm{var}}(\hat{\tau}_{\mathrm{e}}^{\sI}) + \tilde{\beta}/\widehat{\mathrm{var}}(\tilde{\beta})}{1/\widehat{\mathrm{var}}(\hat{\tau}_{\mathrm{e}}^{\sI}) + 1/\widehat{\mathrm{var}}(\tilde{\beta})},
\]
which is the well-known inverse variance weighted estimator in meta-analysis   \citep{lin2010relative}.
The estimator $\hat{\tau}_{\mathrm{e}}$  can be viewed as a calibration estimator where the external summary statistics $\tilde \beta$ are used to 
calibrate the internal data-based efficient estimator $\hat{\tau}_{\mathrm{e}}^{\sI}$. 
Calibration is a standard technique used in survey sampling for efficiency improvement with auxiliary information.
\cite{chen2000unified}, \citet{wang2015semiparametric} and  \citet{yang2020combining} considered calibration with validation data in the contexts of measurement error and confounding adjustment, where the validation dataset contains individual random samples from the  internal data.
In contrast, here we consider the situation where only summary statistics are available from the external   study and the external data
are not necessarily random samples from the internal population.

The estimator  $\hat{\tau}_{\mathrm{e}}$ can also be viewed as a generalization of the confidence density estimator \citep{liu2015multivariate}  which only employs summary statistics $(\hat{\tau}_{\mathrm{e}}^{\sI}, \hat{\beta}_{\mathrm{e}}^{\sI}, \widehat{\Sigma})$  and   $(\tilde \beta, \widehat{\Sigma}_1)$ to estimate $\tau$.

\begin{proposition}\label{CD estimator}
Given $(\hat{\tau}_{\mathrm{e}}^{\sI}, \hat{\beta}_{\mathrm{e}}^{\sI}, \widehat{\Sigma})$  and   $(\tilde \beta, \widehat{\Sigma}_1)$,
then for  some $\hat\beta$, 
\begin{equation}\label{IVW}
(\hat{\tau}_{\mathrm{e}},  \hat\beta) = \arg\min_{\tau, \beta}
\left\{ 
\left(\begin{array}{c}
\hat{\tau}_{\mathrm{e}}^{\sI} - \tau\\
\hat{\beta}_{\mathrm{e}}^{\sI} - \beta
\end{array}
\right)^\T
\widehat{\Sigma}^{-1}
\left(\begin{array}{c}
\hat{\tau}_{\mathrm{e}}^{\sI} - \tau\\
\hat{\beta}_{\mathrm{e}}^{\sI} - \beta
\end{array}
\right)
+ \rho(\tilde{\beta} - \beta)^\T  \widehat{\Sigma}_1^{-1}(\tilde{\beta} - \beta) \right\}.
\end{equation}
\end{proposition}
The confidence density approach of \citet{liu2015multivariate} concerns the estimation of  parameters in a parametric model with summary statistics
whose probability limit has a completely known functional relationship to the parameters of interest.
Here however, we consider the estimation of a general functional in semiparametric models and the availability of internal individual data of the internal population obviates the need to know the explicit functional relationship between the probability limit of the summary statistics and the parameter of interest.  
{ For illustration, we apply our estimation method to the causal inference problem in Example~\ref{causal}.}

\begin{example}[Continuation of Example \ref{causal}]\label{causal2}
Suppose in addition to  individual samples on $Z=(D,X,Y)\sim P_0$ in the internal  study,
we   also  have  the ordinary least squares estimate  $\tilde\beta$  of $\beta= \{E(V V^\T)\}^{-1}E(V Y)$ obtained from the linear regression of $Y$ on $V=(1,X^\T,D)^\T$  in the external study.
Suppose  $P_1=P_0$.
The efficient influence function for $\beta$ using only internal data is $\eta_{\mathrm{e}} = \{E(V V^\T )\}^{-1}V(Y - V^\T\beta)$.
Given $(\hat{\tau}_{\mathrm{e}}^{\sI},\phi_{\mathrm{e}})$ described in Example \ref{causal} and $\hat{\beta}_{\mathrm{e}}^{\sI}$ by regressing $Y$ on $V$ in the internal data, we have
\[
\hat{\tau}_{\mathrm{e}}=\hat{\tau}_{\mathrm{e}}^{\sI} - \frac{m}{m+n}\left[\widehat{E}\big\{\phi_{\mathrm{e}}(Y-V^\T\hat\beta_{\mathrm{e}}^{\sI})V\big\}\right]^\T\left[\widehat{E}\big\{(Y-V^\T\hat\beta_{\mathrm{e}}^{\sI})^2V^\T V\big\} \right]^{-1} \widehat{E}(VV^\T) (\hat{\beta}_{\mathrm{e}}^{\sI} - \tilde{\beta}).
\]
\end{example}

\section{Adaptive Fusion in the Presence of Population Heterogeneity}\label{sec:adaptive fusion}
\subsection{An Adaptive Fusion Estimator}
In practice,  there may well be heterogeneity between populations in different studies. In the presence of population heterogeneity, the external summary statistics may only be partially transportable or untransportable, i.e., Assumption 2 may fail for some or all the components of $\beta(\cdot)$. 
In this case,  the integration of summary statistics as in \eqref{efficient} will introduce bias.
Specifically, we have
$\hat{\tau}_{\mathrm{e}} - \tau \to  
E(\phi_{\mathrm{e}}\eta_{\mathrm{e}}^\T) \{\Sigma_1/\rho + E(\eta_{\mathrm{e}}\eta_{\mathrm{e}}^\T)\}^{-1}h$, 
in probability, where $h = \beta(P_{1}) - \beta(P_{0})$ is the heterogeneity parameter leading to  non-negligible   bias of $\hat{\tau}_{\mathrm{e}}$.
To mitigate this problem, we construct a robust estimator that can effectively use the  transportable  components of the external summary statistics  to improve the efficiency while 
keeping invulnerable to untransportable components.

Let $\mathcal{A}=\{j: \beta_{j}(P_0) = \beta_{j}(P_1), j=1,\ldots,q \}$ denote the set of transportable external summary statistics. 
If such a set is known a priori, an oracle estimator could be obtained by incorporating only this subset of external summary statistics, $\tilde{\beta}_\mathcal{A}$,  utilizing the efficient data-fusion method proposed in \eqref{efficient}.
Let $\hat{\tau}_{\mathrm{orc}} = \hat{\tau}_{\mathrm{e}}^{\mathcal{A}}$ denote such an oracle estimator and $B^{\mathcal{A}} = E(\phi_{\mathrm{e}}\phi_{\mathrm{e}}^\T ) - E(\phi_{\mathrm{e}}\eta_{\mathrm{e}, \mathcal{A}}^\T) \left\{ \Sigma_{1, \mathcal{A}}/\rho + E(\eta_{\mathrm{e}, \mathcal{A}}\eta_{\mathrm{e}, \mathcal{A}}^\T ) \right\}^{-1}E(\phi_{\mathrm{e}}\eta_{\mathrm{e}, \mathcal{A}}^\T )^\T$ denote its asymptotic variance.
In practice, $\mathcal A$ is unknown. One can first select the transportable components based on $\tilde{\beta} - \hat{\beta}_{\mathrm{e}}^{\sI}$ and then construct the data-fused estimator using the selected component. 
However, such a select-and-fuse procedure leads to an estimator that is discontinuous with respect to the observed data. The discontinuity is undesirable and can diminish the numerical stability of the procedure in practice \citep{fan2001variable}.
To resolve the problem, we propose an adaptive fusion estimator that is continuous with respect to the observed data and shares the same asymptotic distribution as the oracle estimator.
Specifically, let 
\[\hat{a}_{j}^{2} = \max\left\{0, 1 - \lambda|\tilde{\beta}_j - \hat{\beta}_{\mathrm{e},j}^{\sI}|^{\alpha}\right\}\]
for $j = 1,\dots, q$ where $\lambda, \alpha > 0$ are tuning parameters. Let $\widehat{A} = {\rm diag}\{\hat{a}_{1}^{2}, \dots, \hat{a}_{q}^{2}\}$ and $\hat{a} = (\hat{a}_{1}, \dots, \hat{a}_{q})$. The adaptive fusion estimator is defined as
\begin{equation}\label{adaptive fusion formula}
	\hat{\tau}_\mathrm{adf} = \hat{\tau}_{\mathrm{e}}^{\sI} - \widehat{\Sigma}_{\phi\eta}\widehat{A} \left\{\left(I - \widehat{A} + \hat{a}\hat{a}^{\T}\right)\odot\left(\widehat{\Sigma}_1/\rho + \widehat{\Sigma}_{\eta\eta}\right)\right\}^{-1}(\hat{\beta}_{\mathrm{e}}^{\sI} - \tilde{\beta}),
\end{equation} 
where $\odot$ denotes the element-wise product and $I$ is the identity matrix. The adaptive fusion estimator $\hat{\tau}_{\rm adf}$ leverages $\{\hat{a}_{j}\}_{j=1}^{q}$ to adaptively control the incorporation of the external summary statistics.
Under regularity conditions, the $\hat{\tau}_{\rm adf}$ possesses the following ``oracle" property.

\begin{theorem}\label{adaptive fusion}
Under Assumption~\ref{beta}, if $\lambda \to \infty$, $\lambda n^{-\alpha/2} \to 0$, and regularity Condition \ref
{cond: regular internal} in Supplementary Material, we have $\hat{\tau}_\mathrm{adf}$ has  the same asymptotic distribution as $\hat{\tau}_{\mathrm{orc}}$  and 
$n^{1/2}(\hat{\tau}_{\mathrm{adf}} - \tau) \rightarrow N(0, B^{\mathcal{A}})$.
\end{theorem}

Theorem \ref{adaptive fusion} shows that the adaptive fusion estimator $\hat{\tau}_\mathrm{adf}$ retains consistency and asymptotic normality even if  $\tilde{\beta}$ contains untransportable components,
and it is asymptotically as efficient  as  the oracle estimator $\hat{\tau}_{\mathrm{orc}}$. 
The asymptotic variance $B^{\mathcal{A}}$ can be consistently estimated by
\[
   \widehat{\Sigma}_{\phi\phi} -  \widehat{\Sigma}_{\phi\eta}\widehat{A} \left\{\left(I - \widehat{A} + \hat{a}\hat{a}^{\T}\right)\odot\left(\widehat{\Sigma}_1/\rho + \widehat{\Sigma}_{\eta\eta}\right)\right\}^{-1}\widehat{A}\widehat{\Sigma}_{\phi\eta}^{\T}.
\]

\citet{zhai2022data} have previously proposed a  penalized constrained maximum likelihood method for the fusion of  internal individual data and   possibly untransportable summary statistics;
however, they focused on inference about parameters in a parametric conditional density model and ignored the uncertainty of external summary statistics, which may lead to efficiency loss as discussed at the end of Section \ref{sec: classic theory}.
In contrast, our method applies to a general functional in semiparametric models and accounts for the uncertainty of external summary statistics,
which can achieve the semiparametric efficiency bound of combining individual data and summary statistics. 
In addition, our estimator enjoys a closed form,  which bypasses sophisticated programming and computation and can be easily implemented { provided that all the efficient influence functions in the estimator  are available and easy-to-compute. Recently, \cite{zhai2024integrating} modified the method of \cite{zhai2022data} and proposed the dPCML method to account for the uncertainty of external summary statistics while accommodating untransportable summary statistics under the conditional density model. However, it does not achieve the semiparametric efficiency bound in general in the presence of untransportable summary statistics. Please refer to the discussions before Remark \ref{remark: taylor} in Section \ref{app: cond density} in Supplementary Material for more details.}  

For the choice of the tuning parameters, we set $\alpha = 4$ and $\lambda = cn^{1/2}$, which meet the requirement of Theorem \ref{adaptive fusion}. 
As a thumb of rule, we adopt the cross-validation method to choose the constant term $c$.
Specifically, we split the full internal data into $K\geq 2$ balanced subsets. Given a candidate set $\mathcal{C}$ for $c$, we propose to select $c$ using the cross-validation procedure summarized in Algorithm \ref{algo: tuning}. We set $K = 3$ and $\mathcal{C} = \{1/5, 1/4, \dots, 1, \dots, 5\}$ in the simulation and real data analysis.
\begin{algorithm}
	\For{$c\in \mathcal{C}$}{
		\For{$k = 1, \dots, K$}
		{
			Compute the internal data-based efficient estimator $\hat{\tau}_{\mathrm{e}, (k)}^{\sI}$ using the observations in the $k$-th subset \;
			Compute the adative fusion estimator $\hat{\tau}_{\mathrm{adf}, (k)}^{(c)}$ based on the external summary statistics $\tilde{\beta}$ and all internal data except for the $k$-th subset using the tuning parameter $\lambda = cn^{1/2}$\;
		}
	}
	Select 
	\[
	c_{*} = \mathop{\arg\min}_{c\in \mathcal{C}} \frac{1}{K}\sum_{k=1}^{K}\left\{\hat{\tau}_{\mathrm{e}, (k)}^{\sI} - \hat{\tau}_{\mathrm{adf}, (k)}^{(c)}\right\}^{2}.
	\]
	\caption{Cross validation method for selecting $c$.}\label{algo: tuning}
\end{algorithm}
 
\subsection{A Re-bootstrap Procedure to Improve Finite Sample Coverage}\label{subsec: reboot}
Theorem \ref{adaptive fusion} establishes the asymptotic distribution of  $\hat{\tau}_{\rm adf}$. The result is obtained by treating the difference $h_j=\beta_{j}(P_{1}) - \beta_{j}(P_{0})$ as fixed for $j = 1,\dots, q$ when the sample size $n$ is large. Thus, the difference $h_j$, if non-zero, asymptotically dominates the estimation error of $\tilde{\beta}_{j} - \hat{\beta}_{\mathrm{e}, j}^{\sI}$ as $n \to \infty$. This dominance ensures that one can consistently determine whether the summary statistics are transportable or not, which is essential for the oracle property in Theorem \ref{adaptive fusion}. However, the asymptotic distribution may not accurately approximate the actual distribution of $\hat{\tau}_{\rm adf}$ when $h_j$ is of a similar magnitude as the estimation error of $\tilde{\beta}_{j} - \hat{\beta}_{\mathrm{e}, j}^{\sI}$ for some $j \in \{1, \dots, q\}$, i.e., $|h_{j}| \asymp n^{-1/2}$. 
{ In real data fusion problems, the observed difference $\tilde{\beta}_{j} - \hat{\beta}_{\mathrm{e}, j}^{\sI}$ is usually of the similar magnitude as its standard error. For example, $\tilde{\beta} - \hat{\beta}_{\mathrm{e}}^{\sI} \approx -0.023$ and its standard error is approximately $0.054$ in the real data example considered in Section \ref{sec: real data}. In this case, it is possible that the mean of $\tilde{\beta}_{j} - \hat{\beta}_{\mathrm{e}, j}^{\sI}$ is of the same magnitude as its standard error, i.e., $h_{j} \asymp n^{-1/2}$.}
The above scenario, referred to as moderate heterogeneity, may lead to undercoverage of the confidence intervals based on the asymptotic distribution in Theorem \ref{adaptive fusion}. See Section \ref{subsec: moderate hetero} for a numerical illustration of this issue. We provide theoretical analyses of this issue in Section \ref{app: asy trans} of Supplementary Material.

In this section, we propose a re-bootstrap procedure to mitigate the undercoverage issue in finite samples. 
We aim to construct the confidence interval for $\tau_{j}$ for some $j = 1,\dots, p$ based on the estimator $\hat{\tau}_{\mathrm{adf}, j}$. For $j = 1,\dots, p$ and $0 < z < 1$, let  $q_{j}(z; h)$ denote the  $z$-quantile of the distribution of $\hat{\tau}_{\mathrm{adf}, j} - \tau_{j}$, which may depend on the heterogeneity parameter $h = (h_{1},\dots,h_{q})^{\T}$.  Consequently, $[\hat{\tau}_{\mathrm{adf},j} - q_{j}(1 - z/2; h), \hat{\tau}_{\mathrm{adf}, j} - q_{j}(z/2; h)]$ forms a valid $1 - z$ confidence interval for $\tau_{j}$. Since $\hat{\tau}_{\mathrm{adf}, j} - \tau_{j}$ is a function of $\hat{\tau}_{\mathrm{e}}^{\sI} - \tau$ and $\hat{\beta}_{\mathrm{e}}^{\sI} - \tilde{\beta}$, the quantile $q_{j}(z; h)$ is thus a functional of the distribution of these two quantities. The moderate heterogeneity does not invalidate the normal approximation for $\hat{\tau}_{\mathrm{e}}^{\sI} - \tau$ and $\hat{\beta}_{\mathrm{e}}^{\sI} - \tilde{\beta}$, but it may affect the mean of $\hat{\beta}_{\mathrm{e}}^{\sI} - \tilde{\beta}$, which is approximately equal to $h$. Therefore, the quantile $q_{j}(z; h)$ can be estimated by the bootstrap quantile $\hat{q}_{j}(z; h)$, which is obtained based on the bootstrap counterparts of $\hat{\tau}_{\mathrm{e}}^{\sI} - \tau$ and $\hat{\beta}_{\mathrm{e}}^{\sI} - \tilde{\beta}$ generated from a joint normal distribution that depends on $h$. See Section \ref{app: detail reboot} of Supplementary Material for more details. 
The remaining challenge for inference is the unknown heterogeneity parameter $h$.
Although $\tilde{\beta} - \hat{\beta}_{\mathrm{e}}^{\sI}$ is a $\sqrt{n}$-consistent estimator of $h$, the estimation error of $\tilde{\beta} - \hat{\beta}_{\mathrm{e}}^{\sI}$ has a comparable scale to the quantile $q_{j}(z, h)$ and hence non-negligible. To improve the robustness against this estimation error, we propose to calculate bootstrap quantiles under multiple reasonable candidate values for the heterogeneity parameter and use the most conservative quantile to construct the confidence interval. To generate these candidate values, we sample from the asymptotic confidence distribution $N(\tilde{\beta} - \hat{\beta}_{\mathrm{e}}^{\sI}, \widehat{\Sigma}_1 / m + \widehat{\Sigma}_{\eta\eta}/n)$ of $h$, which is induced by the asymptotic likelihood function of $h$ as suggested by \citet{xie2013confidence}, and make some calibration to the sampled heterogeneity parameters. Further details are deferred to Section \ref{app: detail reboot} of Supplementary Material. The resulting candidate heterogeneity parameters are denoted as $\hat{h}_{\mathrm{cal}}^{(r)}$ for $r = 1,\dots, \bar{r}$, where $\bar{r}$ is a user-specified integer.  The confidence interval for $\tau_{j}$ is then constructed as 
$\left[\hat{\tau}_{\mathrm{adf}, j} - \max_{r = 1,\dots, \bar{r}}\hat{q}_{j}(1 - z / 2; \hat{h}_{\mathrm{cal}}^{(r)}), \hat{\tau}_{\mathrm{adf}, j} - \min_{r = 1,\dots, \bar{r}}\hat{q}_{j}(z / 2; \hat{h}_{\mathrm{cal}}^{(r)})\right]$.

The above procedure constructs the confidence interval using the most conservative quantile derived from multiple candidate values. The quantiles $q_{j}(z/2; h)$ and $q_{j}(1 - z/2; h)$ are continuous functions of $h$ due to the continuity of the adaptive fusion estimator with respect to the observed data. If the number of candidate values $\bar{r}$ is large, then with high probability, there exists some candidate value that is extremely close to the true heterogeneity parameter $h$ \citep{guo2023causal}, making the resulting quantiles very close to the true quantiles. Thus, the coverage rate can be ensured by adopting the most conservative quantile among those derived from $\bar{r}$ different candidate values. 
On the other hand, the candidate values are designed to lie within a small neighborhood of $h$, ensuring that the quantiles calculated from these values are likely to be similar. This prevents the confidence interval from being overly conservative. The desired properties of the re-bootstrap procedure are confirmed by our numerical results in Section \ref{subsec: moderate hetero}. 
{ The computation cost of the re-bootstrap procedure mainly arises from simulating bootstrap quantiles under multiple candidate heterogeneity parameters. The computation complexity of the re-bootstrap procedure grows linearly in the number of candidate values $\bar{r}$, which is computationally feasible given that $\bar{r}$ is not required to be very large. Simulations in Supplementary Material \ref{supp simulations} show that the resulting confidence interval  only changes slightly with different choices of $\bar{r}$ ($10, 20, 50$) under multiple sample sizes.}

\section{Numerical Experiments}\label{sec: numerical}
\subsection{Simulation with Transportable Summary Statistics}\label{subsec: sim transportable}
We conduct simulation studies to evaluate the performance of the proposed estimators.
We consider different scenarios with transportable and untransportable external summary statistics, respectively.
In Scenario I, a triplet of treatment $D$, outcome $Y$ and covariate $X$ are generated as follows in both the internal and external   data:
\[
\begin{aligned}
&X\sim N(0,0.6),\quad \pr(D=1\mid X)= \text{expit}(1 - X/2),\\
&Y=1 + X + DX^2 + D\varepsilon_1 + (1-D)\varepsilon_0,\quad (\varepsilon_0, \varepsilon_1) \ind (X, D),
\end{aligned}
\]
where $\varepsilon_1\sim N(0,4)$, $\varepsilon_0\sim N(0,0.5)$  and $\text{expit}(x) = 1/\{ 1 + \exp(-x) \}$. 
{ We consider two internal sample sizes $n = 200$ and $500$ and the external sample size $m$ increases from $200$ to $2000$.}
The functional of interest $\tau$ is the treatment effect of $D$ on $Y$.
Following Example \ref{causal2}  we use the internal individual data and the ordinary least squares estimate  $\tilde\beta$  obtained by regressing $Y$ on $(1, X^\T, D)^\T$ in the external study for estimating $\tau$. 
We implement four estimators,  (i) INT: $\hat{\tau}_{\mathrm{e}}^{\sI}$ using only internal data;
 (ii)  PRM:  the primitive estimator ignoring  uncertainty of  $\tilde{\beta}$;
 (iii) EFF: the   data-fused efficient estimator $\hat{\tau}_{\mathrm{e}}$;
 (iv) KNW: the efficient estimator knowing the true value of $\beta$.
The GIM estimator by \cite{zhang2020generalized} does not apply to the causal effect functional, and we do not implement it.

\begin{figure}[h]
\centering
\subcaptionbox{$n = 200$}{
	\includegraphics[width=0.23\textwidth]{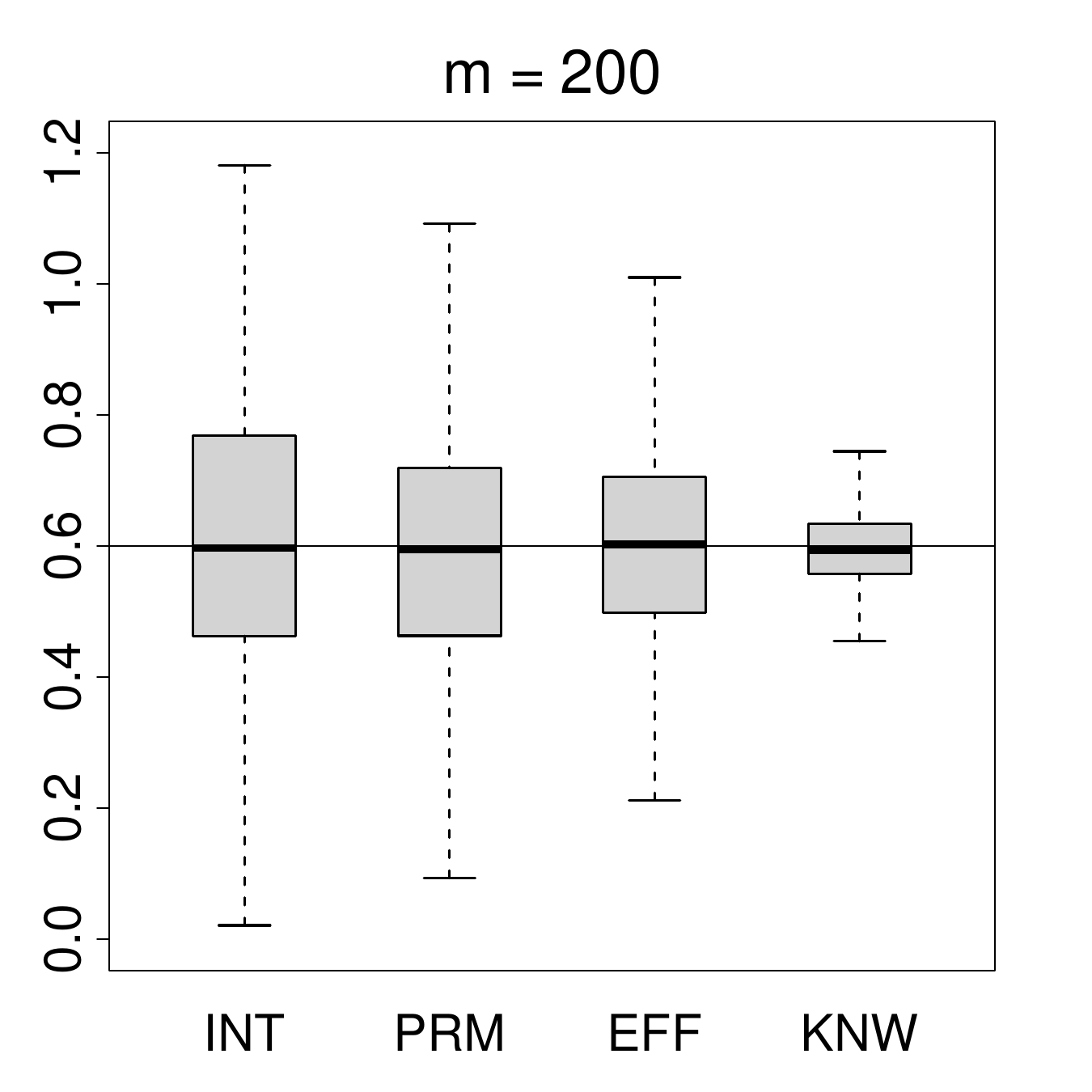}
	\includegraphics[width=0.23\textwidth]{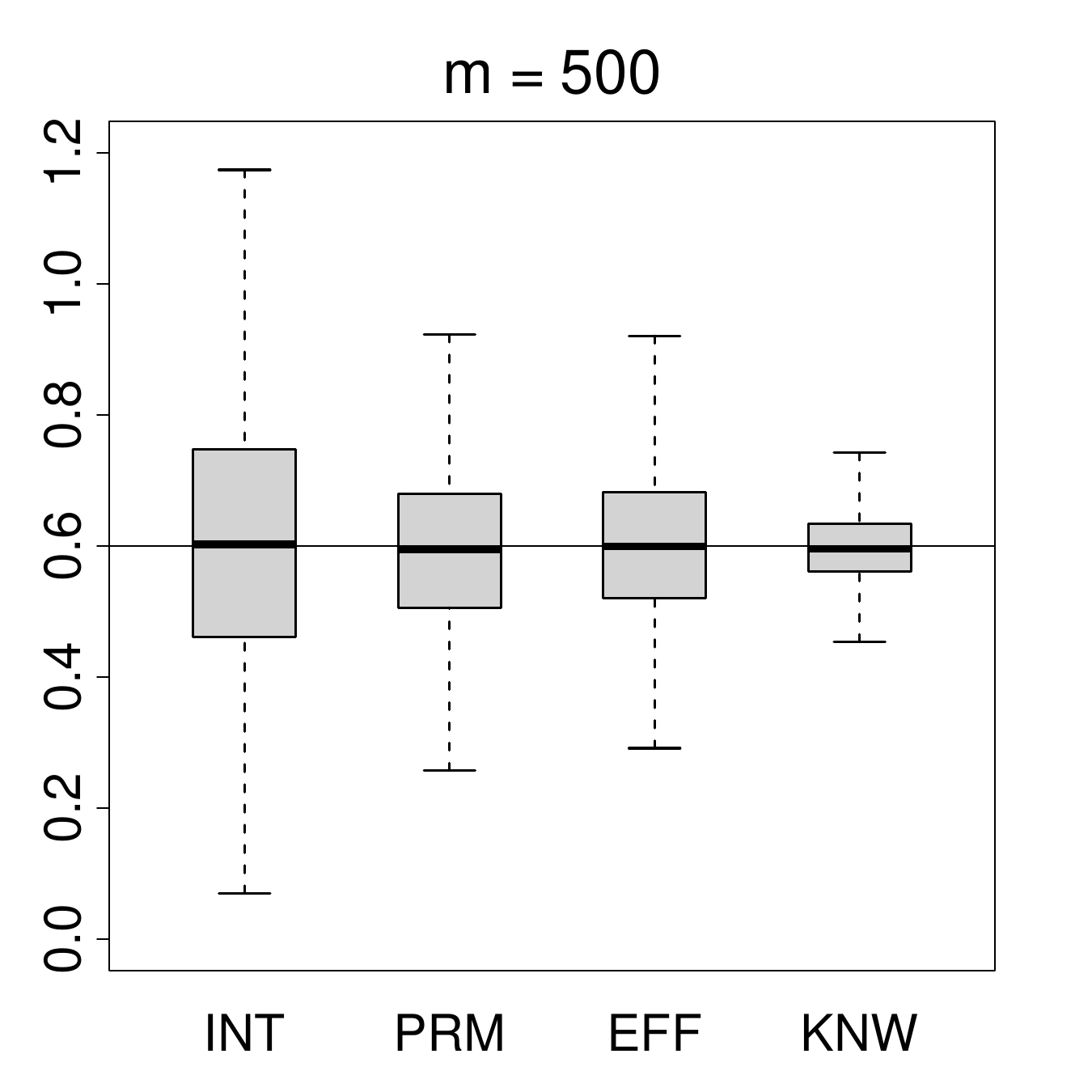}   
	\includegraphics[width=0.23\textwidth]{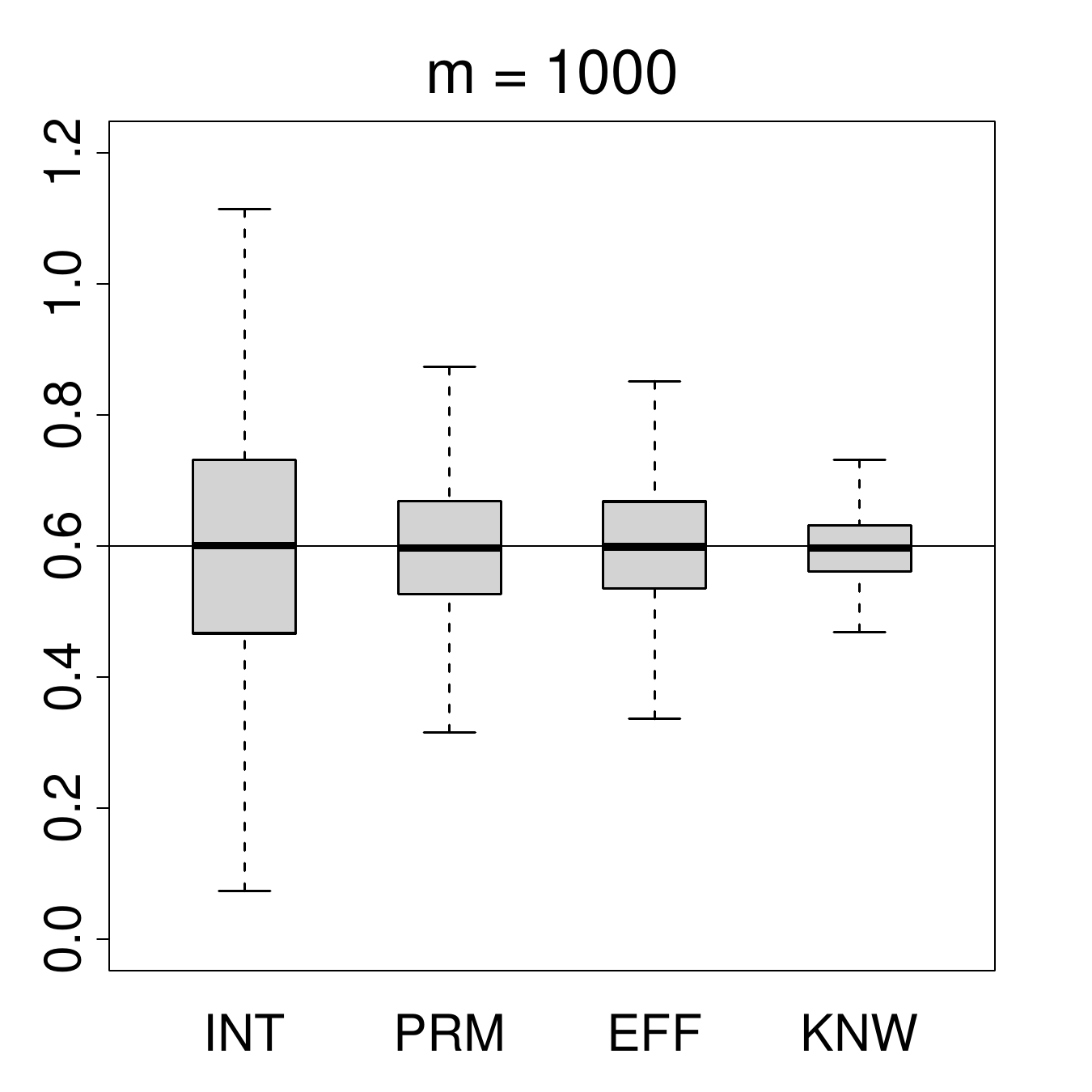}
	\includegraphics[width=0.23\textwidth]{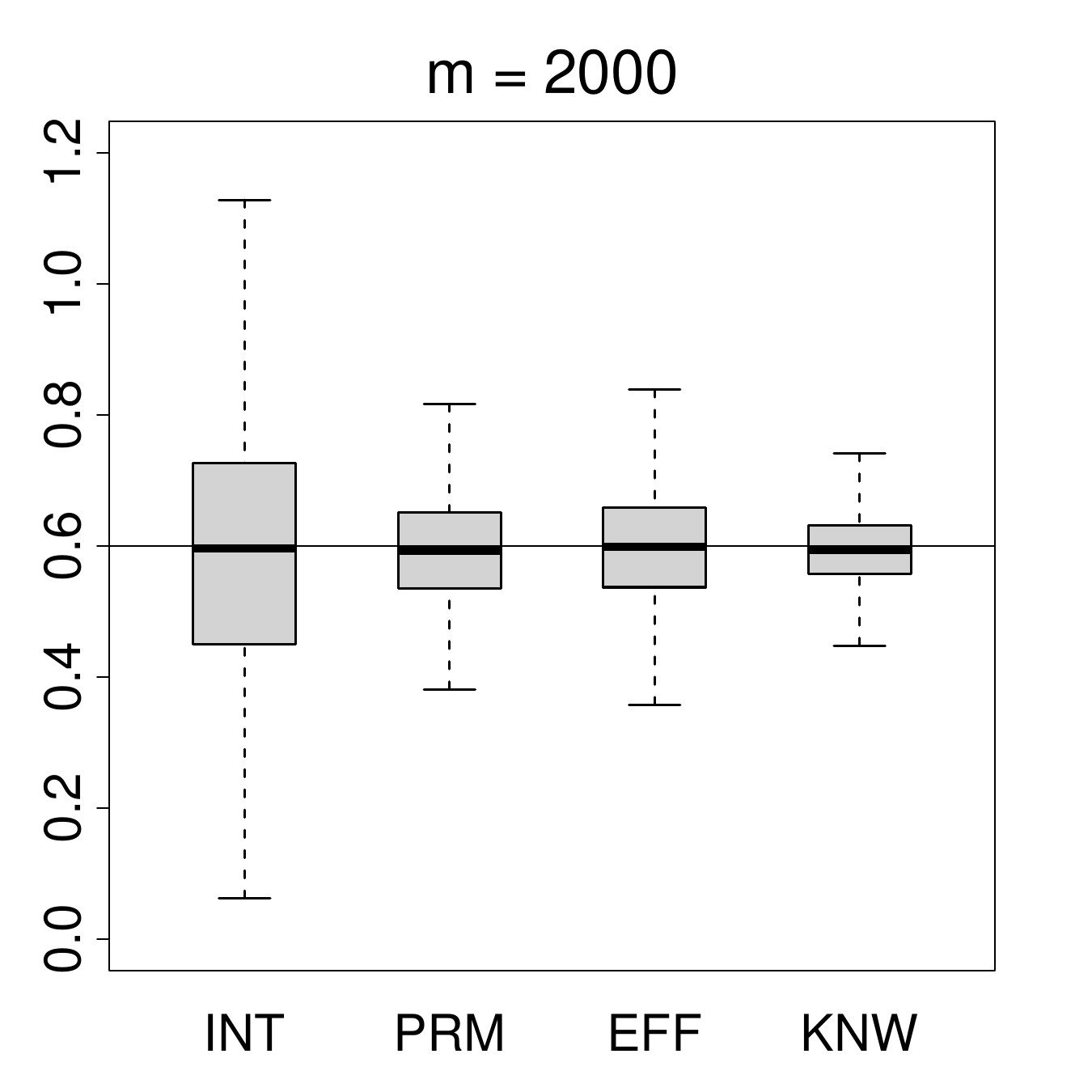}}  
\subcaptionbox{$n = 500$}{
  \includegraphics[width=0.23\textwidth]{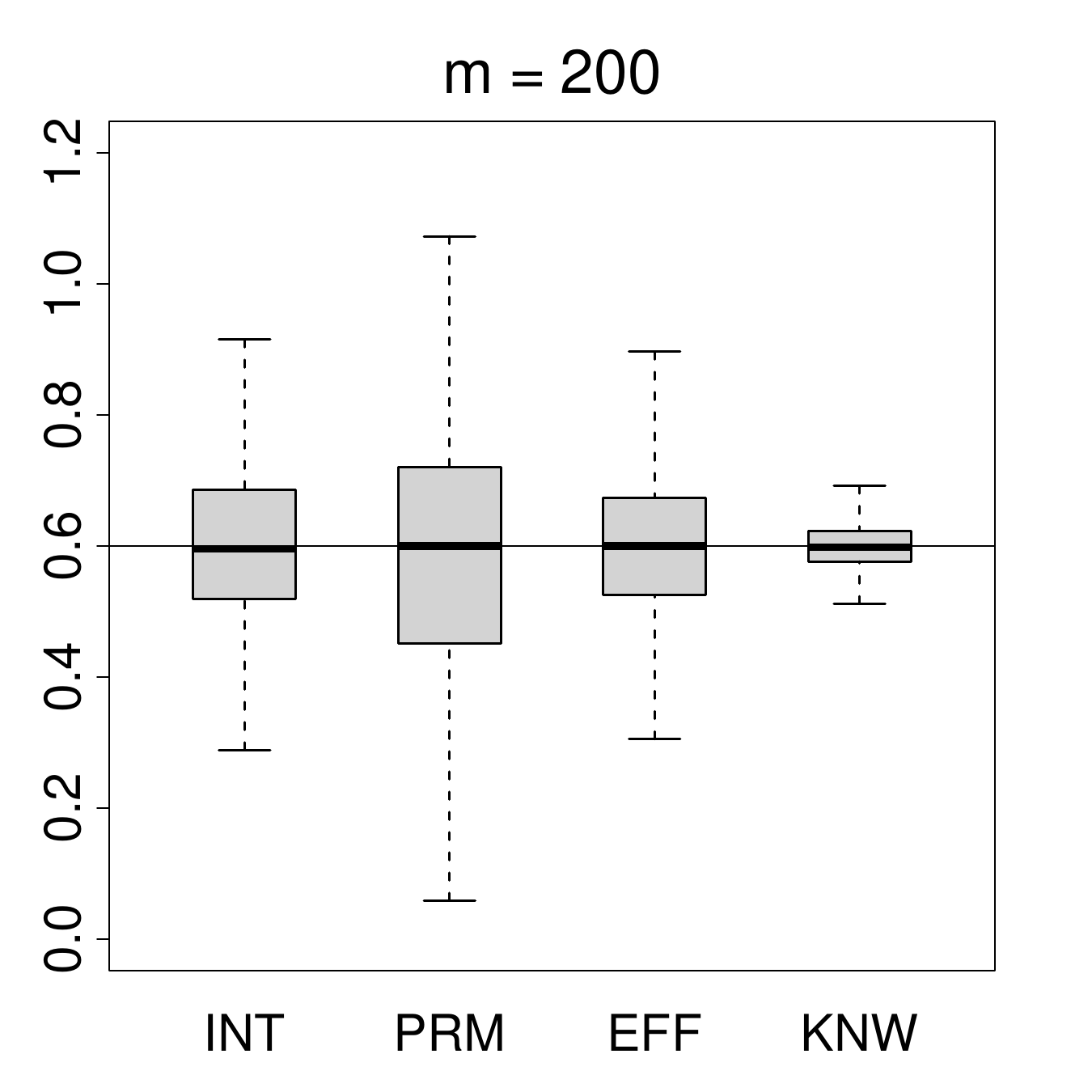}
\includegraphics[width=0.23\textwidth]{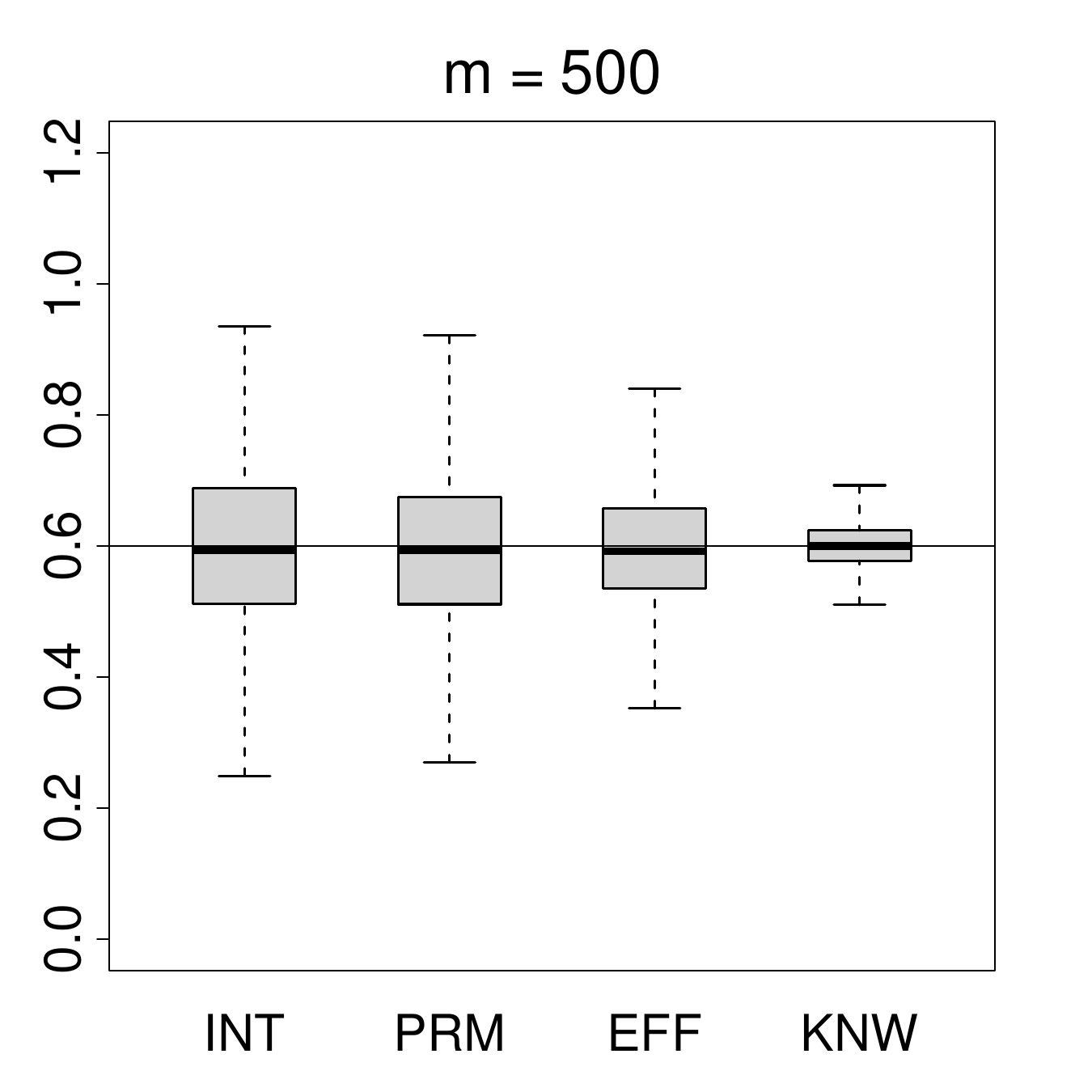}   
  \includegraphics[width=0.23\textwidth]{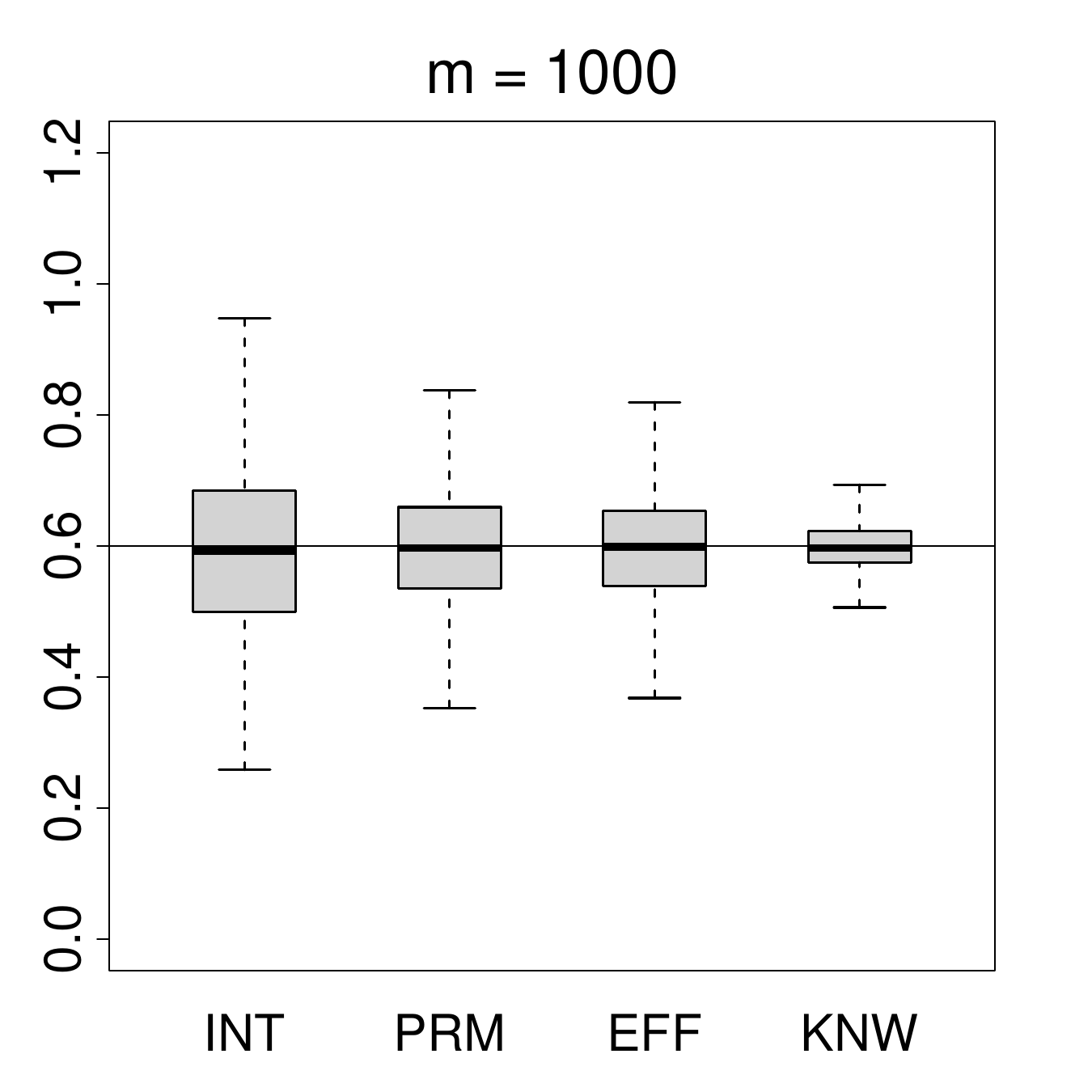}
  \includegraphics[width=0.23\textwidth]{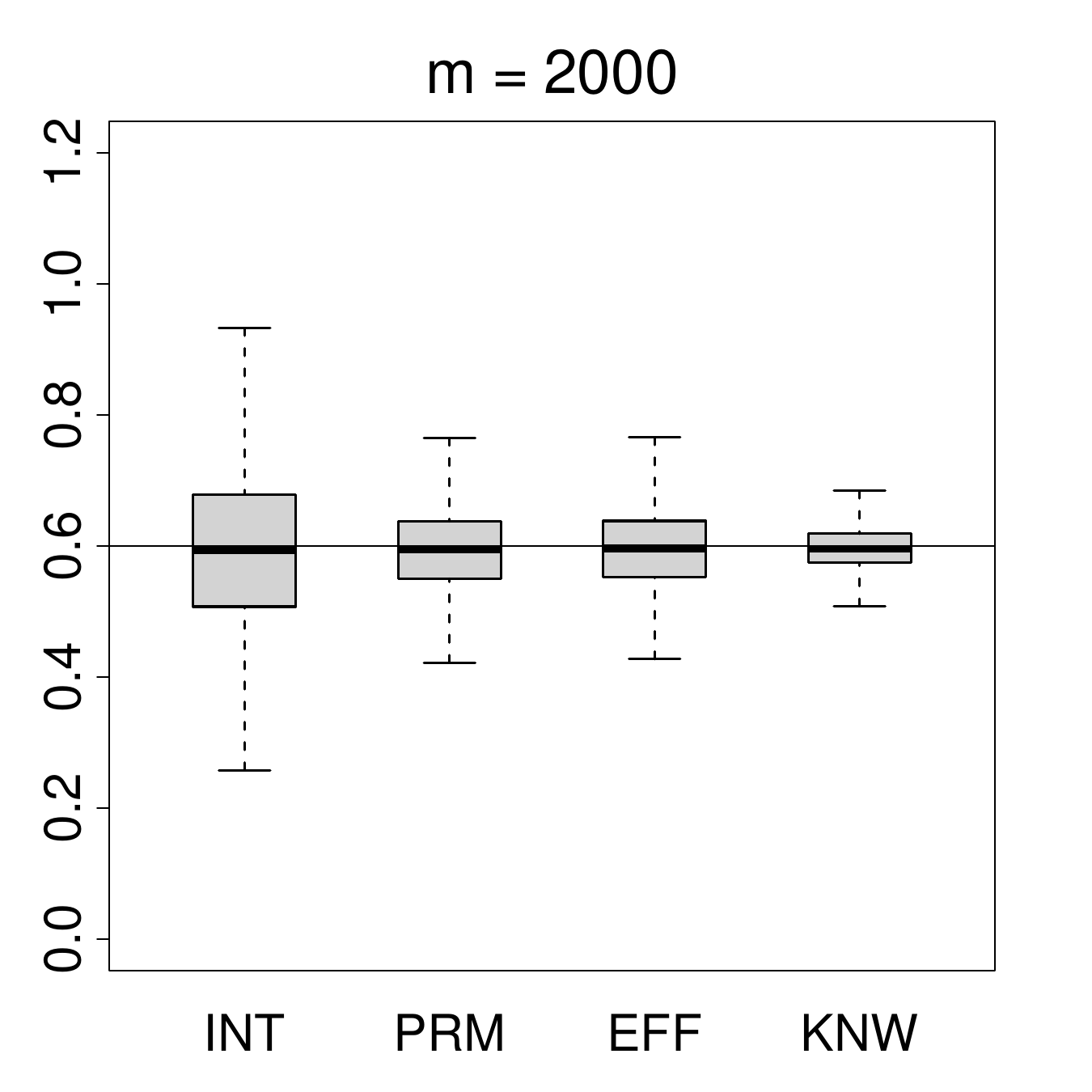}}  
\caption{Boxplots of various estimators in Scenario I. The horizontal line marks the true value.}
\label{fig1}
\end{figure}

We replicate 1000 simulations.
Figure~\ref{fig1} shows the boxplot of different estimators. Table~\ref{MSE1}
shows root mean squared error (RMSE), average standard error and coverage probability of the four estimators under different external sample sizes. 
The PRM estimator improves  efficiency against $\hat{\tau}_{\mathrm{e}}^{\sI}$ only  if the external sample size   is large relative to that of internal data ($m > n$), 
otherwise (e.g., when $n = 500$ and $m = 200$)  efficiency loss emerges.
The EFF estimator outperforms the INT and PRM estimators and ensures efficiency gain under all sample sizes.
The EFF estimator is less efficient than the KNW estimator, but the KNW estimator is not feasible in practice because one does not know $\beta$. 
{ The confidence interval based on the infeasible KNW estimator has the undercoverage issue when $n = 200$, probably due to the gap between the asymptotic and finite-sample distributions when $n$ is small. The undercoverage issue is mitigated when $n = 500$.}
For all other methods, coverage probabilities of the $95\%$ confidence intervals are close to the nominal level.

\begin{table}[h]
\centering
\caption{RMSE, average standard error (ASE) and coverage probability of 95\% confidence interval (CP) for Scenario I. All numbers are multiplied by 100}\label{MSE1}
\resizebox{\textwidth}{!}{
\begin{tabular}{cccccccccccccc}
\toprule
&&\multicolumn{3}{c}{$m=200$}    &\multicolumn{3}{c}{$m=500$}  & \multicolumn{3}{c}{$m=1000$}  & \multicolumn{3}{c}{$m=2000$}   \\
\midrule
\multirow{5}{*}{$n = 200$}
&     &  RMSE    &  ASE & CP      &  RMSE     &  ASE   &  CP    &  RMSE    &  ASE   &  CP   &  RMSE   &  ASE  &  CP   \\  
&INT  &  21.33 &	20.24 &	94.4 &	20.70 &	20.23 &	94.6 &	20.16 &	20.27 &	94.5 &	19.75 &	20.24 &	95.8\\
&PRM  &  18.97 &	20.33 &	96.3 &	13.08 &	13.65 &	95.1 &	10.32 &	10.5 &	94.8 &	8.67 &	8.43 &	94.2\\
&EFF  &  15.37 &	14.90 &	93.6 &	12.01 &	11.89 &	94.1 &	10.09 &	9.88 &	94.3 &	8.69 &	8.24 &	94.2\\
&KNW  &  6.14 &	5.63 &	87.2 &	6.09 &	5.65 &	86.4 &	5.69 &	5.73 &	87.9 &	6.10 &	5.71 &	88.5\\
\midrule
\multirow{5}{*}{$n = 500$}
&&  RMSE    &  ASE & CP      &  RMSE     &  ASE   &  CP    &  RMSE    &  ASE   &  CP   &  RMSE   &  ASE  &  CP   \\    
&INT  &  12.51 &	12.83 &	95.5 &	13.06 &	12.86 &	94.8 &	13.11 &	12.85 &	95.1 &	12.94 &	12.83 &	93.9\\
&PRM  &  19.09 &	19.76 &	96.2 &	12.32 &	12.88 &	96.4 &	9.48 &	9.43 &	95.2 &	6.65 &	7.07 &	96.1\\
&EFF  &  10.90 &	11.00 &	94.8 &	9.53 &	9.42 &	94.8 &	8.22 &	7.93 &	94.3 &	6.40 &	6.51 &	95.2 \\
&KNW  &  3.54 &	3.38 &	91.9 &	3.57 &	3.46 &	91.0 &	3.54 &	3.42 &	91.0 &	3.37 &	3.37 &	92.6\\
\bottomrule
\end{tabular}
}
\end{table}
{ We conduct additional simulations that use neural networks and cross-fitting to construct the data-fused efficient estimator. The results using neural networks are comparable to the results using correctly specified parametric models for both the propensity score model and the outcome regression model. Please refer to Section \ref{supp simulations} in Supplementary Material for more details.}

\subsection{Simulation with Partially Transportable Summary Statistics}
In Scenario II,  we consider data fusion with partially transportable external summary statistics.
We generate $(Y, X_1, X_2)$  in  the internal data and $(Y,X_1, \tilde X_2)$ in the external data as follows:
\[
\begin{aligned}
Y &= X_1 \tau_1 + X_2\tau_2 + \varepsilon_1, \quad  (X_1, X_2)^\T \sim N\left\{\begin{pmatrix}
0 \\ 
0
\end{pmatrix}, \begin{pmatrix}
1 & 0.6\\ 
0.6 & 1
\end{pmatrix}\right\},\\  
\tilde{X}_2 &= X_2 +  \varepsilon_2, \quad  \varepsilon_1\sim N(0,4),  \quad \varepsilon_2\sim N(0,1).
\end{aligned}
\]
where $\tilde{X}_2$ in the external data is viewed as a surrogate of $X_2$ with measurement error. The internal sample size is $n=500$ and the external sample size is $m = 2000$.
Let  $\tilde \beta=(\tilde \beta_1,\tilde \beta_2)$ be the ordinary least squares estimate obtained from the external data by regressing $Y$ on $X_1$ 
and $\tilde{X}_2$ separately. 
We use the internal individual data and  $\tilde \beta$ to estimate $\tau=(\tau_1,\tau_2)^\T$.
We implement five estimation methods: (i) INT: $\hat{\tau}_{\mathrm{e}}^{\sI}$ using only internal data; 
(ii) ORC: the oracle estimator using internal data and only   $\tilde{\beta}_1$; 
(iii) ADF: the adaptive fusion estimator $\hat{\tau}_{\mathrm{adf}}$; 
(iv) EFF: the efficient estimator $\hat{\tau}_{\mathrm{e}}$ in \eqref{efficient} using both $\tilde \beta_1$ and $\tilde \beta_2$;
(v) GIM:  the estimator of  \cite{zhang2020generalized}.

We replicate 1000 simulations.
Figure~\ref{fig:biased} shows the boxplot of different estimators. Table~\ref{MSE2 biased}  shows the RMSE, average standard error, and coverage probability of different estimators.
The GIM and EFF estimators exhibit large RMSE due to the inclusion of the untransportable component of the summary statistics. 
In contrast,    the ADF estimator can adaptively select and use the transportable component of the external summary statistics; as a result,  it performs similarly to the oracle estimator, both showing negligible bias. 
Besides,   the ADF  estimator of $\tau_1$   has smaller variance than the INT estimator while the  ADF  estimator of $\tau_2$  does not enjoy the efficiency gain, which is consistent with our analysis in Example~\ref{linear example}. 
Coverage probabilities of the 95\% confidence intervals are close to the nominal level for INT, ORC and ADF estimators; while for EFF and GIM estimators, the coverage probabilities are close to zero due to the bias introduced by the untransportable component.

We also evaluate the performance of  these estimators when the external data have accurate measurements of $X_2$, 
in  which case,   $\tilde{\beta}_2$  is also transportable. 
Figure~\ref{fig:biased2} shows the boxplot of different estimators. Table~\ref{MSE2 transportable} shows the RMSE, average standard error and coverage probability of the five estimators.
In this setting, the   ORC, ADF, EFF, and GIM estimators exhibit similar performance,  indicating that the ADF estimator has minimal efficiency loss when all the components are transportable.
Overall, we recommend the ADF estimator because it is statistically efficient, computationally convenient and empirically robust against untransportable external summary statistics.

\begin{minipage}{0.45\textwidth}
\begin{table}[H]
\centering
\caption{ RMSE, ASE and CP for Scenario II with partially transportable summary statistics. All numbers are multiplied by 100}\label{MSE2 biased}
\resizebox{1\textwidth}{!}{
\begin{tabular}{ccccccc}
	\toprule
     & \multicolumn{3}{c}{$\tau_1$}       & \multicolumn{3}{c}{$\tau_2$}  \\ 
     \midrule
     &   RMSE    & ASE  &  CP  &  RMSE      &  ASE  & CP        \\
INT &   11.14 &	11.17 &	94.0 &	11.20 &	11.18 &	95.3\\
ORC &   8.33 &	8.32 &	94.8 &	11.20 &	11.12 &	94.7\\
ADF &   8.43 &	8.41 &	95.3 &	11.22 &	11.12 &	94.8\\
EFF &   42.99 &	5.01 &	0.0 &	82.97 &	5.01 &	0.0\\
GIM &   47.42 &	5.28 &	0.0 &	73.45 &	6.56 &	0.0\\
\bottomrule
\end{tabular}
}
\end{table}
\end{minipage}
\hfill
\begin{minipage}{0.45\textwidth}
\begin{table}[H]
\centering
\caption{RMSE, ASE and CP for Scenario II with transportable summary statistics. All numbers are multiplied by 100}\label{MSE2 transportable}
\resizebox{1\textwidth}{!}{
\begin{tabular}{ccccccc}
	\toprule
     & \multicolumn{3}{c}{$\tau_1$}       & \multicolumn{3}{c}{$\tau_2$}  \\ 
     \midrule
     &   RMSE    & ASE  &  CP  &  RMSE      &  ASE  & CP\\
INT  & 10.75 &	11.21 &	95.7 &	11.41 &	11.19 &	94.8\\
ORC  & 6.35 &	6.75 &	96.1 &	6.71 &	6.72 &	94.6\\
ADF  & 6.50 &	6.88 &	96.2 &	7.04 &	6.89 &	94.6\\
EFF  & 6.35 &	6.75 &	96.1 &	6.71 &	6.72 &	94.6\\
GIM  & 6.34 &	6.77 &	96.8 &	6.71 &	6.74 &	94.5\\
\bottomrule
\end{tabular}
}
\end{table}
\end{minipage}

\begin{minipage}{0.45\textwidth}
\begin{figure}[H]
\centering
  \includegraphics[width=0.5\textwidth]{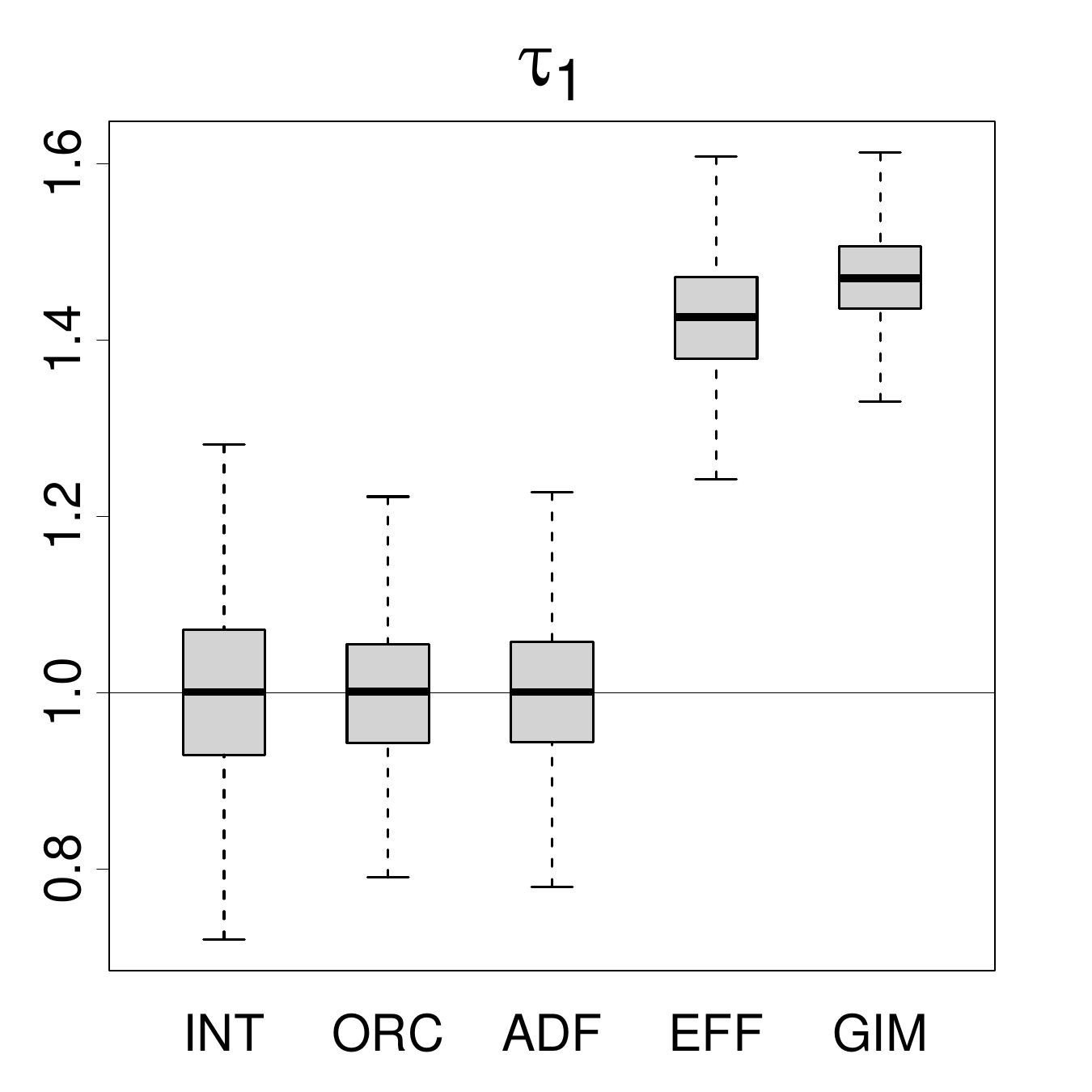}\hfill
  \includegraphics[width=0.5\textwidth]{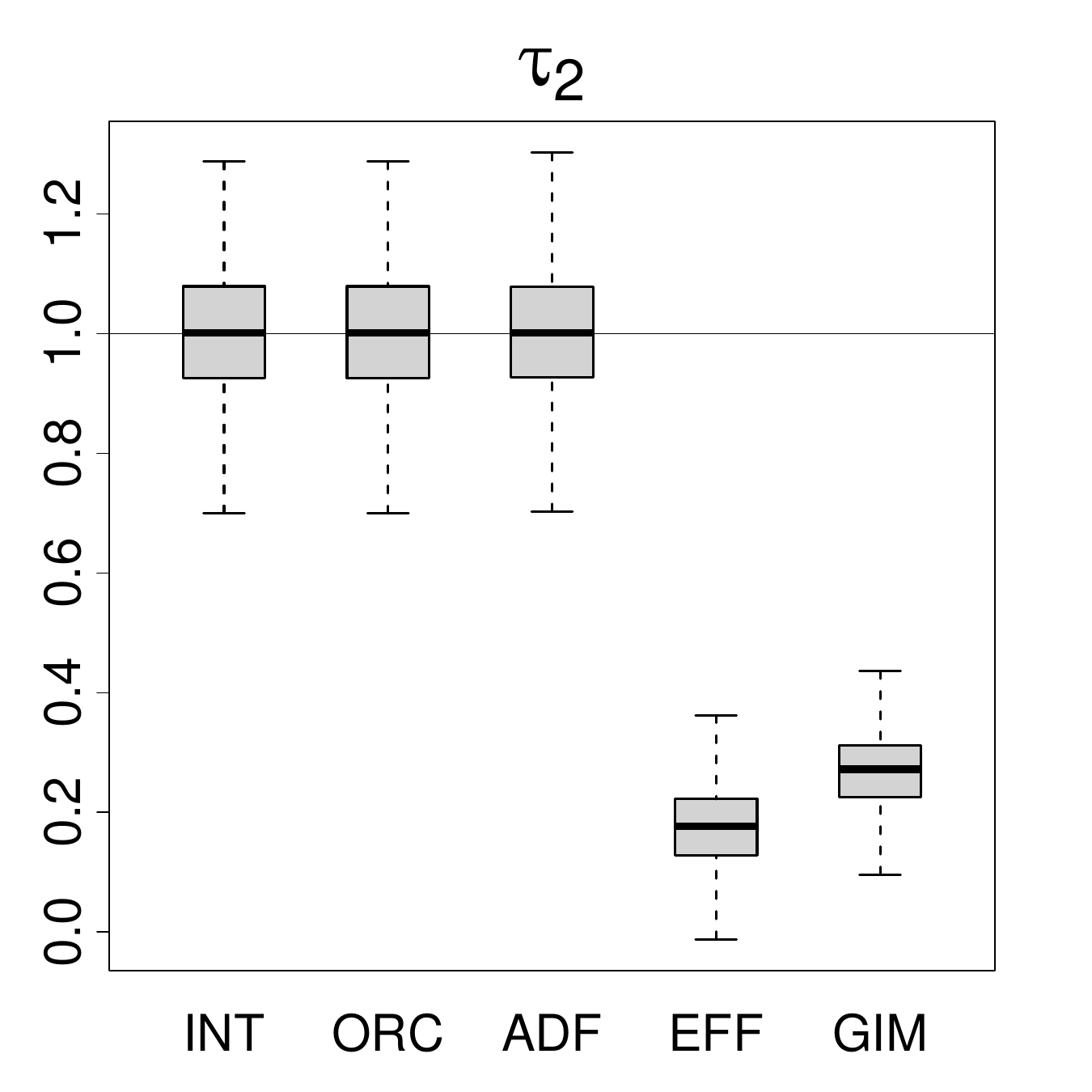}
\caption{Boxplots of various estimators in Scenario II with partially transportable external summary statistics.}\label{fig:biased}
\end{figure}
\end{minipage}
\hfill
\begin{minipage}{0.45\textwidth}
\begin{figure}[H]
\centering
  \includegraphics[width=0.5\textwidth]{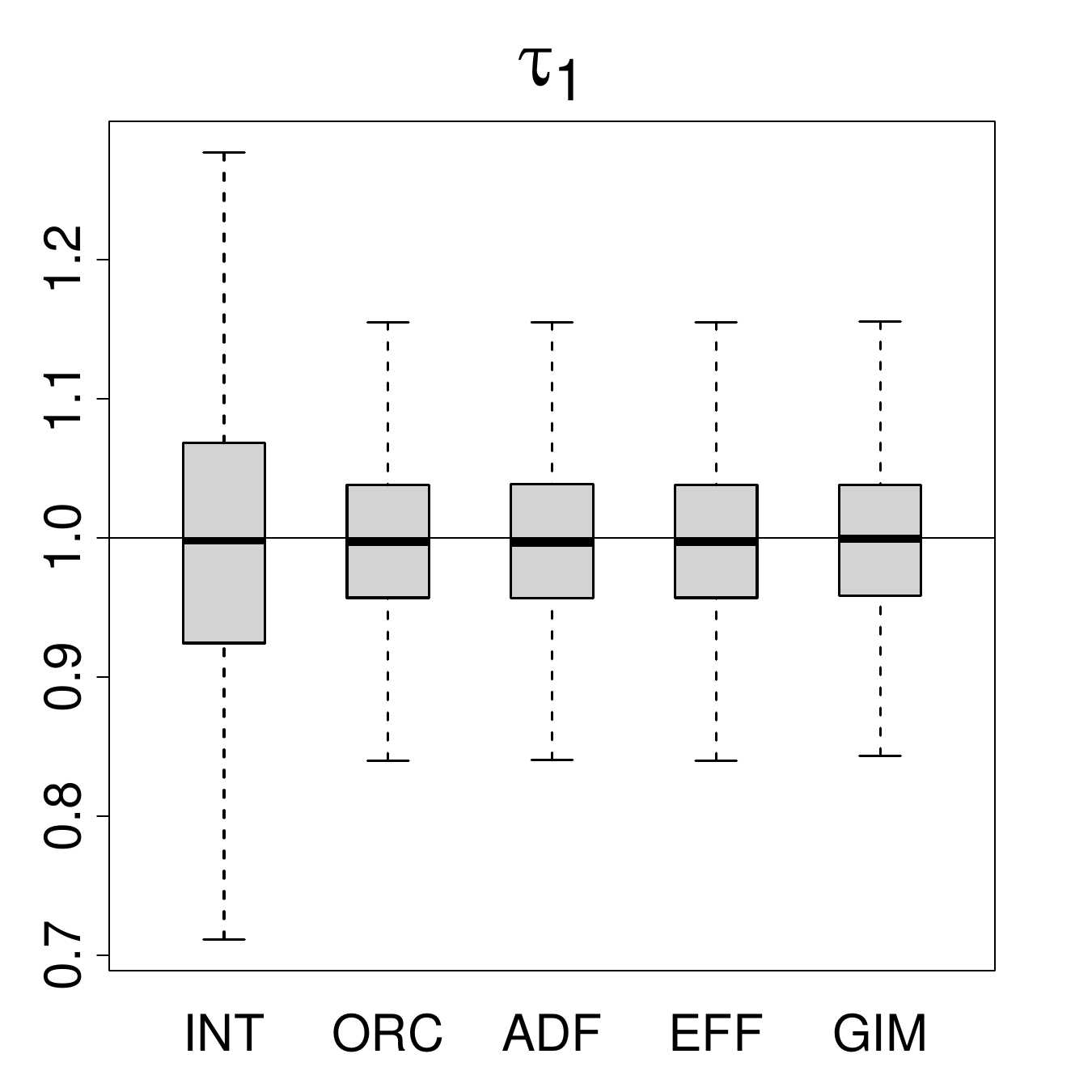}\hfill
  \includegraphics[width=0.5\textwidth]{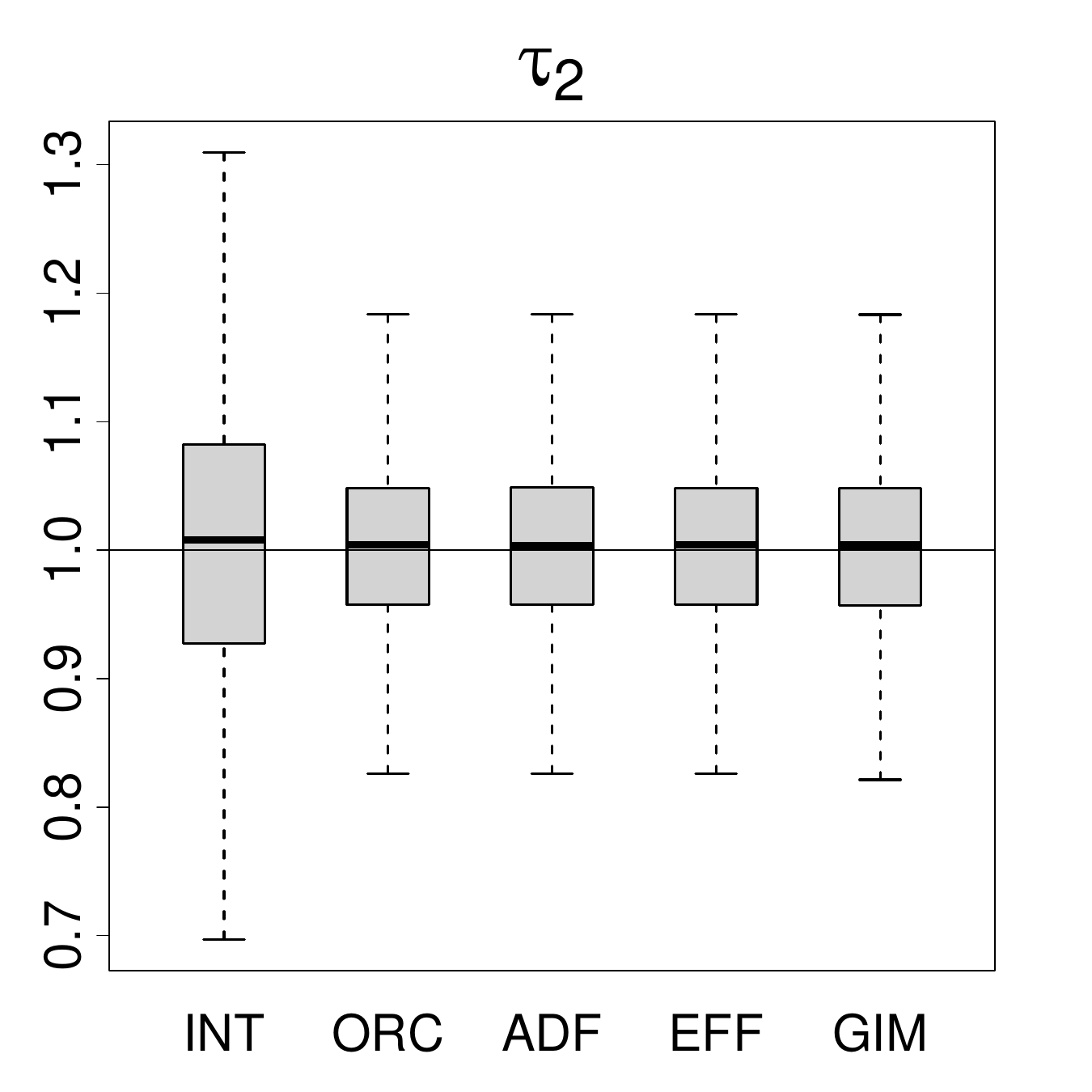}
\caption{Boxplots of various estimators in Scenario II with transportable external summary statistics.}\label{fig:biased2}
\end{figure}
\end{minipage}

\subsection{Simulation with Moderate Heterogeneity}\label{subsec: moderate hetero}
In the presence of moderate heterogeneity, the confidence interval based on the adaptive fusion estimator may have coverage probability below the nominal coverage rate. { In this section, we run simulations under the moderate heterogeneity setting discussed in Section \ref{subsec: reboot} where $\beta_1(P_1) = \beta_1(P_0)$ and  $|\beta_2(P_{1}) - \beta_2(P_{0})| \asymp n^{-1/2}$.} We consider the setting of Scenario II to evaluate the performance of different methods in the presence of moderate heterogeneity. 
Specifically, we let $\tilde{X}_2 = X_2 + \varepsilon_2$, where $\varepsilon_2 \sim N(0, \sigma^2)$ and $\sigma^2=Cn^{-1/2}$. Other settings are the same as in Scenario II in the previous section. 
The results are shown in Table~\ref{MSE2 small bias}.

\begin{table}[h]
	\centering
	\caption{RMSE, average standard error (ASE) and coverage probability (CP) for Scenario II with moderate heterogeneity. All numbers are multiplied by 100 
 }\label{MSE2 small bias}
	\footnotesize
	\begin{tabular}{ccccccccccc}
		\toprule
		& &\multicolumn{3}{c}{$C=0.05$}         &\multicolumn{3}{c}{$C=1$}     & \multicolumn{3}{c}{$C=20$}      \\[5pt]
		\midrule    
		& & RMSE & ASE   & CP               &  RMSE & ASE   & CP               & RMSE & ASE   & CP           \\
		\multirow{5}{*}{$\tau_1$}
		&INT & 11.25 &	11.17 &	94.2 &	11.25 &	11.17 &	94.2 &	11.25 &	11.17 &	94.2\\
		&ORC & 8.27 &	8.30 &	94.8 &	8.27 &	8.3 &	94.8 &	8.27 &	8.30 &	94.8\\
		&ADF & 6.87 &	6.92 &	94.8 &	7.69 &	6.96 &	90.5 &	8.47 &	8.42 &	94.7\\
		&EFF & 6.63 &	6.69 &	94.4 &	7.45 & 6.69 &	91.9 &	40.22 &	6.69 &	0.0\\
		&GIM & 6.62 &	6.73 &	94.2 &	7.51 &	6.61 & 91.6 &	44.85 &	5.32 &	0.0\\
		\midrule
		\multirow{5}{*}{$\tau_2$}
		&INT & 11.27 &	11.17 &	94.9 &	11.27 &	11.17 &	94.9 &	11.27 &	11.17 &	94.9\\
		&ORC & 11.27 &	11.10 &	94.4 &	11.27 &	11.10 &	94.4 &	11.27 &	11.10 &	94.4\\
		&ADF & 6.83 &	6.90 &	95.1 &	9.42 &	7.04 &	86.9 &	11.27 &	11.10 &	94.5\\
		&EFF & 6.60 &	6.69 &	94.5 &	9.37 &	6.69 &	84.7 &	77.88 &	6.69 &	0.0\\
		&GIM & 6.60 &	6.73 &	94.3 &	9.32 &	6.70 &	85.2 &	69.74 &	6.55 &	0.0\\
		\bottomrule
	\end{tabular}
\end{table}

The simulation results show that the ADF estimator has similar or better performance than the INT  and ORC estimators in terms of RMSE and average standard error across all settings. While the
RMSEs of the EFF  and GIM estimators are smaller than that of the ADF estimator when $C = 0.05$, they become significantly larger when $C = 20$.
Confidence intervals based on the INT and ORC estimators can always achieve the nominal level.
However, the confidence interval based on the adaptive fusion estimator may have coverage probability below the nominal level, especially when the bias of the external summary statistics is moderate ($C = 1$). 
In addition, confidence intervals based on the EFF  and GIM estimators tend to fall below the nominal level when $C = 1$ and may even be zero when $C = 20$.

\begin{table}[h]
	\centering
	\caption{Average width (AW) and coverage probability (CP) for Scenario II with moderate heterogeneity. All numbers are multiplied by 100.}\label{CI biased}
	\resizebox{\textwidth}{!}{
	\begin{tabular}{*{17}{c}}
		\toprule
		&&\multicolumn{2}{c}{$C=0.05$}         &\multicolumn{2}{c}{$C=1$}     & \multicolumn{2}{c}{$C=20$} &&&  &\multicolumn{2}{c}{$C=0.05$}         &\multicolumn{2}{c}{$C=1$}     & \multicolumn{2}{c}{$C=20$}    \\[5pt]
		\midrule    
		& &         AW  & CP          &  AW  & CP          & AW & CP  & & & &       AW  & CP          &  AW  & CP          & AW & CP     \\
		\multirow{6}{*}{$\tau_1$} 
		&INT    & 43.8 &	95.0 &	43.80 &	95.0 &	43.80 &	95.0&& \multirow{6}{*}{$\tau_2$}&INT    & 43.92 &	95.4 &	43.92 &	95.4 &	43.92 &	95.4\\
		&ORC    & 32.59 &	95.4 &	32.59 &	95.4 &	32.59 &	95.4&& &ORC    & 43.65 &	95.4 &	43.65 &	95.4 &	43.65 &	95.4\\
		&ADF    & 26.99 &	95.9 &	27.18 &	90.1 &	33.02 &	95.8&& &ADF    & 26.99 &	95.8 &	27.48 &	85.6 &	43.65 &	95.3\\
		&EFF    & 26.24 &	96.5 &	26.24 &	90.5 &	26.24 &	0.0&& &EFF    & 26.27 &	95.6 &	26.27 &	83.0 &	26.27 &	0.0\\
		&GIM    & 26.35 &	96.8 &	25.86 &	89.9 &	20.84 &	0.0&& &GIM    & 26.46 &	95.4 &	26.34 &	83.2 &	25.72 &	0.0\\
		&ReBoot & 35.27 &	98.0 &  36.19 &	95.9 &	40.91 &	98.3&& &ReBoot & 35.21 &	97.0 &	37.62 &	93.5 &	50.60 &	97.9\\
		\bottomrule
	\end{tabular}}
\end{table}

Next, we compare the average width and coverage probability of the confidence intervals based on the INT, ORC, ADF, EFF and GIM estimators with that provided by the re-bootstrap (ReBoot) procedure in Section \ref{subsec: reboot}.  We set $\bar{r} = 10$ in this simulation. The results in Table~\ref{CI biased} show that the ReBoot procedure mitigates the undercoverage issue of the confidence intervals based on the ADF estimator at the cost of some increase in average width. Although the ReBoot procedure is conservative in some cases, it still produces narrower confidence intervals than those based solely on the internal data in most settings.

\section{Real Data Analysis}\label{sec: real data}
We apply the proposed methods to analyze a Helicobacter pylori infection dataset described by \cite{li2023improving}. 
Helicobacter pylori infection is a leading worldwide infectious disease. 
The triple therapy (clarithromycin, amoxicillin, and omeprazole) is a standard treatment for Helicobacter pylori infection. 
The internal study is a two-arm randomized clinical trial conducted at a traditional Chinese medicine hospital.
This trial aims to investigate whether the additional taking of traditional Chinese medicine ($D=1$) has better efficacy than the standard triple therapy treatment ($D=0$) on Helicobacter pylori infection. 
It contains 362 observations, of which 180 patients are assigned to the triple therapy and the rest are assigned to a combination
treatment including both triple therapy and traditional Chinese medicine. 
The external study is a single-arm study conducted at a Western-style hospital, where  110 patients are all assigned to the triple therapy. 
The outcome $Y$ is the post-treatment infection status assessed with the C--14 urea breath test and baseline covariates $X$ include age, gender, height, BMI, occupation, education level, marital status, and information on patients’ symptoms.
The internal and external studies adopt the same inclusion and exclusion criteria and the same treatment protocols. 
The parameter of interest is the average causal effect of the combination treatment against the standard triple therapy treatment, i.e., $\tau=E(Y_1-Y_0)$. 
We illustrate how to use the  individual data from the internal study and an estimator $\tilde \beta$  of $\beta=E(Y_0)$  from the external study to make the inference about $\tau$.

We implement four methods   INT, PRM, EFF and ADF to estimate $\tau$.
Our goal is to test the null hypothesis $\mathbb H_0:\tau \leq 0$ against $\mathbb H_1: \tau >0$ to investigate whether the combination treatment can improve the efficacy. We calculate the estimate of $\tau$ and its standard error and then use z-test to calculate one-sided p-values.
Table~\ref{real data} presents the analysis results. 
The four point estimates are close to each other, all showing a potentially beneficial effect of the additional use of traditional Chinese medicine.
The EFF estimate and the ADF  estimate are   identical,   suggesting transportablity of the external summary statistic; 
this is because the same inclusion criterion and the same treatment protocols are adopted in both the internal and external studies, and it is reasonable to assume that they are from the same population. 
The INT estimate based solely on the internal data does not reject the null hypothesis $\mathbb H_0$ at level $0.1$. Test based on the PRM estimate rejects $\mathbb H_0$ at level $0.1$ but the PRM estimate has an even larger standard error than the INT estimate. 
In contrast, by appropriate integration of the external summary statistic, the EFF and ADF  estimates achieve smaller standard errors and reject $\mathbb H_0$ at level $0.1$. In addition, the re-bootstrap procedure also rejects $\mathbb H_0$ at level $0.1$ with a p-value $0.074$, which suggests that the result is robust to possible moderate heterogeneity.
These results may serve as evidence in favor of the beneficial effect of the additional use of traditional Chinese medicine in the treatment for Helicobacter
pylori infection.

\begin{table}[h!]
	\centering
\caption{Point estimates of $\tau$, standard errors, and $p$-values}\label{real data}
\begin{tabular}{cccc}
\toprule
&  point estimate& standard error & $p$-value  \\[5pt]
\midrule
INT  &  0.0543    & 0.0442  &  0.1100 \\  
PRM&  0.0773    & 0.0520  &  0.0684 \\ 
EFF  &  0.0628    & 0.0394  &  0.0553 \\ 
ADF &  0.0628    & 0.0394  &  0.0554\\
\bottomrule
\end{tabular}
\end{table}

\section{Discussion}

We have focused on the integration of finite-dimensional summary statistics.  
It is of both theoretical and practical interest to study how to integrate  infinite-dimensional external summary curves, 
such as estimates of a density function,  regression curve, or conditional mean, and a trained neural network model.
We plan to pursue this extension in the future.

\section*{Supplementary Material}
\label{SM}
Supplementary Material available online includes proofs of Theorems 1--6 and Propositions 1--3, detailed discusses for the efficiency gain of incorporating an additional external summary statistics,  additional examples and simulation results, the efficient influence functions under a parametric model for the conditional density,  discussions on the scenario when Assumption 2  holds only asymptotically, and additional details about the re-bootstrap procedure.

\clearpage
\begin{center}
    \LARGE \textbf{Supplementary Material} \\
\end{center}
\vspace{2em}

\appendix
\setcounter{section}{0}
\setcounter{equation}{0}
\setcounter{figure}{0}
\setcounter{table}{0}
\setcounter{page}{1} 

\renewcommand{\thesection}{S\arabic{section}}
\renewcommand{\theequation}{S\arabic{equation}}
\renewcommand{\thefigure}{S\arabic{figure}}
\renewcommand{\thetable}{S\arabic{table}}

\section{Proofs of Theorems}
For convenience, we use the following notation throughout the supplement. 
For function $y(x)$, $\dot{y}(x_0)$ represents the derivative of $y(x)$ evaluated at $x=x_0$, $\ddot{y}(x_0)$ represents the second derivative of $y(x)$ evaluated at $x=x_0$.
For a sequence of random variables $U_n$, $L_P(U_n)\rightarrow U$ represents that $U_n$ converges in distribution to $U$ under the law of $P$,  we also denote as $U_n\rightarrow U$ if there is no ambiguity for $P$,
and $\mathrm{avar}(\cdot)$  denotes the asymptotic variance. For an index set $\mathcal{A}$, a vector $v=(v_1, \ldots, v_q)$ and a matrix $G=(g_{ij}), i,j = 1, \ldots,q$, $v_\mathcal{A} = (v_i), i\in \mathcal{A}$ is a sub-vector of $v$ with indices in $\mathcal{A}$; $G_{\mathcal{A}}=(g_{ij}), i,j \in \mathcal{A}$ is a sub-matrix of $G$ with row and column indices in $\mathcal{A}$. As noted in Chapter 25 of \cite{van2000asymptotic}, it suffices to consider one-dimensional submodels when studying semiparametric problems. For simplicity, the term ``parametric submodel" subsequently refers to a one-dimensional parametric submodel.  
Before proving the theorems, we first introduce the following regularity conditions. For any regular  parametric submodel $P_\theta \in \mathcal P_{\rm trans}$ and positive sequence $\{t_{m}\}$ that converges to some $t \geq 0$, let $g_{m}$ be the density of $m^{1/2}\{\tilde{\beta} - \beta(P_1)\}$ under $P_{\theta + t_{m}m^{-1/2}}$.
\begin{condition}\label{cond: regular beta-tilde}
	 For any regular  parametric submodel $P_\theta \in \mathcal P_{\rm trans}$ and positive sequence $\{t_{m}\}$ that converges to some $t \geq 0$, $g_{m}$ is pointwisely bounded and uniformly equicontinuous in the sense that for any $b \in R^{q}$, there exists constant $C_{b}$ such that $\sup_{m}g_{m}(b) < C_{b}$, and for any $\epsilon > 0$, there exist $\delta_{\epsilon}, m_{\epsilon} > 0$ such that $|g_{m}(b_{1}) - g_{m}(b_{2})| < \epsilon$ and $|g_{m}(b_{3})| < \epsilon$ for any $m > m_{\epsilon}$, $\|b_{1} - b_{2}\| < \delta_{\epsilon}$ and $\|b_{3}\| > \delta_{\epsilon}^{-1}$, where $\|\cdot\|$ denotes the Euclidean norm.
\end{condition}

\begin{condition}\label{cond: regular internal}
   $\widehat{\Sigma}_{\phi\eta}$ and $\widehat{\Sigma}_{\eta\eta}$ are consistent estimators for $E(\phi_{\mathrm{e}}\eta_{\mathrm{e}}^\T)$ and $E(\eta_{\mathrm{e}}\eta_{\mathrm{e}}^\T)$, respectively. $\hat{\tau}_{\mathrm{e}}^{\sI}$ and $\hat{\beta}_{\mathrm{e}}^{\sI}$ are RAL estimators based on internal data for $\tau$ and $\beta$ with influence functions $\phi_{\mathrm{e}}$ and $\eta_{\mathrm{e}}$, respectively.
\end{condition}

Next, we introduce the regularity conditon required with multiple external studies. For $s = 1,\dots, S$, let $\mathcal{P}_{s}$ denote collections of models for the $s$-th external data and assume $P_{s} \in \mathcal{P}_{s}$. Define
\[
\mathcal P_{\rm trans}^{[S]}=\left\{\prod_{s=0}^{S} P_{s} \in \prod_{s=0}^{S} \mathcal{P}_{s}: P_0, \dots, P_{S} \text{ such that } \beta_{s}(P_{s}) = \beta_{s}(P_{0}) \text{ for $s = 1,\dots, S$}\right\}.
\] 
For $s = 1,\dots, S$, any regular  parametric submodel $P_\theta \in \mathcal P_{\rm trans}^{[S]}$ and positive sequence $\{t_{m_{s}}\}$ that converges to some $t \geq 0$, let $g_{s, m_{s}}$ be the density of $m_s^{1/2}\{\tilde{\beta}_{s} - \beta_{s}(P_s)\}$ under $P_{\theta + t_{m_{s}}m_{s}^{-1/2}}$. Let $q_{s}$ be the dimension of $\tilde{\beta}_{s}$ for $s = 1,\dots, S$.
\begin{condition}\label{cond: regular beta-tilde multiple}
	 For $s = 1,\dots, S$, any regular  parametric submodel $P_\theta \in \mathcal P_{\rm trans}$ and positive sequence $\{t_{m_{s}}\}$ that converges to some $t \geq 0$, $g_{s, m_{s}}$ is pointwisely bounded and uniformly equicontinuous in the sense that for any $b \in R^{q_{s}}$, there exists constant $C_{b}$ such that $\sup_{m_{s}}g_{s, m_{s}}(b) < C_{b}$, and for any $\epsilon > 0$, there exist $\delta_{\epsilon}, m_{\epsilon} > 0$ such that $|g_{s, m_{s}}(b_{1}) - g_{s, m_{s}}(b_{2})| < \epsilon$ and $|g_{s, m_{s}}(b_{3})| < \epsilon$ for any $m_{s} > m_{\epsilon}$, $\|b_{1} - b_{2}\| < \delta_{\epsilon}$ and $\|b_{3}\| > \delta_{\epsilon}^{-1}$.
\end{condition}
\begin{proof}[Proof of Theorem \ref{convolution}] 
	In fact, this convolution theorem holds for all data-fused regular estimators. 
	In the following proof, we only assume that $T_n(\tilde{\beta})$ is regular as in Definition \ref{RAL estimator}.
	Following the proof in \cite{bickel1993efficient},
	we first prove   the result   in parametric models,
	and then generalize to semiparametric models. The innovation in this proof is that we incorporate the external summary statistics, which is different from the i.i.d case considered in classical semiparametric theory.
	
	For the first step, we denote parametric models $P_0(Z;\theta)$ and $P_1(W;\theta)$ by $P_0^\theta$ and $P_1^\theta$ respectively for notational convenience. 
	Let $l(\theta) = \sum_{i=1}^n \log f(Z_i;\theta) $ be the  log-likelihood for internal data, where $f(Z;\theta)$ is the density function of $P_0^\theta$.
	Under  Assumption~\ref{well defined}, we denote $\tau(\theta) :=\tau(P_0^\theta)$ and  $\beta(\theta) := \beta(P_0^\theta) = \beta(P_1^\theta)$.
	We have  $\phi_{\mathrm{e}}(Z) = \dot{\tau}(\theta)J_\theta^{-1}\partial \log f(Z;\theta)/\partial \theta$, $\eta_{\mathrm{e}}(Z) = \dot{\beta}(\theta)J_\theta^{-1}\partial \log f(Z;\theta)/\partial \theta$, where $J_\theta = E[\{\partial \log f(Z;\theta)/\partial \theta\}\{\partial \log f(Z;\theta)/\partial \theta\}^\T]$. 
	Then we have $M = \dot{\tau}(\theta)J_\theta^{-1} \dot{\beta}^\T(\theta) \left\{\Sigma_1/\rho + \dot{\beta}(\theta) J_\theta^{-1}\dot{\beta}^\T(\theta) \right\}^{-1}$.
	
	Note that $\tilde{\beta}$ is a RAL estimator of $\beta(P_1)$ and $\beta(P_1) = \beta(P_0)$, we have  $\tilde{\beta} - \beta =  m^{-1}\sum_{j=1}^m \tilde{\eta}(W_j) + e_m$ for some influence function $\tilde{\eta}$ with $E_{P_1}\{\tilde{\eta}(W)\} = 0$, $E_{P_1}\{\tilde{\eta}(W)\tilde{\eta}^\T(W)\}$ finite and nonsingular and $e_m = o_P(m^{-1/2})= o_P(n^{-1/2})$.
	Define the sequence $(U_n, V_n) = \big(n^{1/2}\{T_n(\tilde{\beta}) - \tau(\theta)\}, n^{-1/2} [\sum_{i=1}^n \partial \log f(Z_i;\theta)/\partial \theta + m\dot{\beta}^\T(\theta)\Sigma^{-1}_1\{\tilde{\beta} - \beta(\theta) - e_m\}]\big)$. 
	From the regularity of $T_n(\tilde{\beta})$ and the regularity of the parametric submodel, $U_n$ and $V_n$ converge in distribution to $U,V$, respectively. Therefore, they are marginally tight and thus jointly tight. Then for any subsequence $\{n'\}$, there exists a subsequence $\{n''\}$ such that $(U_{n''}, V_{n''})$ jointly converge to $(U, V)$, and $V\sim N(0, \tilde{I}_\theta )$, where $\tilde{I}_\theta = J_\theta + \dot{\beta}(\theta)^\T\rho\Sigma_1^{-1}\dot{\beta}(\theta)$. For convenience, we denote the subsequence $\{n''\}$ by $\{n\}$. 
	
	Under Assumption \ref{beta} and Condition \ref{cond: regular beta-tilde}, the log-likelihood combining internal data and  external summary statistics satisfies
	\[
	\begin{aligned}
		\tilde{l}(\theta) &= l(\theta) - \frac{m}{2}\{\tilde{\beta} - \beta(P_1^{\theta})\}^\T\Sigma_1^{-1}\{\tilde{\beta} - \beta(P_1^\theta)\} - \frac{1}{2}\log\left(|2\pi m^{-1}\Sigma_1|\right) + o_{P}(1),
	\end{aligned}
	\]
	according to Lemma 1 in \cite{boos1985converse}.
	Let $D_n = \tilde{l}(\theta + tn^{-1/2}) - \tilde{l}(\theta)$, then $(U_n, D_n)$ converges to $\big(U, t^\T V - t^\T\tilde{I}_\theta t/2\big)$ by the local asymptotic normality (LAN) property. The regularity of $T_n(\tilde{\beta})$ means that
	\[
	L_{\theta+tn^{-1/2}}\big(n^{1/2}\{ T_n(\tilde{\beta}) - \tau(\theta+tn^{-1/2}) \} \big) \rightarrow U.
	\]
	By the differentiability of $\tau(\theta)$, we have
	\[
	\begin{aligned}
		L_{\theta+tn^{-1/2}}(U_n)  
		&=L_{\theta+tn^{-1/2}}\big(n^{1/2}\{T_n(\tilde{\beta}) - \tau(\theta) \} \big)\\
		&= L_{\theta+tn^{-1/2}}\big(n^{1/2}\{T_n(\tilde{\beta}) - \tau(\theta+tn^{-1/2})\}+n^{1/2}\{\tau(\theta+tn^{-1/2}) - \tau(\theta)\}\big) \\
		&\rightarrow U + \dot{\tau}(\theta)t.
	\end{aligned}
	\]
	Therefore, 
	\begin{equation}\label{S1}
		E_{\theta+tn^{-1/2}}\{\exp(ia^\T U_n)\} \rightarrow \exp\{ia^\T\dot{\tau}(\theta)t\}E^\ast\{\exp(ia^\T U)\},
	\end{equation}
	where $E_\theta(\cdot)$ means the expectation taken under distribution $P_0^\theta\times P_1^\theta$ and $E^\ast(\cdot)$ is taken under the limiting distribution. 
	Besides, by contiguity, we have
	\begin{equation}\label{S2}
		\begin{aligned}
			E_{\theta+tn^{-1/2}}\{\exp(ia^\T U_n)\} &= E_{\theta}\{\exp(ia^\T U_n  + D_n)\} + o(1)\\ 
			&\rightarrow E^\ast\left[\exp\left\{ia^\T U + t^\T V - \frac{1}{2}t^\T\tilde{I}_\theta t\right\}\right].
		\end{aligned}
	\end{equation}
	Then, combining \eqref{S1} and \eqref{S2}, we have 
	\[
	E^\ast\left[\exp\left\{ia^\T U + t^\T V - \frac{1}{2}t^\T\tilde{I}_\theta t\right\}\right] = \exp\{ia^\T \dot{\tau}(\theta)t\}E^\ast\{\exp(ia^\T U)\}.
	\]
	Notice that both sides of the above equation are analytic on the Cartesian product of complex planes and hence the above equation holds on the Cartesian product of complex planes by the uniqueness of analytic continuation. 
	Letting $t = -i\tilde{I}^{-1}(\theta)\dot{\tau}(\theta)^{\T}(a - b)$, we obtain
	\begin{equation}\label{S3}
		\begin{aligned}
			&\quad E^\ast\big(\exp[ia^\T\{U- \dot{\tau}(\theta)\tilde{I}^{-1}_\theta V\} + ib^\T\dot{\tau}(\theta)\tilde{I}^{-1}_\theta V]\big) \\ 
			&= E^\ast\left[\exp\left\{ia^\T U + \frac{1}{2}a^\T\dot{\tau}(\theta)\tilde{I}^{-1}_\theta \dot{\tau}(\theta)^\T a\right\}\right]\exp\left\{-\frac{1}{2}b^\T\dot{\tau}(\theta)\tilde{I}^{-1}_\theta\dot{\tau}(\theta)^\T b\right\}.
		\end{aligned}
	\end{equation}
	Equation~\eqref{S3} gives the characteristic function of $(U_{n''} - \dot{\tau}(\theta)\tilde{I}^{-1}_\theta V_{n''}, \dot{\tau}(\theta)\tilde{I}^{-1}_\theta V_{n''})$'s limiting distribution and holds for every choice of initial subsequence $\{n'\}$. 
	Because the right-hand side of \eqref{S3} is only related to the marginal distribution of $U$,  this equation  holds for the full sequence $\{n\}$, 
	that is, $(U_{n} - \dot{\tau}(\theta)\tilde{I}^{-1}_\theta V_{n}, \dot{\tau}(\theta)\tilde{I}^{-1}_\theta V_{n})$ has a limiting distribution with characteristic function in ~\eqref{S3}.
	In equation~\eqref{S3}, letting $a = 0$, we have
	\begin{equation}\label{S4}
		E^\ast[\exp\{ib^\T\dot{\tau}(\theta)\tilde{I}_\theta^{-1}V \}] = \exp\left\{-\frac{1}{2}b^\T\dot{\tau}(\theta)\tilde{I}^{-1}_\theta \dot{\tau}(\theta)^\T b\right\}.
	\end{equation}
	Further letting $b=0$,   we have
	\begin{equation}\label{S5}
		\begin{aligned}
			&E^\ast\big(\exp[ia^\T\{U- \dot{\tau}(\theta)\tilde{I}^{-1}_\theta V\}\big) 
			= E^\ast\left[\exp\left\{ia^\T U + \frac{1}{2}a^\T\dot{\tau}(\theta)\tilde{I}^{-1}_\theta \dot{\tau}(\theta)^\T a\right\}\right].
		\end{aligned}
	\end{equation}
	Combining the equations~\eqref{S3}, \eqref{S4} and \eqref{S5} yields
	\[
	\begin{aligned}
		&\quad E^\ast\big(\exp[ia^\T\{U- \dot{\tau}(\theta)\tilde{I}^{-1}_\theta V\} + ib^\T\dot{\tau}(\theta)\tilde{I}^{-1}_\theta V]\big) \\
		&=E^\ast\big(\exp[ia^\T\{U- \dot{\tau}(\theta)\tilde{I}^{-1}_\theta V\}]\big)E^\ast[\exp\{ib^\T\dot{\tau}(\theta)\tilde{I}^{-1}_\theta V\}],
	\end{aligned}
	\]  
	which means that $U- \dot{\tau}(\theta)\tilde{I}^{-1}_\theta V$ and $\dot{\tau}(\theta)\tilde{I}^{-1}_\theta  V$ are independent. 
	In addition, we have
	\begin{equation}\label{parametric}
		\begin{pmatrix}
			U_n - \dot{\tau}(\theta)\tilde{I}^{-1}_\theta V_n \\ 
			\dot{\tau}(\theta)\tilde{I}^{-1}_\theta V_n
		\end{pmatrix} \rightarrow \begin{pmatrix}
			U- \dot{\tau}(\theta)\tilde{I}^{-1}_\theta V \\ 
			\dot{\tau}(\theta)\tilde{I}^{-1}_\theta V
		\end{pmatrix}
	\end{equation}
	under the parametric model $P_0^\theta\times P_1^\theta$.
	
	Note that 
	\[
	\begin{aligned}
		&\quad n^{1/2}\dot{\tau}(\theta)\tilde{I}^{-1}_\theta V_n  \\ 
		&= n^{1/2}\dot{\tau}(\theta)\left[J_\theta^{-1} - J^{-1}_\theta \dot{\beta}^\T(\theta)\left\{\Sigma_1/\rho + \dot{\beta}(\theta) J_\theta^{-1}\dot{\beta}^\T(\theta) \right\}^{-1}\dot{\beta}(\theta)J_\theta^{-1}\right]V_n \\
		&= \sum_{i=1}^n \biggl[\dot{\tau}(\theta)J_\theta^{-1} - \dot{\tau}(\theta)J_\theta^{-1}\dot{\beta}^\T(\theta)\left\{\Sigma_1/\rho + \dot{\beta}(\theta) J_\theta^{-1}\dot{\beta}^\T(\theta) \right\}^{-1}\dot{\beta}(\theta) J_\theta^{-1}\biggl]\frac{\partial \log f(Z_i;\theta)}{\partial \theta}\\
		&\quad+ \dot{\tau}(\theta)\left[J_\theta^{-1} - J^{-1}_\theta \dot{\beta}^\T(\theta)\left\{\Sigma_1/\rho + \dot{\beta}(\theta) J_\theta^{-1}\dot{\beta}^\T(\theta) \right\}^{-1}\dot{\beta}(\theta) J_\theta^{-1}\right]\dot{\beta}^\T(\theta) m\Sigma_1^{-1}\{\tilde{\beta} - \beta(\theta) - e_m\}\\
		&= \sum_{i=1}^n \biggl[\dot{\tau}(\theta)J_\theta^{-1} - \dot{\tau}(\theta)J_\theta^{-1}\dot{\beta}^\T(\theta)\left\{\Sigma_1/\rho + \dot{\beta}(\theta) J_\theta^{-1}\dot{\beta}^\T(\theta) \right\}^{-1}\dot{\beta}(\theta) J_\theta^{-1}\biggl]\frac{\partial \log f(Z_i;\theta)}{\partial \theta}\\
		&\quad+ n\dot{\tau}(\theta)J_\theta^{-1} \dot{\beta}^\T(\theta) \left\{\Sigma_1/\rho + \dot{\beta}(\theta) J_\theta^{-1}\dot{\beta}^\T(\theta) \right\}^{-1}\{\tilde{\beta} - \beta(\theta) - e_m\} \\ 
		&= \sum_{i=1}^n \{ \phi_{\mathrm{e}}(Z_{i}) - M\eta_{\mathrm{e}}(Z_{i}) + M(\tilde{\beta} - \beta - e_m)  \}   \qquad\qquad\qquad\qquad\qquad(\text{As $e_m = o_P(n^{-1/2})$})\\ 
		&= \sum_{i=1}^n \{ \phi_{\mathrm{e}}(Z_{i}) - M\eta_{\mathrm{e}}(Z_{i}) \} + nM(\tilde{\beta} - \beta) + o_P(n^{1/2}), \\
	\end{aligned}
	\]
	and we have
	\[
	\begin{aligned}
		&\quad n^{-1}\sum_{i=1}^n \left\{\phi_{\mathrm{e}}(Z_{i}) - M\eta_{\mathrm{e}}(Z_{i})\right\} + M(\tilde{\beta} - \beta) \\
		&= n^{-1/2}\dot{\tau}\tilde{I}_\theta^{-1}V_n + o_P(n^{-1/2}).
	\end{aligned}
	\]
	So \eqref{parametric} implies that 
	\[
	n^{1/2}\left[\begin{matrix}
		T_n(\tilde{\beta}) - \tau(P_0) - n^{-1}\sum_{i=1}^n \left\{\phi_{\mathrm{e}}(Z_{i}) - M\eta_{\mathrm{e}}(Z_{i})\right\} - M(\tilde{\beta} - \beta)\\
		n^{-1}\sum_{i=1}^n \left\{\phi_{\mathrm{e}}(Z_{i}) - M\eta_{\mathrm{e}}(Z_{i})\right\} + M(\tilde{\beta} - \beta)
	\end{matrix}
	\right]\rightarrow \begin{pmatrix}
		U- \dot{\tau}(\theta)\tilde{I}^{-1}_\theta V \\ 
		\dot{\tau}(\theta)\tilde{I}^{-1}_\theta V
	\end{pmatrix},
	\]
	thus the convolution theorem holds in parametric models.

	We generalize the above results to semiparametric models.  
	Let $R_n = n^{-1/2}[\sum_{i=1}^n \{\phi_{\mathrm{e}}(Z_{i}) - M\eta_{\mathrm{e}}(Z_{i})\}] + n^{1/2}M(\tilde{\beta} - \beta - e_m) $. 
	As $\tau$ and $\beta$ are both pathwise differentiable, there exists a sequence of parametric submodels in the semiparametric model $\mathcal P_{\rm trans}=\{P_0\times P_1: P_0, P_1 \text{  satisfy Assumption~\ref{well defined}}\}$ having a uniformly bounded sequence of score $s_j=\partial \log f_j(Z;\theta)/\partial \theta$ such that
	\begin{equation}\label{approximate}
		\|\dot{\tau}(\theta;j) J_{\theta,j}^{-1}s_j - \phi_{\mathrm{e}}\|_{P_0}\rightarrow 0, \|\dot{\beta}(\theta;j) J_{\theta,j}^{-1}s_j - \eta_{\mathrm{e}}\|_{P_0} \rightarrow 0, ~\text{as $j \rightarrow \infty$ },
	\end{equation}
	where $f_j(Z;\theta)$ is the density function of $j$-th parametric submodel, $\dot{\tau}(\theta;j), \dot{\beta}(\theta;j)$ are the derivative of $\tau, \beta$ in the $j$-th parametric submodel respectively, $J_{\theta,j}$ is the Fisher information for $\theta$ in the $j$-th parametric submodel, and   $\|f\|_{P_0} = (\int f^2dP_0)^{1/2}$. 
	Let
	\[
	\begin{aligned}
		V_{nj} =  n^{-1/2} \biggl[\sum_{i=1}^n s_j(Z_i) + m\dot{\beta}^\T(\theta;j)\Sigma^{-1}_1\{\tilde{\beta} - \beta(\theta) - e_m\}\biggl],
	\end{aligned}
	\]
	and suppose $U_n, V_{nj}$ marginally converge to $U$ and $V_j$, respectively. 
	From   \eqref{parametric} for parametric submodels, we have
	\[
	\begin{pmatrix}
		U_n - \dot{\tau}(\theta;j)\tilde{I}^{-1}_{\theta,j} V_{nj} \\ 
		\dot{\tau}(\theta;j)\tilde{I}^{-1}_{\theta,j} V_{nj}
	\end{pmatrix} \rightarrow \begin{pmatrix}
		U- \dot{\tau}(\theta;j)\tilde{I}^{-1}_{\theta,j} V_j \\ 
		\dot{\tau}(\theta;j)\tilde{I}^{-1}_{\theta,j} V_j
	\end{pmatrix},
	\]
	where $\tilde{I}_{\theta,j} = J_{\theta,j} + \dot{\beta}(\theta;j)^\T\rho\Sigma_1^{-1}\dot{\beta}(\theta;j)$. 
	Also, we have
	\[
	\begin{aligned}
		&\quad E^\ast\big(\exp[ia^\T\{U- \dot{\tau}(\theta;j)\tilde{I}^{-1}_{\theta,j} V_j\} + ib^\T\dot{\tau}(\theta;j)\tilde{I}^{-1}_{\theta,j} V_j]\big) \\
		&=E^\ast\big(\exp[ia^\T\{U- \dot{\tau}(\theta;j)\tilde{I}^{-1}_{\theta,j} V_j\}]\big)E^\ast[\exp\{ib^\T\dot{\tau}(\theta;j)\tilde{I}^{-1}_{\theta,j} V_j\}].
	\end{aligned}
	\]  
	For $(U_n, R_n)$, we know $U_n$ and $R_n$ marginally converges to a limiting distribution. For any subsequence $\{n'\}$, there exists a subsequence $\{n''\}$ such that $(U_{n''}, R_{n''})$ has a joint limiting distribution $(U, R)$. For ease of notation, we  denote the subsequence $\{n''\}$ by $\{n\}$. We have 
	\[
	\begin{aligned}
		&\quad\sup_n E_\theta\{R_n - \dot{\tau}(\theta;j)\tilde{I}^{-1}_{\theta,j} V_{nj} \}^{\otimes 2} \\
		&= E_\theta\{\phi_{\mathrm{e}} - M\eta_{\mathrm{e}} - \dot{\tau}(\theta;j)\tilde{I}^{-1}_{\theta,j} s_j \}^{\otimes 2} \\\ 
		&\quad+ nE_\theta\{ [M - \dot{\tau}(\theta;j)J_{\theta,j}^{-1} \dot{\beta}^\T(\theta;j) \{\Sigma_1/\rho + \dot{\beta}(\theta;j) J_{\theta,j}^{-1}\dot{\beta}^\T(\theta;j) \}^{-1}] (\tilde{\beta} - \beta-e_m) \}^{\otimes 2} \\ 
		&=E_\theta\{\phi_{\mathrm{e}} - M\eta_{\mathrm{e}} - \dot{\tau}(\theta;j)\tilde{I}^{-1}_{\theta,j} s_j \}^{\otimes 2} \\\ 
		&\quad+ n/mE_\theta\{ [M - \dot{\tau}(\theta;j)J_{\theta,j}^{-1} \dot{\beta}^\T(\theta;j) \{\Sigma_1/\rho + \dot{\beta}(\theta;j) J_{\theta,j}^{-1}\dot{\beta}^\T(\theta;j) \}^{-1}] \tilde{\eta} \}^{\otimes 2},
	\end{aligned}
	\]
	where $E_\theta(X)^{\otimes 2} = E_\theta(XX^\T)$. 
	From   \eqref{approximate}, we know when $j\rightarrow +\infty$,
	\[
	\begin{aligned}
		E\{\dot{\tau}(\theta;j) J_{\theta,j}^{-1} \dot{\beta}^\T(\theta;j) \} =E\{ \dot{\tau}(\theta;j) J_{\theta,j}^{-1}s_j (\dot{\beta}(\theta;j) J_{\theta,j}^{-1}s_j)^\T  \} &\rightarrow E\{\phi_{\mathrm{e}} \eta^\T_{\mathrm{e}} \},  \\ 
		E\{\dot{\beta}(\theta;j) J_{\theta,j}^{-1} \dot{\beta}^\T(\theta;j) \} = E\{ \dot{\beta}(\theta;j) J_{\theta,j}^{-1}s_j (\dot{\beta}(\theta;j) J_{\theta,j}^{-1}s_j)^\T  \} &\rightarrow E\{\eta_{\mathrm{e}} \eta^\T_{\mathrm{e}} \}, \\ 
		\dot{\tau}(\theta;j)J_{\theta,j}^{-1} \dot{\beta}^\T(\theta;j) \left\{\Sigma_1/\rho + \dot{\beta}(\theta;j) J_{\theta,j}^{-1}\dot{\beta}^\T(\theta;j) \right\}^{-1} &\rightarrow M.
	\end{aligned}
	\]
	Therefore, $\sup_n E_\theta\{R_n - \dot{\tau}(\theta;j)\tilde{I}^{-1}_{\theta,j} V_{nj} \}^{\otimes 2} \rightarrow 0$ as $j\rightarrow \infty$. Then we must have
	\[
	L(U-R, R) = \lim_{j\rightarrow \infty} L(U - \dot{\tau}(\theta;j)\tilde{I}^{-1}_{\theta,j} V_j, \dot{\tau}(\theta;j)\tilde{I}^{-1}_{\theta,j} V_j),
	\]  
	and 
	\[
	\begin{aligned}
		&\quad E^\ast[\exp\{ia^\T(U- R) + ib^\T R\}] \\ 
		&=\lim_{j\rightarrow\infty} E^\ast\big(\exp[ia^\T\{U- \dot{\tau}(\theta;j)\tilde{I}^{-1}_{\theta,j} V_j\} + ib^\T\dot{\tau}(\theta;j)\tilde{I}^{-1}_{\theta,j} V_j]\big) \\
		&=\lim_{j\rightarrow \infty} E^\ast\big(\exp[ia^\T\{U- \dot{\tau}(\theta;j)\tilde{I}^{-1}_{\theta,j} V_j\}]\big) \lim_{j\rightarrow \infty}E^\ast[\exp\{ib^\T\dot{\tau}(\theta;j)\tilde{I}^{-1}_{\theta,j} V_j\}] \\ 
		&= E^\ast[\exp\{ia^\T(U- R)\}]E\{\exp(ib^\T R)\}.
	\end{aligned}
	\]  
	This  implies that $U - R$ and $R$ are independent. 
	Note that $L(U-R, R)$ is the same for any subsequence $\{n''\}$, so we have 
	\begin{equation}\label{S8}
		(U_n - R_n, R_n) \rightarrow (U-R, R).    
	\end{equation}
	Let $\Delta_0 = U -R$  and $S_0 = R$. Then, it holds that $S_0 \sim N(0, B)$ where $B = E(\phi_{\mathrm{e}}\phi_{\mathrm{e}}^\T ) - E(\phi_{\mathrm{e}}\eta_{\mathrm{e}}^\T) \left\{ \Sigma_1/\rho + E(\eta_{\mathrm{e}}\eta_{\mathrm{e}}^\T ) \right\}^{-1}E(\phi_{\mathrm{e}}\eta_{\mathrm{e}}^\T )^\T$.
	Finally, because $R_n = n^{-1/2}[\sum_{i=1}^n \{\phi_{\mathrm{e}}(Z_{i}) - M\eta_{\mathrm{e}}(Z_{i})\}] + n^{1/2}M(\tilde{\beta} - \beta) + o_P(1)$, \eqref{S8} can be written as 
	\[
	n^{1/2}\left[\begin{matrix}
		T_n(\tilde{\beta}) - \tau(P_0) - n^{-1}\sum_{i=1}^n \left\{\phi_{\mathrm{e}}(Z_{i}) - M\eta_{\mathrm{e}}(Z_{i})\right\} - M(\tilde{\beta} - \beta)\\
		n^{-1}\sum_{i=1}^n \left\{\phi_{\mathrm{e}}(Z_{i}) - M\eta_{\mathrm{e}}(Z_{i})\right\} + M(\tilde{\beta} - \beta)
	\end{matrix}
	\right]\rightarrow \begin{pmatrix}
		\Delta_0\\
		S_0
	\end{pmatrix}.
	\]
	This completes the proof.
\end{proof} 

\begin{proof}[Proof of Theorem \ref{efficiency theorem}]
	From Theorem \ref{convolution}, for a regular estimator $T_n(\tilde{\beta})$ we have  $n^{1/2}\{T_n(\tilde{\beta}) - \tau\} \rightarrow U  = \Delta_0 + S_0$ and  $\mathrm{var}(U) = \mathrm{\Sigma_{[S]}}(\Delta_0) + \mathrm{var}(S_0)\geq \mathrm{var}(S_0) =  B$. 
	The equality holds when $\Delta_0$ is constant.  
\end{proof}

{ 
\begin{proof}[Proof of Theorem \ref{individual external}]
		The proof is similar to the proof of Theorem 1. To prove this theorem, we denote  $\phi_{\mathrm{e}}$ as the efficient influence function for $\tau$ in $P_0$, $\eta^{\sI}_{\mathrm{e}}$ the efficient influence function for $\beta$ in $P_0$ and $\eta_{\mathrm{e}}$ the efficient influence function for $\beta$ in $P_0$. We first prove the result in parametric model.   We denote parametric models $P_0(Z;\theta_0)$ and $P_1(W;\theta_1)$ by $P_0^{\theta_0}$ and $P_1^{\theta_1}$ respectively for notational convenience. 
	Let $l(\theta_0) = \sum_{i=1}^n \log f(Z_i;\theta_0) $ and $l_{\text{ext}}(\theta_1) = \sum_{j=1}^m \log g(W_i;\theta_1) $ be the  log-likelihood for internal data and external data, where $f(Z;\theta_0)$ is the density function of $P_0^{\theta_0}$ and $g(W_i;\theta_1)$ is the density function of $P_1^{\theta_1}$. 
	We denote $\theta=(\theta_0, \theta_1)$, $\tau(\theta_0) = \tau(P_0^{\theta_0})$ and $\dot{\tau}(\theta) = (\dot{\tau}(\theta_0), 0)$.  
	Under Assumption 2, we have $\beta_{0}(\theta_0) :=\beta(P_0^{\theta_0}) = \beta(P_1^{\theta_1})=: \beta_{1}(\theta_1)$. This equality imposes a restriction on the parameter space, thereby constraining the direction of perturbation.
    To keep the regularity of the  parametric model, we assume $\theta_0$ and $\theta_1$ are functions of $\theta^\ast$ such that $P_0^{\theta^\ast}\times P_1^{\theta^\ast}$ is a regular parametric model. 
	We have  $\phi_{\mathrm{e}} = \dot{\tau}(\theta_0)J_{\theta_0}^{-1}\partial \log f(Z;\theta_0)/\partial \theta_0$, $\eta^{\sI}_{\mathrm{e}} = \dot{\beta}_{0}(\theta_0)J_{\theta_0}^{-1}\partial \log f(Z;\theta_0)/\partial \theta_0$, $\eta_{\mathrm{e}} = \dot{\beta}_{1}(\theta_1)H_{\theta_1}^{-1}\partial \log g(W;\theta_1)/\partial \theta_1$, where 
	\begin{equation*}
	\begin{aligned}
		J_{\theta_0} &= E[\{\partial \log f(Z;\theta_0)/\partial \theta_0\}\{\partial \log f(Z;\theta_0)/\partial \theta_0\}^\T], \\ 
		H_{\theta_1} &= E[\{\partial \log g(W;\theta_1)/\partial \theta_1\}\{\partial \log g(W;\theta_1)/\partial \theta_1\}^\T].
	\end{aligned}
	\end{equation*}
	The likelihood for the whole data is:
	\begin{equation*}
		\tilde{l}(\theta) = \sum_{i=1}^n \log f(Z_i;\theta_0) + \sum_{k=1}^m \log g(W_k;\theta_1).
	\end{equation*}

	We define the sequence
	\begin{equation*}
	\begin{aligned}
	 	U_n &= n^{1/2}\{T_n(\tilde{\beta}) - \tau(\theta_0)\}, \\ 
	 	V_n &= n^{-1/2}\left(\sum_{i=1}^n \partial \log f(Z_i;\theta_0)/\partial \theta_0, \sum_{k=1}^m \partial \log g(W_k;\theta_1)/\partial \theta_1\right).
	\end{aligned} 	
	\end{equation*}
	From the regularity of $T_n(\tilde{\beta})$ and the regularity of the parametric submodel, $U_n$ and $V_n$ converge in distribution to $U,V$, respectively. Therefore, they are marginally tight and thus jointly tight. Then for any subsequence $\{n'\}$, there exists a subsequence $\{n''\}$ such that $(U_{n''}, V_{n''})$ jointly converge to $(U, V)$, and $V\sim N(0, \tilde{I}_\theta )$, where 
    \[\tilde{I}_\theta = \begin{pmatrix}
		J_{\theta_0} & 0 \\ 
		0 & \rho H_{\theta_1}
	\end{pmatrix}.
    \]For convenience, we denote the subsequence $\{n''\}$ by $\{n\}$.

    Let $t_{n} = (t_{0n}^{\T}, t_{1n}^{\T})^{\T}$ be a sequence such that $t_n \rightarrow t$ for some $t$ and $\beta_{0}(\theta_{0} + t_{0n} / \sqrt{n}) = \beta_{1}(\theta_{1} + t_{1n} / \sqrt{n})$ for any $n$. Let $D_n = \tilde{l}(\theta + t_{n} / \sqrt{n} ) - \tilde{l}(\theta)$. Then, $(U_n, D_n)$ converges to $\big(U, t^\T V - t^\T\tilde{I}_\theta t/2\big)$ by the local asymptotic normality (LAN) property. The regularity of $T_n(\tilde{\beta})$ means that
	\[
	L_{\theta+ t_{n} / \sqrt{n}}\big(n^{1/2}\{ T_n(\tilde{\beta}) - \tau(\theta+ t_{n} / \sqrt{n}) \} \big) \rightarrow U.
	\]
	By the differentiability of $\tau(\theta)$, we have
	\[
	\begin{aligned}
		L_{\theta+ t_{n} / \sqrt{n}}(U_n)  
		&=L_{\theta+ t_{n} / \sqrt{n}}\big(n^{1/2}\{T_n(\tilde{\beta}) - \tau(\theta) \} \big)\\
		&= L_{\theta+ t_{n} / \sqrt{n}}\big(n^{1/2}\{T_n(\tilde{\beta}) - \tau(\theta+ t_{n} / \sqrt{n})\}+n^{1/2}\{\tau(\theta+ t_{n} / \sqrt{n}) - \tau(\theta)\}\big) \\
		&\rightarrow U + \dot{\tau}(\theta)t.
	\end{aligned}
	\]
	Therefore, 
	\begin{equation}\label{S9}
		E_{\theta+tn^{-1/2}}\{\exp(ia^\T U_n)\} \rightarrow \exp\{ia^\T\dot{\tau}(\theta)t\}E^\ast\{\exp(ia^\T U)\},
	\end{equation}
	where $E_\theta(\cdot)$ means the expectation taken under distribution $P_0^\theta\times P_1^\theta$ and $E^\ast(\cdot)$ is taken under the limiting distribution. 
	Besides, by contiguity, we have
	\begin{equation}\label{S10}
		\begin{aligned}
			E_{\theta+tn^{-1/2}}\{\exp(ia^\T U_n)\} &= E_{\theta}\{\exp(ia^\T U_n  + D_n)\} + o(1)\\ 
			&\rightarrow E^\ast\left[\exp\left\{ia^\T U + t^\T V - \frac{1}{2}t^\T\tilde{I}_\theta t\right\}\right].
		\end{aligned}
	\end{equation}
	Then, combining \eqref{S9} and \eqref{S10}, we have 
	\[
	E^\ast\left[\exp\left\{ia^\T U + t^\T V - \frac{1}{2}t^\T\tilde{I}_\theta t\right\}\right] = \exp\{ia^\T \dot{\tau}(\theta)t\}E^\ast\{\exp(ia^\T U)\}.
	\] 
	Due to the constraint $\beta_{0}(\theta_{0} + t_{0n} / \sqrt{n}) = \beta_{1}(\theta_{1} + t_{1n} / \sqrt{n})$, the direction $t$ can not take over the full space, we choose $t = -i\tilde{I}_\theta^{-1}\left( I - A  \right) \dot{\tau}^{\T}(\theta_0)(a - b)$, where $I$ is identity matrix, $A = B^{\T}\left\{B \tilde{I}_\theta^{-1} B^{\T} \right\}^{-1} B \tilde{I}_\theta^{-1}$ and $B = (\dot{\beta}_{0}(\theta_0), -\dot{\beta}_{1}(\theta_1))$. The term $I-A$ represents the constraint imposed by Assumption 2. Then, we obtain
	\begin{equation}\label{individual external 1}
		\begin{aligned}
			&\quad E^\ast\big(\exp[ia^\T\{U- \dot{\tau}(\theta)\left( I - A^{\T}  \right)\tilde{I}^{-1}_\theta V\} + ib^\T\dot{\tau}(\theta)\left( I - A^{\T}  \right)\tilde{I}^{-1}_\theta V]\big) \\ 
			&= E^\ast\left[\exp\left\{ia^\T U + \frac{1}{2}a^\T\dot{\tau}(\theta)\left( I - A^{\T}  \right)\tilde{I}^{-1}_\theta \dot{\tau}^{\T}(\theta) a\right\}\right]\exp\left\{-\frac{1}{2}b^\T\dot{\tau}(\theta) \left( I - A^{\T}  \right)\tilde{I}^{-1}_\theta \dot{\tau}^{\T}(\theta) b\right\}.
		\end{aligned}
	\end{equation}
	Equation~\eqref{individual external 1} gives the characteristic function of $(U_{n''} - \dot{\tau}(\theta)\left( I - A^{\T}  \right)\tilde{I}^{-1}_\theta V_{n''}, \dot{\tau}(\theta)\left( I - A^{\T}  \right)\tilde{I}^{-1}_\theta V_{n''})$'s limiting distribution and holds for every choice of initial subsequence $\{n'\}$. 
	Because the right-hand side of \eqref{individual external 1} is only related to the marginal distribution of $U$,  this equation  holds for the full sequence $\{n\}$, 
	that is, $(U_{n} - \dot{\tau}(\theta)\left( I - A^{\T}  \right)\tilde{I}^{-1}_\theta V_{n}, \dot{\tau}(\theta)\left( I - A^{\T}  \right) \tilde{I}^{-1}_\theta V_{n})$ has a limiting distribution with characteristic function in ~\eqref{individual external 1}.
	In equation~\eqref{individual external 1}, letting $a = 0$, we have
	\begin{equation}\label{individual external 2}
		E^\ast[\exp\{ib^\T\dot{\tau}(\theta)\left( I - A^{\T}  \right)\tilde{I}_\theta^{-1}V \}] = \exp\left\{-\frac{1}{2}b^\T\dot{\tau}(\theta)\left( I - A^{\T}  \right)\tilde{I}^{-1}_\theta \dot{\tau}^{\T}(\theta) b\right\}.
	\end{equation}
	Further letting $b=0$,   we have
	\begin{equation}\label{individual external 3}
		\begin{aligned}
			&E^\ast\big(\exp[ia^\T \{U- \dot{\tau}(\theta)\left( I - A^{\T}  \right)\tilde{I}^{-1}_\theta V \}]\big) \\ 
			=& E^\ast\left[\exp\left\{ia^\T U + \frac{1}{2}a^\T\dot{\tau}(\theta) \left( I - A^{\T}  \right)\tilde{I}^{-1}_\theta \dot{\tau}^{\T}(\theta) a\right\}\right]. \\ 
		\end{aligned}
	\end{equation}
	Combining the equations~\eqref{individual external 1}, \eqref{individual external 2} and \eqref{individual external 3} yields
	\[
	\begin{aligned}
		&\quad E^\ast\big(\exp[ia^\T\{U- \dot{\tau}(\theta) \left( I - A^{\T}  \right) \tilde{I}^{-1}_\theta V\} + ib^\T\dot{\tau}(\theta) \left( I - A^{\T}  \right) \tilde{I}^{-1}_\theta V]\big) \\
		&=E^\ast\big(\exp[ia^\T\{U- \dot{\tau}(\theta) \left( I - A^{\T}  \right) \tilde{I}^{-1}_\theta V\}]\big)E^\ast[\exp\{ib^\T\dot{\tau}(\theta) \left( I - A^{\T}  \right)\tilde{I}^{-1}_\theta V\}],
	\end{aligned}
	\]  
	which means that $U- \dot{\tau}(\theta) \left( I - A^{\T}  \right) \tilde{I}^{-1}_\theta V$ and $\dot{\tau}(\theta) \left( I - A^{\T}  \right) \tilde{I}^{-1}_\theta  V$ are independent. 
	In addition, we have
	\begin{equation}\label{parametric2}
		\begin{pmatrix}
			U_n - \dot{\tau}(\theta) \left( I - A^{\T}  \right) \tilde{I}^{-1}_\theta V_n \\ 
			\dot{\tau}(\theta) \left( I - A^{\T}  \right) \tilde{I}^{-1}_\theta V_n
		\end{pmatrix} \rightarrow \begin{pmatrix}
			U- \dot{\tau}(\theta) \left( I - A^{\T}  \right) \tilde{I}^{-1}_\theta V \\ 
			\dot{\tau}(\theta) \left( I - A^{\T}  \right)\tilde{I}^{-1}_\theta V
		\end{pmatrix}
	\end{equation}
	under the parametric model $P_0^{\theta_0}\times P_1^{\theta_1}$.

	Let $M = E(\phi_{\mathrm{e}}^{\sI}(\eta_{\mathrm{e}}^{\sI })^\T )\left\{ E(\eta_{\mathrm{e}}(\eta_{\mathrm{e}}^{\sE })^\T)/\rho + E(\eta_{\mathrm{e}}^{\sI}(\eta_{\mathrm{e}}^{\sI })^\T  \right\}^{-1}$. Note that
	\[
	\begin{aligned}
		&\quad n^{1/2}\dot{\tau}(\theta)\left( I - A^{\T}  \right)\tilde{I}^{-1}_\theta V_n  \\ 
		&= n^{1/2}\dot{\tau}(\theta)\left[I - \tilde{I}^{-1}_\theta \begin{pmatrix}
			\dot{\beta}^\T(\theta_0) \\ 
			-\dot{\beta}^\T(\theta_1)
		\end{pmatrix} \left\{(\dot{\beta}_{0}(\theta_0), -\dot{\beta}_{1}(\theta_1))\tilde{I}^{-1}_\theta (\dot{\beta}_{0}(\theta_0), -\dot{\beta}_{1}(\theta_1))^{\T} \right\}^{-1}(\dot{\beta}_{0}(\theta_0), -\dot{\beta}_{1}(\theta_1))\right] \tilde{I}_\theta^{-1}V_n \\
		&= (\dot{\tau}(\theta_0)J_{\theta_0}^{-1}, 0) \left[I - \begin{pmatrix}
			\dot{\beta}^\T(\theta_0) \\ 
			-\dot{\beta}^\T(\theta_1)
		\end{pmatrix}\left\{E(\eta_{\mathrm{e}}(\eta_{\mathrm{e}}^{\sE })^\T)/\rho + E(\eta_{\mathrm{e}}^{\sI}(\eta_{\mathrm{e}}^{\sI })^\T ) \right\}^{-1}(\dot{\beta}_{0}(\theta_0), -\dot{\beta}_{1}(\theta_1)) \tilde{I}_\theta^{-1}\right]V_n\\
		&= \biggl[\dot{\tau}(\theta_0)J_{\theta_0}^{-1} - \dot{\tau}(\theta_0)J_{\theta_0}^{-1}\dot{\beta_0}^\T(\theta)\left\{E(\eta_{\mathrm{e}}(\eta_{\mathrm{e}}^{\sE })^\T)/\rho + E(\eta_{\mathrm{e}}^{\sI}(\eta_{\mathrm{e}}^{\sI })^\T ) \right\}^{-1}\dot{\beta}_{0}(\theta_0) J_{\theta_0}^{-1}\biggl] \sum_{i=1}^n\frac{\partial \log f(Z_i;\theta_0)}{\partial \theta_0}\\
		&\quad+ \biggl[\dot{\tau}(\theta_0)J_{\theta_0}^{-1}\dot{\beta_0}^\T(\theta)\left\{E(\eta_{\mathrm{e}}(\eta_{\mathrm{e}}^{\sE })^\T)/\rho + E(\eta_{\mathrm{e}}^{\sI}(\eta_{\mathrm{e}}^{\sI })^\T ) \right\}^{-1}\dot{\beta}_{1}(\theta_1) H_{\theta_1}^{-1}/\rho\biggl] \sum_{k=1}^m\frac{\partial \log g(W_k;\theta_1)}{\partial \theta_1} \\ 
		&= \sum_{i=1}^n \{ \phi_{\mathrm{e}}(Z_i) - M\eta^{\sI}_{\mathrm{e}}(Z_i) \} + \frac{1}{\rho}\sum_{k=1}^mM \eta_{\mathrm{e}}(W_k). \\ 
	\end{aligned}
	\]
	Thus, \eqref{parametric2} implies that 
	\[
	n^{1/2}\left[\begin{matrix}
		T_n(\tilde{\beta}) - \tau(P_0) - n^{-1} \left[ \sum_{i=1}^n \{ \phi_{\mathrm{e}}(Z_i) - M\eta^{\sI}_{\mathrm{e}}(Z_i) \} + \frac{1}{\rho}\sum_{k=1}^mM \eta_{\mathrm{e}}(W_k)\right] \\
		n^{-1} \left[ \sum_{i=1}^n \{ \phi_{\mathrm{e}}(Z_i) - M\eta^{\sI}_{\mathrm{e}}(Z_i) \} + \frac{1}{\rho}\sum_{k=1}^mM \eta_{\mathrm{e}}(W_k)\right]
	\end{matrix}
	\right]\rightarrow \begin{pmatrix}
		U- \dot{\tau}(\theta)\tilde{I}^{-1}_\theta V \\ 
		\dot{\tau}(\theta)\tilde{I}^{-1}_\theta V
	\end{pmatrix},
	\]
	thus the convolution theorem holds in parametric models.

	We generalize the above results to semiparametric models.  
	We denote $$R_n = n^{-1/2}\left[ \sum_{i=1}^n \{ \phi_{\mathrm{e}}(Z_i) - M\eta^{\sI}_{\mathrm{e}}(Z_i) \} + \frac{1}{\rho}\sum_{k=1}^mM \eta_{\mathrm{e}}(W_k)\right], $$ 
	where $M = E(\phi_{\mathrm{e}}(\eta^{\sI}_{\mathrm{e}})^\T) \left\{ E(\eta_{\mathrm{e}}(\eta_{\mathrm{e}})^\T)/\rho + E(\eta^{\sI}_{\mathrm{e}}(\eta^{\sI}_{\mathrm{e}})^\T ) \right\}^{-1}$. 
	As $\tau$ and $\beta$ are both pathwise differentiable, there exists a sequence of parametric submodels in the semiparametric model $\mathcal P_{\rm trans}=\{P_0\times P_1: P_0, P_1 \text{  satisfy Assumption 2}\}$ having a uniformly bounded sequence of score $s_j = (s^{(0)}_j, s^{(1)}_j)$, where $s^{(0)}_j=\partial \log f_j(Z;\theta_0)/\partial \theta_0$ and $s^{(1)}_j=\partial \log g_j(W;\theta_1)/\partial \theta_1$ such that
	\begin{equation}\label{approximate2}
	\begin{aligned}
		&\|\dot{\tau}(\theta_0;j) J_{\theta_0,j}^{-1}s^{(0)}_j - \phi_{\mathrm{e}}\|_{P_0}\rightarrow 0, \|\dot{\beta}(\theta_0;j) J_{\theta_0,j}^{-1}s^{(0)}_j - \eta^{\sI}_{\mathrm{e}}\|_{P_0} \rightarrow 0, ~\text{as $j \rightarrow \infty$ }, \\ 
		&\|\dot{\beta}(\theta_1;j) J_{\theta_1,j}^{-1}s^{(1)}_j - \eta_{\mathrm{e}}\|_{P_1} \rightarrow 0, ~\text{as $j \rightarrow \infty$ }
	\end{aligned}
	\end{equation}
	where $f_j(Z;\theta_0)$ is the density function of $j$-th parametric submodel for $P_0$ and $g_j(W;\theta_1)$ is the density function of $j$-th parametric submodel for $P_1$, $\dot{\tau}(\theta_0;j), \dot{\beta}(\theta_0;j)$ and $\dot{\beta}(\theta_1;j)$ are the derivatives of $\tau, \beta_{0}(\theta_0)$ and $\beta_{1}(\theta_1)$ in the $j$-th parametric submodel respectively, $J_{\theta_0,j}$ is the Fisher information for $\theta_0$ in the $j$-th parametric submodel of $P_0$, $H_{\theta_1,j}$ is the Fisher information for $\theta_1$ in the $j$-th parametric submodel of $P_1$, and   $\|f\|_{P_0} = (\int f^2dP_0)^{1/2}, \|g\|_{P_1} = (\int g^2dP_1)^{1/2}$. 
	Let
	\[
	\begin{aligned}
		V_{nj} &= n^{-1/2} \Biggl[\sum_{i=1}^n s^{(0)}_j(Z_i),  \sum_{k=1}^m s^{(1)}_j(W_k) \Biggl],
	\end{aligned}
	\]
	and suppose $U_n, V_{nj}$ marginally converge to $U$ and $V_j$, respectively. 
	From   \eqref{parametric2} for parametric submodels, we have
	\[
	\begin{pmatrix}
		U_n - \dot{\tau}(\theta;j) \left( I - A_j^{\T}  \right) \tilde{I}^{-1}_{\theta,j} V_{nj} \\ 
		\dot{\tau}(\theta;j) \left( I - A_j^{\T}  \right)\tilde{I}^{-1}_{\theta,j} V_{nj}
	\end{pmatrix} \rightarrow \begin{pmatrix}
		U- \dot{\tau}(\theta;j) \left( I - A_j^{\T}  \right) \tilde{I}^{-1}_{\theta,j} V_j \\ 
		\dot{\tau}(\theta;j) \left( I - A_j^{\T}  \right) \tilde{I}^{-1}_{\theta,j} V_j
	\end{pmatrix},
	\]
	where $\tilde{I}_{\theta,j} = \begin{pmatrix}
		J_{\theta_0,j} & 0 \\ 
		0 & \rho H_{\theta_1, j}
	\end{pmatrix}$ and $A_j = B_j^{\T}\left\{B_j \tilde{I}_{\theta,j}^{-1} B_j^{\T} \right\}^{-1} B_j \tilde{I}_{\theta,j}^{-1}$ and $B_j = (\dot{\beta}(\theta_0;j), -\dot{\beta}(\theta_1;j))$. 
	Also, we have
	\[
	\begin{aligned}
		&\quad E^\ast\big(\exp[ia^\T\{U- \dot{\tau}(\theta;j) \left( I - A_j^{\T}  \right) \tilde{I}^{-1}_{\theta,j} V_j \} + ib^\T\dot{\tau}(\theta;j) \left( I - A_j^{\T}  \right) \tilde{I}^{-1}_{\theta,j} V_j ]\big) \\
		&=E^\ast\big(\exp[ia^\T\{U- \dot{\tau}(\theta;j) \left( I - A_j^{\T}  \right) \tilde{I}^{-1}_{\theta,j} V_j \}]\big)E^\ast[\exp\{ib^\T\dot{\tau}(\theta;j) \left( I - A_j^{\T}  \right) \tilde{I}^{-1}_{\theta,j} V_j\}].
	\end{aligned}
	\]  
	For $(U_n, R_n)$, we know $U_n$ and $R_n$ marginally converges to a limiting distribution. For any subsequence $\{n'\}$, there exists a subsequence $\{n''\}$ such that $(U_{n''}, R_{n''})$ has a joint limiting distribution $(U, R)$. And we denote $M_j = \dot{\tau}(\theta_0;j)J^{-1}_{\theta_0,j}\dot{\beta}^{\T}(\theta_0;j)\left\{\dot{\beta}(\theta_0;j)J^{-1}_{\theta_0,j} \dot{\beta}^{\T}(\theta_0;j) + \frac{1}{\rho} \dot{\beta}(\theta_1;j)H^{-1}_{\theta_1,j} \dot{\beta}^{\T}(\theta_1;j)  \right\}^{-1}$. For ease of notation, we  denote the subsequence $\{n''\}$ by $\{n\}$. We have 
	\[
	\begin{aligned}
		&\quad\sup_n E_\theta\{R_n - \dot{\tau}(\theta;j) \left( I - A_j^{\T}  \right) \tilde{I}^{-1}_{\theta,j} V_{nj} \}^{\otimes 2} \\
		&= E_\theta\{\phi_{\mathrm{e}} - M\eta_{\mathrm{e}} - \dot{\tau}(\theta_0;j)J^{-1}_{\theta_0,j} s^{(0)}_j - M_j\dot{\beta}(\theta_0;j)J^{-1}_{\theta_0,j} s^{(0)}_j \}^{\otimes 2} \\\ 
		&\quad+ n/m \frac{1}{\rho^2}E_\theta\{  M \eta_{\mathrm{e}}(W_k)  - M_j \dot{\beta}(\theta_1;j)H^{-1}_{\theta_1,j}s^{(1)}_j  \}^{\otimes 2},
	\end{aligned}
	\]
	where $E_\theta(X)^{\otimes 2} = E_\theta(XX^\T)$. 
	From   \eqref{approximate2}, we know when $j\rightarrow +\infty$,
	\[
	\begin{aligned}
		E\{\dot{\tau}(\theta_0;j) J_{\theta_0,j}^{-1} \dot{\beta}^\T(\theta_0;j) \} =E\{ \dot{\tau}(\theta_0;j) J_{\theta_0,j}^{-1}s^{(0)}_j (\dot{\beta}(\theta_0;j) J_{\theta,j}^{-1}s^{(0)}_j)^\T  \} &\rightarrow E\{\phi_{\mathrm{e}} (\eta^{\sI}_{\mathrm{e}})^\T \},  \\ 
		E\{\dot{\beta}(\theta_0;j) J_{\theta_0,j}^{-1} \dot{\beta}^\T(\theta_0;j) \} = E\{ \dot{\beta}(\theta_0;j) J_{\theta_0,j}^{-1}s^{(0)}_j (\dot{\beta}(\theta_0;j) J_{\theta,j}^{-1}s^{(0)}_j)^\T  \} &\rightarrow E\{\eta^{\sI}_{\mathrm{e}} (\eta^{\sI}_{\mathrm{e}})^\T \}, \\ 
		E\{\dot{\beta}(\theta_1;j) H_{\theta_1,j}^{-1} \dot{\beta}^\T(\theta_1;j) \} = E\{ \dot{\beta}(\theta_1;j) H_{\theta_1,j}^{-1}s^{(1)}_j (\dot{\beta}(\theta_1;j) J_{\theta,j}^{-1}s^{(1)}_j)^\T  \} &\rightarrow E\{\eta^{\sI}_{\mathrm{e}} (\eta^{\sI}_{\mathrm{e}})^\T \},
	\end{aligned}
	\]
	so we have $M_j \rightarrow M$.
	Therefore, $\sup_n E_\theta\{R_n - \dot{\tau}(\theta;j) \left( I - A_j^{\T}  \right) \tilde{I}^{-1}_{\theta,j} V_{nj} \}^{\otimes 2} \rightarrow 0$ as $j\rightarrow \infty$. Then we must have
	\[
	L(U-R, R) = \lim_{j\rightarrow \infty} L(U - \dot{\tau}(\theta;j) \left( I - A_j^{\T}  \right) \tilde{I}^{-1}_{\theta,j} V_j, \dot{\tau}(\theta;j) \left( I - A_j^{\T}  \right) \tilde{I}^{-1}_{\theta,j} V_j),
	\]  
	and 
	\[
	\begin{aligned}
		&\quad E^\ast[\exp\{ia^\T(U- R) + ib^\T R\}] \\ 
		&=\lim_{j\rightarrow\infty} E^\ast\big(\exp[ia^\T\{U- \dot{\tau}(\theta;j) \left( I - A_j^{\T}  \right) \tilde{I}^{-1}_{\theta,j} V_j\} + ib^\T\dot{\tau}(\theta;j) \left( I - A_j^{\T}  \right) \tilde{I}^{-1}_{\theta,j} V_j]\big) \\
		&=\lim_{j\rightarrow \infty} E^\ast\big(\exp[ia^\T\{U- \dot{\tau}(\theta;j) \left( I - A_j^{\T}  \right) \tilde{I}^{-1}_{\theta,j} V_j\}]\big) \lim_{j\rightarrow \infty}E^\ast[\exp\{ib^\T\dot{\tau}(\theta;j) \left( I - A_j^{\T}  \right) \tilde{I}^{-1}_{\theta,j} V_j\}] \\ 
		&= E^\ast[\exp\{ia^\T(U- R)\}]E\{\exp(ib^\T R)\}.
	\end{aligned}
	\]  
	This  implies that $U - R$ and $R$ are independent. 
	Note that $L(U-R, R)$ is the same for any subsequence $\{n''\}$, so we have 
	\begin{equation}\label{S16}
		(U_n - R_n, R_n) \rightarrow (U-R, R).    
	\end{equation}
	Let $\Delta_0 = U -R$  and $S_0 = R$. Then, it holds that $S_0 \sim N(0, B)$ where $B = E(\phi_{\mathrm{e}}\phi_{\mathrm{e}}^\T ) - E(\phi_{\mathrm{e}}(\eta^{\sI}_{\mathrm{e}})^\T) \left\{ E(\eta_{\mathrm{e}}(\eta_{\mathrm{e}})^\T)/\rho + E(\eta^{\sI}_{\mathrm{e}}(\eta^{\sI}_{\mathrm{e}})^\T ) \right\}^{-1}E(\phi_{\mathrm{e}}(\eta^{\sI}_{\mathrm{e}})^\T )^\T$.
	Finally, plugging in the formula of $R_n$, \eqref{S16} can be written as 
	\[
	n^{1/2}\left[\begin{matrix}
		T_n(\tilde{\beta}) - \tau(P_0) - n^{-1} \left[ \sum_{i=1}^n \{ \phi_{\mathrm{e}}(Z_i) - M\eta^{\sI}_{\mathrm{e}}(Z_i) \} + \frac{1}{\rho}\sum_{k=1}^mM \eta_{\mathrm{e}}(W_k)\right] \\
		n^{-1} \left[ \sum_{i=1}^n \{ \phi_{\mathrm{e}}(Z_i) - M\eta^{\sI}_{\mathrm{e}}(Z_i) \} + \frac{1}{\rho}\sum_{k=1}^mM \eta_{\mathrm{e}}(W_k)\right]
	\end{matrix}
	\right]\rightarrow \begin{pmatrix}
		\Delta_0\\
		S_0
	\end{pmatrix}.
	\]

	Then the efficiency bound for $\tau$ is $B$, which completes the proof.

\end{proof}
}

\begin{proof}[Proof of Theorem \ref{efficient estimator}]
	Under Assumptions \ref{beta}--\ref{well defined}, and regularity Condition \ref
    {cond: regular internal}, we have
    \[
      \hat{\tau}_{\mathrm{e}} - \tau = \hat{E}\{\phi_{\mathrm{e}}- M\eta_{\mathrm{e}} + M(\tilde{\beta} - \beta)\} + o_{P}(n^{-1/2}).
    \]
    Thus, $n^{1/2}(\hat{\tau}_{\mathrm{e}} - \tau) \to N(0, B)$ according to Assumption \ref{beta}, Condition \ref{cond: regular internal} and the central limit theorem.
	Therefore, $\hat{\tau}_{\mathrm{e}}$ achieves the efficiency bound in \eqref{efficiency bound} of  Theorem \ref{efficiency theorem}.
\end{proof}

\begin{proof}[Proof of Theorem \ref{adaptive fusion}]
Without loss of generality, assume $\mathcal{A} = \{1, \dots, q_{1}\}$ for some positive integer $q_{1}$. Recall that $\lambda \to \infty$ and $\lambda n^{-\alpha/2} \to 0$. Then, $\hat{a}_{j} \to 1$ in probability if $j \in \mathcal{A}$ and $\hat{a}_{j} \to 0$ if $j \not\in \mathcal{A}$. We have
\[
   \widehat{\Sigma}_{\phi\eta}\widehat{A} \to 
   \Big(
E(\phi_{\mathrm{e}}\eta_{\mathrm{e}, \mathcal{A}}^\T) ~~~ 0
   \Big)
\]
and
\[
  \left(I - \widehat{A} + \hat{a}\hat{a}^{\T}\right)\odot\left(\widehat{\Sigma}_1/\rho + \widehat{\Sigma}_{\eta\eta}\right) \to
  \begin{pmatrix}
  	\Sigma_{1, \mathcal{A}} / \rho + E(\eta_{\mathrm{e}, \mathcal{A}}\eta_{\mathrm{e}, \mathcal{A}}^\T) & 0\\
  	0 & D_{\eta, \mathcal{A}^{c}}
  \end{pmatrix}  
\]
in probability, where $D_{\eta, \mathcal{A}^{c}}$ is the diagonal matrix consisting of the diagonal elements of $\Sigma_{1, \mathcal{A}^{c}} / \rho + E(\eta_{\mathrm{e}, \mathcal{A}^{c}}\eta_{\mathrm{e}, \mathcal{A}^{c}}^\T)$ and $\mathcal{A}^{c} = \{j: j \not\in \mathcal{A}\}$.
This implies that
\[
   \begin{aligned}
   	  \hat{\tau}_\mathrm{adf} 
   	  & = \hat{\tau}_{\mathrm{e}}^{\sI} - 
   	  	\Big(E(\phi_{\mathrm{e}}\eta_{\mathrm{e}, \mathcal{A}}^\T) ~~~ 0
       \Big)
   	  \begin{pmatrix}
   	  	\Sigma_{1, \mathcal{A}} / \rho + E(\eta_{\mathrm{e}, \mathcal{A}}\eta_{\mathrm{e}, \mathcal{A}}^\T) & 0\\
   	  	0 & D_{\eta, \mathcal{A}^{c}}
   	  \end{pmatrix}^{-1}
   	  (\hat{\beta}_{\mathrm{e}}^{\sI} - \tilde{\beta})
   	  + o_{P}\left(n^{-1/2}\right)\\
   	  & = \hat{\tau}_{\mathrm{e}}^{\sI} - 
   	  E(\phi_{\mathrm{e}}\eta_{\mathrm{e}, \mathcal{A}}^\T)\left\{\Sigma_{1, \mathcal{A}} / \rho + E(\eta_{\mathrm{e}, \mathcal{A}}\eta_{\mathrm{e}, \mathcal{A}}^\T)\right\}^{-1}
   	  (\hat{\beta}_{\mathrm{e}}^{\sI,\mathcal{A}} - \tilde{\beta}_{\mathcal{A}})
   	  + o_{P}\left(n^{-1/2}\right)\\
   	  & = \hat{\tau}_{\mathrm{orc}} + o_{P}\left(n^{-1/2}\right). 
   \end{aligned}
\] 
\end{proof}

\section{Proofs of Propositions}
We need the following lemma for proof of Proposition \ref{characterization}.

\begin{lemma}[Le Cam's third lemma] \label{lecam} 
	Suppose $P_n, Q_n$ are two sequences of probability measures and $X_n$ is a sequence of random variables, if 
	\[
	L_{P_n}\left(X_n, \log\frac{dQ_n}{dP_n} \right) \rightarrow N\left\{\begin{pmatrix}
		\mu\\ 
		-\frac{1}{2}\sigma^2
	\end{pmatrix}, \begin{pmatrix}
		\Sigma & \varphi\\ 
		\varphi^\T   &  \sigma^2
	\end{pmatrix} \right\},
	\]
	then  $L_{Q_n}(X_n) \rightarrow N(\mu+\varphi, \Sigma)$.
\end{lemma}

\begin{proof}[Proof of Proposition \ref{characterization}]
	By   Taylor's expansion, we have $\gamma(\tilde{\beta}) = \xi(P_0)(\tilde{\beta} - \beta) + o\big( \tilde{\beta} - \beta \big)$. As $\tilde{\beta} - \beta = \tilde{\beta} - \beta(P_1)=  1/m\sum_{j=1}^m \tilde{\eta}(W_j) + e_m$ with $E_{P_1}\{\tilde{\eta}(W)\} = 0$, $E_{P_1}\{\tilde{\eta}(W)\tilde{\eta}^\T(W)\}$ finite and nonsingular and $e_m = o_P(m^{-1/2}) = o_P(n^{-1/2})$, we have $\tilde{\beta} - \beta = O_P(m^{-1/2}) = O_P(n^{-1/2})$. 
	Therefore $T_n(\tilde{\beta}) = \tau + \frac{1}{n}\sum_{i=1}^n\psi(Z_i) + \xi(\tilde{\beta} - \beta) + o_P(n^{-1/2})$.
	
	For the ease of notation, we denote $P_0(Z;\theta)$ and $P_1(W;\theta)$ by $P_0^\theta$ and $P_1^\theta$ respectively. 
	Suppose that $P_0^\theta\times P_1^\theta$ is a parametric submodel in semiparametric model   $\mathcal{P}_0\times \mathcal{P}_1=\{P_0\times P_1: P_0, P_1 \text{  satisfy Assumption~\ref{well defined}}\}$.  
	Let $l(\theta) = \sum_{i=1}^n \log f(Z_i;\theta) $ be the  log-likelihood for internal data, where $f(Z;\theta)$ is the density function of $P_0^\theta$. 
	We denote $\tau(\theta) :=\tau(P_0^\theta), \beta(\theta) := \beta(P_0^\theta)=\beta(P_1^\theta)$.    
	
	Under Assumption \ref{beta} and Condition \ref{cond: regular beta-tilde}, the log-likelihood combining internal data and  external summary statistics satisfies
	\[
	\begin{aligned}
		\tilde{l}(\theta) &= l(\theta) - \frac{m}{2}\{\tilde{\beta} - \beta(P_1^{\theta})\}^\T\Sigma_1^{-1}\{\tilde{\beta} - \beta(P_1^\theta)\} - \frac{1}{2}\log\left(|2\pi m^{-1}\Sigma_1|\right) + o_{P}(1),
	\end{aligned}
	\]
	according to Lemma 1 in \cite{boos1985converse}.
	Letting  
	\[(U_n, V_n) = \left( n^{1/2}\{T_n(\tilde{\beta}) - \tau(\theta)\}, n^{-1/2} \left[\sum_{i=1}^n \partial \log f(Z_i; \theta)/\partial \theta + m\dot{\beta}^\T(\theta)\Sigma^{-1}_1\{\tilde{\beta} - \beta(\theta) - e_m\}\right] \right),\] 
	then by the central limit theorem we have $(U_n, V_n) \rightarrow (U, V)$,  
	\[
	\begin{pmatrix}
		U \\ 
		V
	\end{pmatrix} \sim N\left\{0, \begin{pmatrix}
		E(\psi\psi^\T) + \xi\Sigma_1/\rho \xi^\T ~~& E\{\psi\frac{\partial \log f(Z_i;\theta)}{\partial \theta} \} + \xi\dot{\beta}(\theta)\\ 
		[E\{\psi\frac{\partial \log f(Z_i;\theta)}{\partial \theta} \} + \xi\dot{\beta}(\theta)]^\T ~~ &  J_\theta + \dot{\beta}^\T(\theta)\rho\Sigma_1^{-1}\dot{\beta}(\theta)
	\end{pmatrix} \right\},
	\]
	where $J_\theta = E[\{\partial \log f(Z_i;\theta)/\partial \theta\}^\T\{\partial \log f(Z_i;\theta)/\partial \theta\}]$. 
	Letting $\tilde{I}_\theta = J_\theta+\dot{\beta}^\T(\theta)\rho\Sigma_1^{-1}\dot{\beta}(\theta)$,
	$(U_n, D_n) = \big(n^{1/2}\{T_n(\tilde{\beta}) - \tau(\theta)\}, \tilde{l}(\theta + tn^{-1/2}) - \tilde{l}(\theta)\big)$, we have $L_{\theta}(U_n, D_n) \rightarrow \big(U, t^\T V - t^\T\tilde{I}_\theta t/2\big)$ by the LAN property, where $L_{\theta}(U_n) \rightarrow U$ represents that $U_n$ converges in distribution to $U$ under the law of $P_0^\theta\times P_1^\theta$. 
	Then, by Le Cam's third lemma (Lemma \ref{lecam}), we have
	\[
	L_{\theta + tn^{-1/2}}\big(n^{1/2}\{T_n(\tilde{\beta}) - \tau(\theta)\}\big) = L_{\theta + tn^{-1/2}}(U_n) \rightarrow N\big( E^\ast(U^\T V) t, E^\ast(U^\T U)\big),
	\]
	where $E^\ast(\cdot)$ is taken under the limiting distribution.
	By the regularity of $T_n$, we have
	\[
	L_{\theta+tn^{-1/2}}\big(n^{1/2}\{T_n(\tilde{\beta}) - \tau(\theta + tn^{-1/2})\}\big) \rightarrow N\big(0, E^\ast(U^\T U)\big),
	\]
	and $n^{1/2}\{\tau(\theta + tn^{-1/2}) - \tau(\theta)\} \rightarrow E^\ast(U^\T V) t$, where $E^\ast(U^\T V) = E\{\psi\partial\log f(Z;\theta)/\partial \theta\}$  $+ \xi\dot{\beta}(\theta)$. 
	So we have $E\{\psi\partial\log f(Z;\theta)/\partial \theta\} + \xi\dot{\beta}(\theta) = \dot{\tau}(\theta)$ for all parametric submodels.
	Then we get
	\[
	\begin{aligned}
		E\{(\psi + \xi\eta_{\mathrm{e}} )\partial\log f(Z;\theta)/\partial \theta \}= \dot{\tau}(\theta).\\
	\end{aligned}
	\]
	As a result, $\psi + \xi\eta_{\mathrm{e}}$ is an influence function for $\tau$.
	Letting  $\phi = \phi(Z) = \psi + \xi\eta_{\mathrm{e}}$ denote an arbitrary influence function for $\tau$,   then $\psi= \phi - \xi\eta_{\mathrm{e}}$,
	which gives  the representation~\eqref{complex} in Proposition~\ref{characterization}.
	
	From the linearization of $T_n(\tilde{\beta})$, we can   calculate the asymptotic variance:
	\[
	\begin{aligned}
		\mathrm{avar}(T_n(\tilde{\beta})) &= \mathrm{var}(\phi - \xi\eta_{\mathrm{e}}) + \xi n\mathrm{var}(\tilde{\beta})\xi^\T\\
		& \rightarrow  E(\phi\phi^\T) + \xi E(\eta_{\mathrm{e}}\eta_{\mathrm{e}}^\T)\xi^\T -2\xi E(\eta_{\mathrm{e}}\phi^\T) + \xi\Sigma_1\xi^\T/\rho.
	\end{aligned}
	\]
	Because $T_{n, \mathrm{e}}(\beta)$ is the efficient estimator for $\tau$ when $\beta$ is known, the corresponding $\phi$ is the efficient influence function $\phi_{\mathrm{e}}$ for $\tau$, and $\xi = E(\phi_{\mathrm{e}}\eta_{\mathrm{e}}^\T)\{E(\eta_{\mathrm{e}}\eta_{\mathrm{e}}^\T)\}^{-1}$. Plugging these two terms into the above formula we obtain the asymptotic variance of $T_{n, \mathrm{e}}(\beta)$.
	
\end{proof}

	

\begin{proof}[Proof of Proposition 2]
	Consider the following exponential family distribution,
	\[
	p(y_i;\theta_i, \psi) = \exp\left[\left\{y_i\theta_i - b(\theta_i)\right\}/a(\psi)\right]c(y_i,\psi),
	\]
	where $\theta_i, \psi$ are parameters and $a(\cdot), b(\cdot), c(\cdot, \cdot)$ are known functions. 
	It is well known that $E(Y_i) = \dot{b}(\theta_i), \mathrm{var}(Y_i) = a(\psi)\ddot{b}(\theta_i)$. 
	Denote $X = (X_1^\T, X_2^\T)$, $v_i = X_{i1}^\T\tau  + X_{i2}^\T \zeta$, $\mu_i = E(Y_i\mid X_i) = g^{-1}(v_i) = \dot{b}(\theta_i)$, where $g(\cdot)$ is the link function. 
	The efficient influence function for $(\tau^\T, \zeta^\T)^\T$ is
	\[
	\phi_{\mathrm{e}} = \{E(X WX^\T )\}^{-1}X W\frac{dv}{d\mu}\left\{Y-g^{-1}(X_1^\T\tau  + X_2^\T \zeta)\right\}/a(\psi), 
	\]
	where $W = \mathrm{var}(Y\mid X)/\{a^2(\psi) \ddot{b}^2(\theta) \dot{g}^2(\mu)\} = 1/\{a(\psi) \ddot{b}(\theta) \dot{g}^2(\mu) \}$ and $dv/d\mu = \dot{g}(\mu)$. 
	Solving the estimating equation $\sum_{i}[ X_i w_i (dv_i/d\mu_i )\{Y_i - g^{-1}(X_{i1}^\T\tau  + X_{i2}^\T \zeta)\} ] = 0$ results in an efficient estimator of $(\tau^\T, \zeta^\T)^\T$, where  $w_{i} = 1/\{a(\psi) \ddot{b}(\theta_i)\dot{g}^2(\mu_i) \}$ and  $dv_i/d\mu_i = \dot{g}(\mu_i)$.

	When $g(\cdot)$ is the canonical link function, that is $g(\cdot) = \dot{b}^{-1}(\cdot)$, then $\theta_i = \dot{b}^{-1}(E(Y_i\mid X_i)) =v_i= X_{i1}^\T\tau  + X_{i2}^\T \zeta $ and $\dot{g}(\mu_i) = 1/\ddot{b}(\theta_i) =1/ \ddot{b}(X_{i1}^\T\tau  + X_{i2}^\T \zeta)$. 
	The efficient influence function for $(\tau^\T, \zeta^\T)^\T$ simplifies to 
	\[
	\phi_{\mathrm{e}} = \{E(X WX^\T)\}^{-1}X\{Y-g^{-1}(X_1^\T\tau  + X_2^\T \zeta)\}/a(\psi), 
	\] 
	where $W = \ddot{b}(X_1^\T\tau  + X_2^\T \zeta)/a(\psi)$. 
	Consider a nested working model $g^{-1}(X_2^\T\beta)$ for $E(Y\mid X_2)$, where $X_2$ is $q$-dimensional. 
	The functional $\beta$ is defined as the solution to estimating equation $E[ X_2\{Y - g^{-1}(X_2^\T\beta) \} ] = 0$. 
	An influence function $\eta_1$ for $\beta$ is 
	\[
	\eta_1 = \{E(X_2\tilde{W}X_2^\T)\}^{-1}X_2\{Y-g^{-1}(X_2^\T\beta)\}/a(\tilde{\psi}),
	\]
	where $\tilde{W} = \ddot{b}(X_2^\T\beta)/a(\tilde{\psi})$. 
	We have
	\[
	\begin{aligned}
		&\quad E(\phi_{\mathrm{e}}\eta_{\mathrm{e}}^\T) \\ 
		&= E(\phi_{\mathrm{e}}\eta_1^\T)\\ 
		&= \{E(X WX^\T)\}^{-1} E\left[X\{Y-g^{-1}(X^\T\tau)\}\{Y-g^{-1}(X_2^\T\beta)\}X_2^\T \right] \{E(X_2\tilde{W}X_2^\T)\}^{-1}/\{a(\psi)a(\tilde{\psi})\}\\
		&= \{E(X WX^\T)\}^{-1} E(X WX_2^\T)\{E(X_2\tilde{W}X_2^\T)\}^{-1}\frac{a(\psi)}{a(\tilde{\psi})}\\
		&= \begin{pmatrix}
			0_{(p-q)\times q} \\ 
			I_{q\times q}\\
		\end{pmatrix}\{E(X_2\tilde{W}X_2^\T)\}^{-1} \frac{a(\psi)}{a(\tilde{\psi})}\\
		&= \begin{pmatrix}
			0_{(p-q)\times q}  \\ 
			\{E(X_2\tilde{W}X_2^\T)\}^{-1}
		\end{pmatrix} \frac{a(\psi)}{a(\tilde{\psi})}.
	\end{aligned}
	\]
	So from the efficiency bound \eqref{efficiency bound}, there is no efficiency gain for $\tau$.
	
\end{proof}

\begin{proof}[Proof of Proposition 3]
	Denote the objective function in equation \eqref{IVW} by $Q$, and $(\hat{\tau}_{\mathrm{CD}}, \hat{\beta}_{\mathrm{CD}})$  the minimizer of $Q$, we   prove that $\hat{\tau}_{\mathrm{CD}} = \hat{\tau}_{\mathrm{e}}$.
	
	Note  that $(\hat{\tau}_{\mathrm{CD}}, \hat{\beta}_{\mathrm{CD}})$ satisfies the first-order conditions:
	\[
	\begin{aligned}
		&\frac{\partial Q}{\partial (\tau,\beta)} = -2 \widehat{\Sigma}^{-1}\begin{pmatrix}
			\hat{\tau}_{\mathrm{e}}^{\sI} - \tau \\ 
			\hat{\beta}_{\mathrm{e}}^{\sI} - \beta
		\end{pmatrix} - 2\begin{pmatrix}
			0 & 0\\
			0 & \rho \widehat{\Sigma}_1^{-1}
		\end{pmatrix}\begin{pmatrix}
			0 \\ 
			\tilde{\beta} - \beta
		\end{pmatrix} = 0\\
		&\Rightarrow \left\{ \widehat{\Sigma}^{-1} + \begin{pmatrix}
			0 &0\\
			0 & \rho\widehat{\Sigma}_1^{-1}
		\end{pmatrix} \right\}\begin{pmatrix}
			\tau \\ 
			\beta
		\end{pmatrix} = \widehat{\Sigma}^{-1}\begin{pmatrix}
			\hat{\tau}_{\mathrm{e}}^{\sI}\\ 
			\hat{\beta}_{\mathrm{e}}^{\sI}
		\end{pmatrix} + \begin{pmatrix}
			0 & 0\\
			0 & \rho \widehat{\Sigma}_1^{-1}
		\end{pmatrix}\begin{pmatrix}
			0\\ 
			\tilde{\beta}
		\end{pmatrix}.\\
	\end{aligned}
	\]
	So we have 
	\[
	\begin{aligned}
		(\hat{\tau}_{\mathrm{CD}}^\T, \hat{\beta}_{\mathrm{CD}}^\T)^\T &= \left\{ \widehat{\Sigma}^{-1} + \begin{pmatrix}
			0 &0\\
			0 & \rho\widehat{\Sigma}_1^{-1}
		\end{pmatrix} \right\}^{-1}\left\{\widehat{\Sigma}^{-1} \begin{pmatrix}
			\hat{\tau}_{\mathrm{e}}^{\sI} \\ 
			\hat{\beta}_{\mathrm{e}}^{\sI}
		\end{pmatrix} + \begin{pmatrix}
			0 & 0\\
			0 & \rho \widehat{\Sigma}_1^{-1}
		\end{pmatrix} \begin{pmatrix}
			0 \\ 
			\tilde{\beta}
		\end{pmatrix} \right\}\\
		&= \left\{\widehat{\Sigma}- \widehat{\Sigma}\begin{pmatrix}
			0 & 0\\
			0 & \{ \widehat{\Sigma}_1/\rho + \widehat{\Sigma}_{\eta\eta}^{\mathrm{e}}\}^{-1}
		\end{pmatrix}\widehat{\Sigma} \right\}\left\{\widehat{\Sigma}^{-1}\begin{pmatrix}
			\hat{\tau}_{\mathrm{e}}^{\sI} \\ 
			\hat{\beta}_{\mathrm{e}}^{\sI}
		\end{pmatrix} + \begin{pmatrix}
			0 & 0\\
			0 & \rho \widehat{\Sigma}_1^{-1}
		\end{pmatrix}\begin{pmatrix}
			0 \\ 
			\tilde{\beta}
		\end{pmatrix} \right\}\\
		&= \left\{I - \widehat{\Sigma} \begin{pmatrix}
			0 & 0\\
			0 & \{ \widehat{\Sigma}_1/\rho + \widehat{\Sigma}_{\eta\eta}^{\mathrm{e}}\}^{-1}
		\end{pmatrix} \right\}\begin{pmatrix}
			\hat{\tau}_{\mathrm{e}}^{\sI} \\ 
			\hat{\beta}_{\mathrm{e}}^{\sI}
		\end{pmatrix}\\
		&\quad+ \widehat{\Sigma}\left\{\begin{pmatrix}
			0 & 0\\
			0 & \rho \widehat{\Sigma}_1^{-1}
		\end{pmatrix} - \begin{pmatrix}
			0 & 0\\
			0 & \{ \widehat{\Sigma}_1/\rho + \widehat{\Sigma}_{\eta\eta}^{\mathrm{e}} \}^{-1}
		\end{pmatrix}\widehat{\Sigma} \begin{pmatrix}
			0 & 0\\
			0 & \rho \widehat{\Sigma}_1^{-1}
		\end{pmatrix}  \right\}\begin{pmatrix}
			0 \\ 
			\tilde{\beta}
		\end{pmatrix}\\
		&= \begin{pmatrix}
			\hat{\tau}_{\mathrm{e}}^{\sI} - \widehat{\Sigma}_{\phi\eta}^{\mathrm{e}}\{ \widehat{\Sigma}_1/\rho + \widehat{\Sigma}_{\eta\eta}^{\mathrm{e}} \}^{-1}\hat{\beta}_{\mathrm{e}}^{\sI}\\
			\hat{\beta}_{\mathrm{e}}^{\sI} - \widehat{\Sigma}_{\eta\eta}^{\mathrm{e}}\{ \widehat{\Sigma}_1 /\rho + \widehat{\Sigma}_{\eta\eta}^{\mathrm{e}} \}^{-1}\hat{\beta}_{\mathrm{e}}^{\sI}\\
		\end{pmatrix}\\ 
		&\quad+ \widehat{\Sigma}\begin{pmatrix}
			0 &0\\
			0 & \{ \widehat{\Sigma}_1/\rho + \widehat{\Sigma}_{\eta\eta}^{\mathrm{e}}\}^{-1}
		\end{pmatrix}\begin{pmatrix}
			0 \\ 
			\tilde{\beta}
		\end{pmatrix} \\
		&= \begin{pmatrix}
			\hat{\tau}_{\mathrm{e}}^{\sI} - \widehat{\Sigma}_{\phi\eta}^{\mathrm{e}}\{ \widehat{\Sigma}_1/\rho + \widehat{\Sigma}_{\eta\eta}^{\mathrm{e}}\}^{-1}(\hat{\beta}_{\mathrm{e}}^{\sI} - \tilde{\beta})\\
			\hat{\beta}_{\mathrm{e}}^{\sI} - \widehat{\Sigma}_{\eta\eta}^{\mathrm{e}}\{ \widehat{\Sigma}_1/\rho + \widehat{\Sigma}_{\eta\eta}^{\mathrm{e}}\}^{-1}(\hat{\beta}_{\mathrm{e}}^{\sI} - \tilde{\beta})\\
		\end{pmatrix}.
	\end{aligned}
	\]
	As a result, $\hat{\tau}_{\mathrm{e}} = \hat{\tau}_{\mathrm{CD}}$, which completes the proof of Proposition 4.

	When $\widehat{\Sigma}_1$ is not available, \citet{liu2015multivariate} proposed to use a working positive-definite matrix $\Omega$  for $\widehat{\Sigma}_1$. 
	Denote the resulting estimator by $\hat{\tau}_{\mathrm{CD}}^\Omega$, we have
	\[
	\hat{\tau}_{\mathrm{CD}}^\Omega = \hat{\tau}_{\mathrm{e}}^{\sI} - \widehat{\Sigma}_{\phi\eta}^{\mathrm{e}}\{ \Omega/\rho + \widehat{\Sigma}_{\eta\eta}^{\mathrm{e}}\}^{-1}(\hat{\beta}_{\mathrm{e}}^{\sI} - \tilde{\beta}).
	\]
	Note that $\hat{\tau}_{\mathrm{CD}}^\Omega$ is also  consistent to $\tau$. 
	As $n^{1/2}((\hat{\tau}_{\mathrm{e}}^{\sI}- \tau)^\T, (\hat{\beta}_{\mathrm{e}}^{\sI} - \beta)^\T)^\T \rightarrow N(0, \Sigma)$, we have
	\[
	\begin{aligned}
		&\quad \mathrm{avar}(\hat{\tau}_{\mathrm{CD}}^\Omega)  \\ 
		&= \mathrm{avar}(\hat{\tau}_{\mathrm{e}}^{\sI} - \widehat{E}(\phi_{\mathrm{e}}\eta_{\mathrm{e}}^\T ) \{ \Omega/\rho + \widehat{\Sigma}_{\eta\eta}^{\mathrm{e}} \}^{-1}\hat{\beta}_{\mathrm{e}}^{\sI}) + \mathrm{avar}(\widehat{\Sigma}_{\phi\eta}^{\mathrm{e}} \{ \Omega/\rho + \widehat{\Sigma}_{\eta\eta}^{\mathrm{e}} \}^{-1}\tilde{\beta})\\
		&= E(\phi_{\mathrm{e}}\phi_{\mathrm{e}}^\T) - 2E(\phi_{\mathrm{e}}\eta_{\mathrm{e}}^\T)\{\Omega/\rho + E(\eta_{\mathrm{e}}\eta_{\mathrm{e}}^\T) \}^{-1}E(\phi_{\mathrm{e}}\eta_{\mathrm{e}}^\T)^\T\\
		&\quad + E(\phi_{\mathrm{e}}\eta_{\mathrm{e}}^\T)\{\Omega/\rho + E(\eta_{\mathrm{e}}\eta_{\mathrm{e}}^\T)\}^{-1} \{\Sigma_1/\rho + E(\eta_{\mathrm{e}}\eta_{\mathrm{e}}^\T) \} \{\Omega/\rho + E(\eta_{\mathrm{e}}\eta_{\mathrm{e}}^\T) \}^{-1} E(\phi_{\mathrm{e}}\eta_{\mathrm{e}}^\T)^\T\\
		&= E(\phi_{\mathrm{e}}\phi_{\mathrm{e}}^\T) \\ 
		& \quad - E(\phi_{\mathrm{e}}\eta_{\mathrm{e}}^\T)\left\{\Omega/\rho + E(\eta_{\mathrm{e}}\eta_{\mathrm{e}}^\T)\right\}^{-1}\left\{(2\Omega - \Sigma_1)/\rho + E(\eta_{\mathrm{e}}\eta_{\mathrm{e}}^\T) \right\}\left\{\Omega/\rho + E(\eta_{\mathrm{e}}\eta_{\mathrm{e}}^\T)\right\}^{-1}E(\phi_{\mathrm{e}}\eta_{\mathrm{e}}^\T)^\T.
	\end{aligned}
	\]
	We next show that $\mathrm{avar}(\hat{\tau}_{\mathrm{CD}}^\Omega) - \mathrm{avar}(\hat{\tau}_{\mathrm{e}})$ is semi-positive definite, where $\mathrm{avar}(\hat{\tau}_{\mathrm{e}})$ is   the efficiency bound in \eqref{efficiency bound}.
	\[
	\begin{aligned}
		&\quad \mathrm{avar}(\hat{\tau}_{\mathrm{CD}}^\Omega) - \mathrm{avar}(\hat{\tau}_{\mathrm{e}})  \\ 
		&= E(\phi_{\mathrm{e}}\eta_{\mathrm{e}}^\T)\{ \Sigma_1/\rho + E(\eta_{\mathrm{e}}\eta_{\mathrm{e}}^\T) \}^{-1}E(\phi_{\mathrm{e}}\eta_{\mathrm{e}}^\T)^\T \\
		&\quad -E(\phi_{\mathrm{e}}\eta_{\mathrm{e}}^\T) \{\Omega/\rho + E(\eta_{\mathrm{e}}\eta_{\mathrm{e}}^\T) \}^{-1} \{2\Omega/\rho - \Sigma_1/\rho + E(\eta_{\mathrm{e}}\eta_{\mathrm{e}}^\T) \} \{\Omega/\rho + E(\eta_{\mathrm{e}}\eta_{\mathrm{e}}^\T) \}^{-1}E(\phi_{\mathrm{e}}\eta_{\mathrm{e}}^\T)^\T \\
		&= E(\phi_{\mathrm{e}}\eta_{\mathrm{e}}^\T)\left[\left\{ \Sigma_1/\rho + E(\eta_{\mathrm{e}}\eta_{\mathrm{e}}^\T) \right\}^{-1} - \left\{ \Omega/\rho + E(\eta_{\mathrm{e}}\eta_{\mathrm{e}}^\T) \right\}^{-1}\right]E(\phi_{\mathrm{e}}\eta_{\mathrm{e}}^\T)^\T \\
		&\quad + E(\phi_{\mathrm{e}}\eta_{\mathrm{e}}^\T)\left\{\Omega/\rho + E(\eta_{\mathrm{e}}\eta_{\mathrm{e}}^\T)\right\}^{-1}\left\{(\Sigma_1 - \Omega)/\rho \right\}\left\{\Omega/\rho + E(\eta_{\mathrm{e}}\eta_{\mathrm{e}}^\T)\right\}^{-1}E(\phi_{\mathrm{e}}\eta_{\mathrm{e}}^\T)^\T. \\
	\end{aligned}
	\]
	We have
	\[
	\begin{aligned}
		&\quad \left\{ \Sigma_1/\rho + E(\eta_{\mathrm{e}}\eta_{\mathrm{e}}^\T) \right\}^{-1} \\ 
		&= \left\{ \Omega/\rho + E(\eta_{\mathrm{e}}\eta_{\mathrm{e}}^\T) + (\Sigma_1 - \Omega)/\rho \right\}^{-1} \\
		&= \left\{ \Omega/\rho + E(\eta_{\mathrm{e}}\eta_{\mathrm{e}}^\T) \right\}^{-1} \\ 
		&\quad - \left\{ \Omega/\rho + E(\eta_{\mathrm{e}}\eta_{\mathrm{e}}^\T) \right\}^{-1}\left[\left\{(\Sigma_1 - \Omega)/\rho \right\}^{-1} - \left\{ \Omega/\rho + E(\eta_{\mathrm{e}}\eta_{\mathrm{e}}^\T) \right\}^{-1} \right]^{-1}\left\{ \Omega/\rho + E(\eta_{\mathrm{e}}\eta_{\mathrm{e}}^\T) \right\}^{-1},
	\end{aligned}
	\]
	and 
	\[
	\begin{aligned}
		&\quad \left[\left\{(\Sigma_1 - \Omega)/\rho \right\}^{-1} - \left\{ \Omega/\rho + E(\eta_{\mathrm{e}}\eta_{\mathrm{e}}^\T) \right\}^{-1} \right]^{-1} \\ 
		&= (\Sigma_1 - \Omega)/\rho -  (\Sigma_1 - \Omega)/\rho\left\{\Sigma_1/\rho + E(\eta_{\mathrm{e}}\eta_{\mathrm{e}}^\T)\right\}^{-1}(\Sigma_1 - \Omega)/\rho.
	\end{aligned}
	\]
	Therefore, 
	\[
	\begin{aligned}
		&\quad \mathrm{avar}(\hat{\tau}_{\mathrm{CD}}^\Omega) - \mathrm{avar}(\hat{\tau}_{\mathrm{e}}) \\ 
		&= E(\phi_{\mathrm{e}}\eta_{\mathrm{e}}^\T)\left\{\Omega/\rho + E(\eta_{\mathrm{e}}\eta_{\mathrm{e}}^\T)\right\}^{-1}\left((\Sigma_1 - \Omega)/\rho -\left[\left\{(\Sigma_1 - \Omega)/\rho \right\}^{-1} - \left\{ \Omega/\rho + E(\eta_{\mathrm{e}}\eta_{\mathrm{e}}^\T) \right\}^{-1} \right]^{-1} \right) \\
		&\quad \times \left\{\Omega/\rho + E(\eta_{\mathrm{e}}\eta_{\mathrm{e}}^\T)\right\}^{-1}E(\phi_{\mathrm{e}}\eta_{\mathrm{e}}^\T)^\T \\
		&= E(\phi_{\mathrm{e}}\eta_{\mathrm{e}}^\T)\left\{\Omega/\rho + E(\eta_{\mathrm{e}}\eta_{\mathrm{e}}^\T)\right\}^{-1}(\Sigma_1 - \Omega)/\rho\left\{\Sigma_1/\rho + E(\eta_{\mathrm{e}}\eta_{\mathrm{e}}^\T)\right\}^{-1}\\ 
		&\quad \times(\Sigma_1 - \Omega)/\rho\left\{\Omega/\rho + E(\eta_{\mathrm{e}}\eta_{\mathrm{e}}^\T)\right\}^{-1}E(\phi_{\mathrm{e}}\eta_{\mathrm{e}}^\T)^\T\\
		&= D\left\{\Sigma_1/\rho + E(\eta_{\mathrm{e}}\eta_{\mathrm{e}}^\T)\right\}^{-1}D^\T,
	\end{aligned}
	\]
	where $D = E(\phi_{\mathrm{e}}\eta_{\mathrm{e}}^\T)\left\{\Omega/\rho + E(\eta_{\mathrm{e}}\eta_{\mathrm{e}}^\T)\right\}^{-1}(\Sigma_1 - \Omega)/\rho$. 
	So $\mathrm{avar}(\hat{\tau}_{\mathrm{CD}}^\Omega) - \mathrm{avar}(\hat{\tau}_{\mathrm{CD}})$ is semi-positive definite. 
	
	According to the asymptotic variance formula of $\mathrm{avar}(\hat{\tau}_{\mathrm{CD}}^\Omega)$, efficiency paradox will not occur if  $\left\{(2\Omega - \Sigma_1)/\rho + E(\eta_{\mathrm{e}}\eta_{\mathrm{e}}^\T) \right\}$ is a positive definite matrix.  
	As mentioned by \citet{liu2015multivariate}, sometimes  a consistent estimate of the diagonal elements of $\Sigma_1$ is available; in this case we let $\Omega$ be such a diagonal matrix, then $2\Omega - \Sigma_1$ is positive-definite for $q = 1,2$, but this no longer holds for  $q \geq 3$. 
	For example, let
	\[
	\Sigma_1 = \begin{pmatrix}
		2 & 1 & 1 \\ 
		1 & 2 & 1 \\ 
		1 & 1 & 1 \\
	\end{pmatrix}, \Omega = \begin{pmatrix}
		2 & 0 & 0 \\ 
		0 & 2 & 0 \\ 
		0 & 0 & 1 \\
	\end{pmatrix}, 2\Omega - \Sigma_1 = \begin{pmatrix}
		2 & -1 & -1 \\ 
		-1 & 2 & -1 \\ 
		-1 & -1 & 1 \\
	\end{pmatrix},
	\]
	then $2\Omega - \Sigma_1$ has eigenvalues $3, 1\pm 2^{1/2}$, which is not positive-definite.
	As a result, when $\rho$ is small, $\left\{(2\Omega - \Sigma_1)/\rho + E(\eta_{\mathrm{e}}\eta_{\mathrm{e}}^\T) \right\}$ will not be positive-definite.
	
\end{proof}

\section{Detailed Discussion for Summary Statistics from Multiple External Studies}
\label{supp discussion}

In Theorem~\ref{multi-source} in the main text, we obtain the efficiency bound 
\begin{equation}\label{bound S}
	E(\phi_{\mathrm{e}}\phi_{\mathrm{e}}^\T) - E(\phi_{\mathrm{e}}\eta_{\mathrm{e}, [S]}^\T) \left\{ \Sigma_{[S]} + E(\eta_{\mathrm{e}, [S]}\eta_{\mathrm{e}, [S]}^\T ) \right\}^{-1}E(\phi_{\mathrm{e}}\eta_{\mathrm{e}, [S]}^\T )^\T,
\end{equation}
with summary statistics from $S$ external studies, where $\Sigma_{[S]} = \mathrm{diag}(\Sigma_1/\rho_1, \ldots, \Sigma_S/\rho_S)$. Suppose the summary statistics $\tilde{\beta}_{S + 1}$ is available from the $(S+1)$-th external study with sample size $m_{S+1}$, $m_{S+1}/n\rightarrow \rho_{S+1}\in(0,\infty)$, $m_{S+1}^{1/2}\{\tilde{\beta}_{S+1} - \beta_{S+1}(P_{S+1})\} \rightarrow N(0, \Sigma_{S+1})$, and 
$\beta_{S+1}(P_{S+1}) = \beta_{S+1}(P_{0})$. 
Let $\eta_{\mathrm{e}, S+1}$ be the efficient influence function for $\beta_{S+1}(P_0)$ based on the internal data, $\eta_{\mathrm{e}, [S+1]} = (\eta_{\mathrm{e},1}^\T, \ldots, \eta_{\mathrm{e},S+1}^\T)^\T$, and  $\Sigma_{[S+1]} = \mathrm{diag}(\Sigma_1/\rho_1, \ldots, \Sigma_{S+1}/\rho_{S+1})$. Then, the efficiency bound with summary statistics from $S+1$ external studies is 
\begin{equation}\label{bound S+1}
	E(\phi_{\mathrm{e}}\phi_{\mathrm{e}}^\T) - E(\phi_{\mathrm{e}}\eta_{\mathrm{e}, [S+1]}^\T) \left\{ \Sigma_{[S+1]} + E(\eta_{\mathrm{e}, [S+1]}\eta_{\mathrm{e}, [S+1]}^\T ) \right\}^{-1}E(\phi_{\mathrm{e}}\eta_{\mathrm{e}, [S+1]}^\T )^\T.
\end{equation}
Let $\epsilon_{[S]}$ and $\epsilon_{S+1}$ are independent normal random vectors with mean zero and variances $\Sigma_{[S]}$ and $\Sigma_{S+1} / \rho_{S+1}$, respectively, and $\epsilon_{[S + 1]} = (\epsilon_{[S]}^{\T}, \epsilon_{S+1})^{\T}$. Then, it is not hard to verify that \eqref{bound S} and \eqref{bound S+1} are the variances of the residuals of projecting $\phi_{\rm e}$ onto the linear spaces spanned by $\eta_{[S]} + \epsilon_{[S]}$ and $\eta_{[S+1]} + \epsilon_{[S+1]}$, respectively.
After some matrix algebra, one can find that \eqref{bound S} - \eqref{bound S+1} equals to
\begin{equation}\label{efficiency gain}
	 E(\phi_{\rm e}e_{S+1}^{\T})\{E(e_{S+1}e_{S+1}^{\T}) + \Sigma_{S+1}/\rho_{S+1}\}^{-1}E(\phi_{\rm e}e_{S+1}^{\T})^{\T} \geq 0,
\end{equation} 
where $e_{S+1} = \eta_{{\rm e}, S+1} - E(\eta_{{\rm e}, S+1}\eta_{\mathrm{e}, [S]}^\T) \left\{ \Sigma_{[S]} + E(\eta_{\mathrm{e}, [S]}\eta_{\mathrm{e}, [S]}^\T ) \right\}^{-1}(\eta_{{\rm e}, [S]} + \epsilon_{[S]})$ is the residual of projecting $\eta_{{\rm e}, S+1}$ onto the linear space spanned by $\eta_{{\rm e},[S]} + \epsilon_{[S]}$. This implies that adding a summary statistic doesn't increase the efficiency bound. From \eqref{efficiency gain}, it can be seen that the efficiency gain of adding $\tilde{\beta}_{S+1}$ into analysis depends on the covariance $E(\phi_{\rm e}e_{S+1}^{\T})$ between $e_{S+1}$ and $\phi_{\rm e}$ and the variances $E(e_{S+1}e_{S+1}^{\T})$ and $\Sigma_{S+1}/\rho_{S+1}$. 
In study design, an estimate or approximation of \eqref{efficiency gain} can serve as a criterion for choosing the summary statistics to incorporate.

\section{Additional Examples} \label{Supp examples}

The following is an example concerning marginal and joint regressions.

\begin{example}\label{linear example}
Suppose the internal data are random samples of $(X_1, X_2, Y) \sim P_0$ and the external individual  data are random samples of $(X_1, X_2,  Y) \sim P_1$, with $P_0 = P_1$, $Y = X_1\tau_1 + X_2\tau_2 + \varepsilon$, $X_1,X_2$ having mean zero, $E(\varepsilon\mid X_1, X_2) = 0$ and $\mathrm{var}(\varepsilon\mid X_1, X_2) = \sigma^2$. 
The external summary statistics are $\tilde{\beta} = (\tilde{\beta}_1, \tilde{\beta}_2)^\T$, where $\tilde{\beta}_1$ and $\tilde{\beta}_2$ are the ordinary least squares coefficients  obtained  by  regressing $Y$ on $X_1$ and $X_2$ separately with the external data.
This happens in genome-wide association studies where researchers provide summary statistics of separate univariate regression coefficients of a quantitative trait ($Y$) on each centered genotype ($X$) \citep{zhu2017bayesian}. 
The efficient influence function for $\beta$ is 
$\eta_{\mathrm{e}} = (\eta_{\mathrm{e}, 1}, \eta_{\mathrm{e}, 2})^\T$, where $ \eta_{\mathrm{e}, 1} = \{E(X_1^2)\}^{-1}X_1(Y - X_1\beta_1)$
and $\eta_{\mathrm{e}, 2} =  \{E(X_2^2)\}^{-1}X_2(Y - X_2\beta_2)$.
Denoting $X=(X_1, X_2)^\T$, the data-fused efficiency bound for $\tau=(\tau_1, \tau_2)^\T$ is  
\[\sigma^2 \{E(XX^\T)\}^{-1} - \frac{\rho \sigma^4}{1 + \rho} 
\mathrm{diag}\{1/E(X_1^2),1/E(X_2^2)\}
\{E(\eta_{\mathrm{e}}\eta_{\mathrm{e}}^\T) \}^{-1} 
\mathrm{diag}\{1/E(X_1^2), 1/E(X_2^2)\}.\] 
If   only $\tilde{\beta}_1$ is available and without loss of generality we assume $X_1, X_2$  have  unity variance and correlation coefficient $\kappa$, 
the data-fused efficiency bound for $\tau$ is
\[
\sigma^2\begin{pmatrix}
\frac{1}{1-\kappa^2}  &  -\kappa \\ 
-\kappa    &    \frac{1}{1-\kappa^2}
\end{pmatrix} - \frac{\rho}{1+\rho}\begin{pmatrix}
\sigma^4/ \mathrm{var} (\eta_{\mathrm{e},1}) & 0\\
0  &  0\\
\end{pmatrix}.
\] 
There is no efficiency gain for estimating $\tau_2$ from external marginal regression estimate $\tilde{\beta}_1$.
\cite{zhang2020generalized} have obtained this result under a special case where the distribution of $\varepsilon$ is  $N(0,1)$.
Nevertheless,  we note that the efficiency gain for estimating $\tau_2$ emerges when  $\varepsilon$ is heteroscedastic; that is,  $\mathrm{var}(\varepsilon\mid X_1, X_2)$ is not a constant.
Details of this example and a simulation are provided below.
\end{example}
\emph{Details of Example \ref{linear example}}.
Let $X = (X_1, X_2)^\T$. 
According to \citet{tsiatis2006semiparametric}, the nuisance tangent space is $\{s(\varepsilon, X):  E\{s(\varepsilon, X)\varepsilon\mid X\} = 0 \}$. 
In addition, the efficient influence function for $\tau$ is $\phi_{\mathrm{e}} = [E\{X V^{-1}(X)X^\T\}]^{-1}X V^{-1}(X)(Y- X^\T\tau)$, where $V(X) $ $ = E(\varepsilon^2\mid X) = \sigma^2$. Thus, $\phi_{\mathrm{e}} = \{E(X X^\T)\}^{-1}X(Y- X^\T\tau)$.  
For $\beta  $ $= (\beta_1, \beta_2)^\T $ $= (\{E(X_1^2)\}^{-1}EX_1Y, \{E(X_2^2)\}^{-1}EX_2Y )^\T $, the efficient influence function is $\eta_{\mathrm{e}} =(\eta_{\mathrm{e}, 1}, \eta_{\mathrm{e}, 2})^\T $.

We have $E(\phi_{\mathrm{e}}\eta_{\mathrm{e}}^\T) = \sigma^2 \mathrm{diag}(\{E(X_1^2) \}^{-1}, \{E(X_2^2) \}^{-1} )$. 
So the data-fused efficiency bound for $\tau$ is
\[
\sigma^2\{E(X^\T X)\}^{-1} - \frac{\rho}{1+\rho}\sigma^4 \begin{pmatrix}
	\{E(X_1^2) \}^{-1} & 0 \\ 
	0 & \{E(X_2^2) \}^{-1} 
\end{pmatrix}\{E(\eta_{\mathrm{e}}\eta_{\mathrm{e}}^\T) \}^{-1}\begin{pmatrix}
	\{E(X_1^2) \}^{-1} & 0 \\ 
	0 & \{E(X_2^2) \}^{-1} 
\end{pmatrix}.
\] 
In particular, suppose we only have summary statistics $\tilde{\beta}_1$ with   efficient influence function  $\eta_{\mathrm{e},1}$.
When $X_1, X_2$  have unity variance  and correlation  coefficient $\kappa$,  the above variance bound reduces to 
\[
\sigma^2\begin{pmatrix}
	\frac{1}{1-\kappa^2}  &  -\kappa \\ 
	-\kappa    &    \frac{1}{1-\kappa^2}
\end{pmatrix} - \frac{\rho}{1+\rho}\begin{pmatrix}
	\sigma^4/ \mathrm{var} (\eta_{\mathrm{e},1}) & 0\\
	0  &  0\\
\end{pmatrix}.
\]

If $V(X)$ is not a constant, we have
\[
E(\phi_{\mathrm{e}}\eta_{\mathrm{e},1}) = [E\{X V^{-1}(X)X^\T\}]^{-1}E(X X_1) \{E(X_1^2)\}^{-1},
\]
and $\tilde{\beta}_1$ may improve the estimation efficiency for $\tau_2$. 
We illustrate with   a simulation. 
The data generating process is as follows:
\[
\begin{aligned}
	\begin{pmatrix}
		X_1\\
		X_2
	\end{pmatrix} \sim N\left\{\begin{pmatrix}
		0\\
		0
	\end{pmatrix}, \begin{pmatrix}
		1    & 0.5 \\
		0.5 &    1 \\
	\end{pmatrix}\right\}, Y =  X^\T \tau + \varepsilon, \text{where~} \tau = (1, 1)^\T.
\end{aligned}
\]
We set $V(X) = 1 + (X_1 + 3X_2)^2$ and   $n=500, m = 2000$. 
For estimation, we obtain $\hat{\tau}_{\mathrm{e}}^{\sI}$ by weighted least squares regression with true $V(X)$ and $\hat{\beta}_{\mathrm{e}}^{\sI}$ by regressing $Y$ on $X_1$ using only internal data. 
The data-fused efficient estimator is the same as in Example \ref{linear example}, that is,
\[
\hat{\tau}_{\mathrm{e}} = \hat{\tau}_{\mathrm{e}}^{\sI} - \rho(1 + \rho)^{-1} \widehat{\mathrm{cov}}(\hat{\tau}_{\mathrm{e}}^{\sI}, \hat{\beta}_{\mathrm{e}}^{\sI})\{ \widehat{\mathrm{var}}(\hat{\beta}_{\mathrm{e}}^{\sI})\}^{-1}(\hat{\beta}_{\mathrm{e}}^{\sI} - \tilde{\beta}_1).
\]
We replicate 1000 simulations.
Figure \ref{heterogeneous} and Table~\ref{MSE hetero} show  bias and root mean squared error  of INT and EFF estimators.
The data-fused efficient  estimator of  $\tau_2$ has smaller variability than the weighted least squares  estimator using only internal data, 
which is evidence  that  $\tilde{\beta}_1$ brings   efficiency gain for estimating $\tau_2$.

\begin{minipage}{0.4\textwidth}
	\begin{figure}[H]
		\centering
		\includegraphics[width=0.8\textwidth]{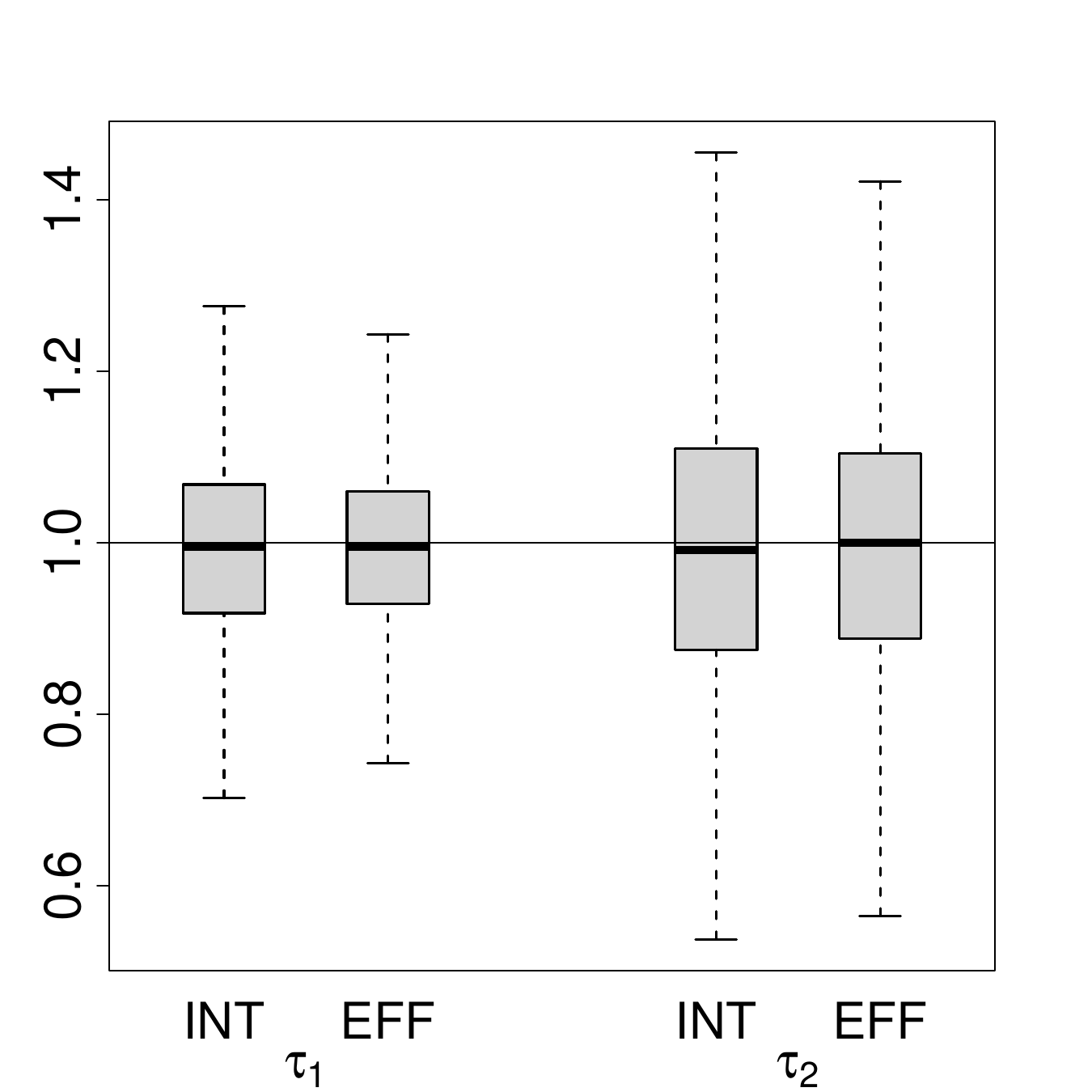}   
		\caption{Boxplot under the heteroscedastic error setting.}
		\label{heterogeneous}
	\end{figure}
\end{minipage}
\hfill
\begin{minipage}{0.4\textwidth}
	\begin{table}[H]
		\centering
		\caption{RMSE for heteroscedastic error setting. All numbers are multiplied by 100} \label{MSE hetero}
		\begin{tabular}{ccc}
			\toprule
			&  $\tau_1$ & $\tau_2$ \\[5pt]
			\midrule
			INT  &  7.60    & 12.23  \\  
			EFF  &  6.74    & 11.49  \\ 
			\bottomrule
		\end{tabular}
	\end{table}
\end{minipage}

\begin{example}\label{semisupervised}
Suppose  $Z=(X, Y)\sim P_0$ in the internal study and  $W = X\sim P_1$ in the external study with $P_1$ equal to the marginal distribution of $P_0$. 
This is a common setting in semi-supervised learning or missing data analysis where $Z$ is the labeled/complete data and $W$ is the unlabeled/incomplete data.
Consider the estimation of $\tau = E(Y)$. 
Suppose individual samples in the internal study and $\tilde \beta$, the sample mean of $X$ in external data, are available.
Using only the internal data, the efficiency bound for $\tau$ is  $\mathrm{var}(Y)$. 
Treating $\tilde{\beta}$ as the true underlying value $\beta$, 
then the plug-in estimator is $T_{n, \mathrm{e}}(\tilde{\beta})=n^{-1}\sum_{i=1}^n \{ Y_i - \hat{\zeta}(X_i - \tilde{\beta}) \}$, where $\hat{\zeta}$ is the   least squares  coefficient of $X$ in the  linear   regression of  $Y$ on $X$. The asymptotic variance of $T_{n, \mathrm{e}}(\tilde{\beta})$ is
 $\mathrm{var}(Y) - (1-1/\rho)\mathrm{cov}^2(X,Y)\{\mathrm{var}(X)\}^{-1}$. 
If $\rho < 1$ and $\mathrm{cov}(X,Y) \neq 0$,  the efficiency paradox emerges.
\end{example}

\emph{Details of Example \ref{semisupervised}}.
This example illustrates the efficiency paradox.
Using only internal data, the efficient influence function for $\tau$ is $Y - \tau$, so the efficiency bound for $\tau$ is $\mathrm{var}(Y)$. If we treat $\tilde{\beta}$ as the true $\beta$, then $T_{n, \mathrm{e}}(\tilde{\beta})=n^{-1}\sum_{i=1}^n \{ Y_i - \hat{\zeta}(X_i - \tilde{\beta}) \}$ with $\hat{\zeta} \rightarrow \mathrm{cov}(X,Y)\{\mathrm{var}(X)\}^{-1}$. The linearization representation of $T_{n, \mathrm{e}}(\tilde{\beta})$ is
\[
\begin{aligned}
	T_{n, \mathrm{e}}(\tilde{\beta}) &= \tau + \frac{1}{n}\sum_{i=1}^n \{ Y_i - \mathrm{cov}(X,Y)\{\mathrm{var}(X)\}^{-1}(X_i - \beta) \} \\ 
	&\quad+  \mathrm{cov}(X,Y)\{\mathrm{var}(X)\}^{-1}(\tilde{\beta} - \beta) + o_P(n^{-1/2}),
\end{aligned}
\]
and its asymptotic variance is $\mathrm{var}(Y) - (1-1/\rho)\mathrm{cov}^2(X,Y)\{\mathrm{var}(X)\}^{-1}$.

\begin{example}[Continuation of Example \ref{semisupervised}]\label{mean}  
We have $\hat\tau_\mathrm{e}^{\sI}= \bar Y$ and $\hat\beta_\mathrm{e}^{\sI} =\bar X$,  the sample mean of $Y$ and $X$, respectively, in the internal data.
The data-fused  efficient estimator   is  $\hat \tau_\mathrm{e}  =\bar{Y} - \rho(1+\rho)^{-1}\hat{\zeta} (\bar{X} - \tilde{\beta})  $, 
where $\hat{\zeta}$ is the   least squares  coefficient of $X$ in the  linear   regression of  $Y$ on $X$. 
This recovers the semi-supervised least squares estimator given by  \cite{zhang2019semi}. 
\end{example}

{ \section{Additional Simulations} \label{supp simulations}

Under Scenario I in Section \ref{subsec: sim transportable} in the main text, we conduct additional simulations that use neural networks to estimate the nuisance parameter in the data-fused efficient influence function and construct the estimator using cross-fitting. We use three-layer neural networks to estimate the propensity score model and the outcome regression model. In each layer, there are 8 neurons.  The activation function is rectified linear unit (ReLU). We split the data into 4 parts to perform cross-fitting. The sample size is $n=1000$ and $m=2000$. The boxplots are shown in Figure \ref{fig: NN}.
\begin{figure}
    \centering
    \includegraphics[width=0.3\textwidth]{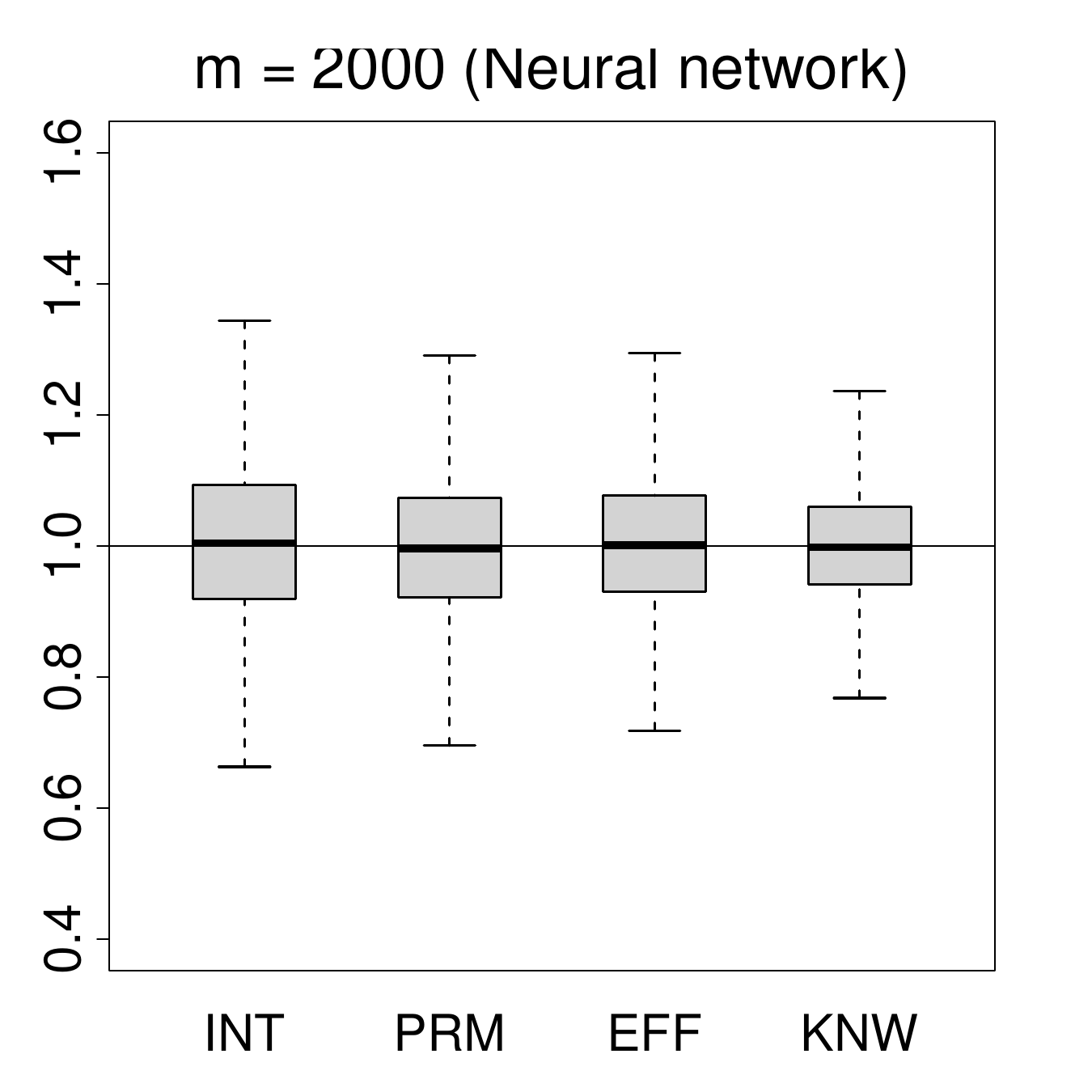}
    \includegraphics[width=0.3\textwidth]{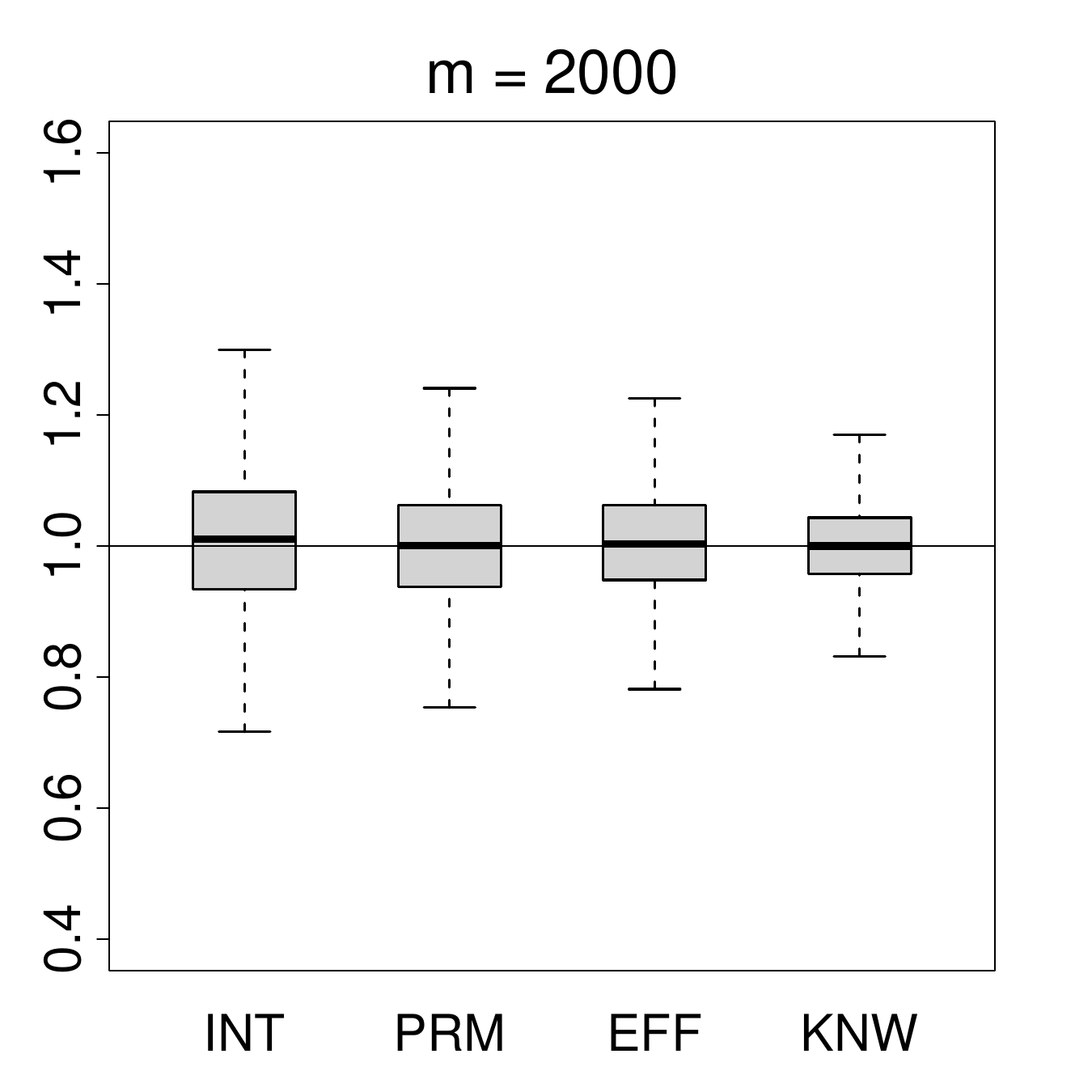}
    \includegraphics[width=0.3\textwidth]{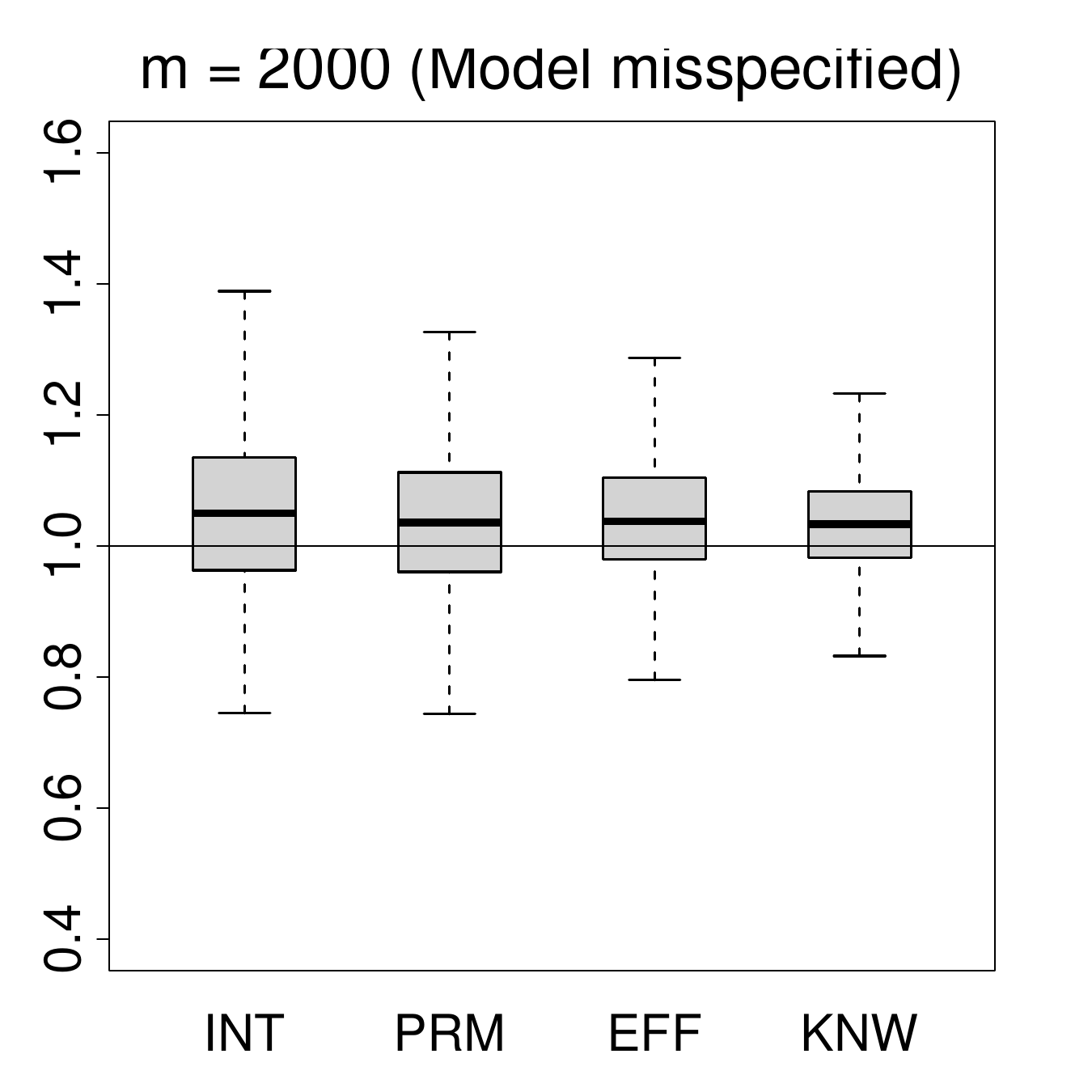}
    \caption{Simulation results for scenario I using neural networks. The plot on the left uses neural networks to fit the propensity score and outcome regression model, the plot in the middle uses correctly specified propensity score and outcome regression model, the plot on the right uses misspecified propensity score and outcome regression model.}\label{fig: NN}
\end{figure}

The results using neural networks are comparable to the results using correctly specified parametric models for both the propensity score model and the outcome regression model. For model misspecification,  we use a generalized linear model with a Cauchit link function to fit the propensity score model and a linear model to fit the outcome regression model. Misspecification of both the propensity score model and the outcome regression model leads to a biased estimate.

We conduct simulations under Scenario II to evaluate the sensitivity of the re-bootstrap confidence intervals to the number of candidate heterogeneity parameters ($\bar{r}=10, 20, 50$) and sample sizes ($n=500, 1000, 2000$). As shown in Table~\ref{CI biased sensitivity}, the re-bootstrap confidence intervals exhibit stable performance across varying numbers of candidate heterogeneity parameters and sample sizes.

	\begin{table}[h]
	\centering
	\caption{Average width (AW) and coverage probability (CP) for re-bootstrap procedure in Scenario II with moderate heterogeneity. All numbers are multiplied by 100.}\label{CI biased sensitivity}
	\footnotesize
	\resizebox{\textwidth}{!}{%
	\begin{tabular}{cc|cccccc|cccccc|cccccc}
		\toprule
		&&\multicolumn{6}{c|}{$\bar{r} = 10$} & \multicolumn{6}{c|}{$\bar{r} = 20$} &  \multicolumn{6}{c}{$\bar{r} = 50$}\\
		&&\multicolumn{2}{c}{$C=0.05$}       &\multicolumn{2}{c}{$C=1$}     & \multicolumn{2}{c|}{$C=20$}  &\multicolumn{2}{c}{$C=0.05$}         &\multicolumn{2}{c}{$C=1$}     & \multicolumn{2}{c|}{$C=20$} &\multicolumn{2}{c}{$C=0.05$}         &\multicolumn{2}{c}{$C=1$}     & \multicolumn{2}{c}{$C=20$}    \\[5pt]
		\midrule    
		& $n$&         AW  & CP          &  AW  & CP          & AW & CP&         AW  & CP          &  AW  & CP          & AW & CP&         AW  & CP          &  AW  & CP          & AW & CP           \\
		\multirow{3}{*}{$\tau_1$} 
		&500 & 37.24 & 98.4    & 38.07 & 97.2     & 40.62 & 98.5  & 38.18 & 97.5    & 38.82 & 97.2     & 41.37 & 97.1   & 39.88 & 98.2    & 40.56 & 97.0     & 42.69 & 97.6    \\
		&1000 & 25.05 & 98.0    & 25.48 & 96.5     & 27.84 & 97.2  & 26.05 & 97.9    & 26.74 & 97.0     & 28.60 & 97.3   & 27.05 & 97.7    & 27.70 & 96.8     & 29.42 & 97.9    \\
		&2000 & 17.47 & 96.9    & 17.84 & 95.7     & 19.72 & 96.4  & 17.95 & 98.1    & 18.36 & 96.8     & 20.11 & 98.2   & 19.18 & 98.5    & 19.90 & 97.3     & 21.12 & 97.7    \\
		\midrule
		\multirow{3}{*}{$\tau_2$}
		&500 & 37.10 & 97.8    & 39.34 & 94.1     & 47.72 & 97.4  & 38.07 & 97.5    & 40.11 & 95.4     & 48.71 & 97.7   & 39.97 & 97.9    & 42.10 & 95.3     & 49.73 & 96.8    \\
		&1000 & 24.93 & 97.6     & 26.28 & 93.2    & 33.87 & 96.4  & 25.84 & 97.6    & 27.76 & 93.5     & 34.55 & 96.6   & 27.10 & 98.7    & 28.93 & 95.0     & 35.27 & 97.7    \\
		&2000 & 17.44 & 97.1    & 18.44 & 92.2     & 24.20 & 96.9  & 17.90 & 97.5    & 19.01 & 92.5     & 24.63 & 98.1   & 18.97 & 98.4    & 20.48 & 94.6     & 25.23 & 97.8    \\
		\bottomrule
	\end{tabular}
	}
\end{table}

}

{\section{Efficient Influence Functions under a Parametric Conditional Density Model}\label{app: cond density}
	Let $Y$ and $X$ be the outcome and covariates, respectively.
	Suppose the internal observations $\{(X_{i}^{\sI}, Y_{i}^{\sI})\}_{i = 1}^{n}$ are i.i.d. observations of $(X, Y)$ from $P_{0}$ and the conditional density of $Y \mid X = x$ has the parametric form $f(y\mid x; \tau)$ under $P_{0}$. Let $X_{1}$ be a subvector of $X$. Let $\{(X_{1i}, Y_{i})\}_{i = 1}^{m}$ be i.i.d. observations of $(X_{1}, Y)$ from $P_{1}$.  Let $S_{\tau} = \partial f(Y\mid X; \tau)/\partial\tau$ be the score function of $\tau$ evaluated at the true parameter. Then, the efficient influence function $\phi_{\mathrm{e}}$ for $\tau$ based on the internal data is $E(S_{\tau}S_{\tau}^{\T})^{-1}S_{\tau}$ according to the maximum likelihood estimation theory \citep[Chapter 3 of][]{tsiatis2006semiparametric}, where $E(\cdot)$ denote the expectation with respect to $P_0$. In addition, the maximum likelihood estimator based on the conditional likelihood is efficient for $\tau$ based on the internal data.
	
	Suppose the external summary statistics $\tilde{\beta}$ is a subvector of the solution $\tilde{\psi}$ of the estimating equation $m^{-1} \sum_{i = 1}^{m}\varpi(Y_{i}, X_{1i}; \psi^{\dag}) = 0$, where $\varpi(\cdot, \cdot; \psi^{\dag})$ is an estimating function and $\psi^{\dag}$ is a $q_{*}$-dimensional vector with $q_{*} \geq q$.  Without loss of generality, we write $\tilde{\psi} = (\tilde{\zeta}^{\T}, \tilde{\beta}^{\T})^{\T}$.
	For any distribution $P$ of $(Y, X)$, let $\psi(P) = (\zeta(P)^{\T}, \beta(P)^{\T})^{\T}$ be the solution of the population-level estimating equation $E_{P}\{\varpi(Y, X_{1}; \psi)\} = 0$ and $\psi = \psi(P_{0})$, where the subscript $P$ indicates that the expectation is taken under $P$. We write $\varpi(Y, X_{1}; \psi)$ and $\partial \varpi(Y, X_{1}; \psi)/\partial\psi$ as  as $\varpi$ and $\partial \varpi/\partial\psi$ for short. Next, we derive the efficient influence function for the functional $\beta(P_{0})$ based on the internal data. Standard calculations \citep{tsiatis2006semiparametric} can show that the tangent space under the conditional density model consists of functions with the form $\Lambda S_{\tau} + S_{X}$ where $\Lambda$ is an arbitrary $q_{*}\times p$ matrix, $S_{\tau}$ is the score function of the conditional density of $Y\mid X$ satisfying $E(S_{\tau}\mid X) = 0$, and $S_{X}$ is an arbitrary $q_{*}$-dimensional function of $X$ satisfying $E(S_{X}) = 0$. Subsequently, we show that the efficient influence function of $\psi$ is 
	\[
	    \eta_{\rm par} = \{E(\partial \varpi/\partial\psi)\}^{-1}\{E(\varpi S_{\tau})E(S_{\tau}S_{\tau}^{\T})^{-1}S_{\tau} + E(\varpi\mid X)\}.
	\]
	Note that $\eta_{\rm par} = \Lambda S_{\tau} + \{E(\partial \varpi/\partial\psi)\}^{-1}E(\varpi\mid X)$ with $\Lambda = \{E(\partial \varpi/\partial\psi)\}^{-1}E(\varpi S_{\tau})$ and $E\{E(\varpi\mid X)\} = E(\varpi) = 0$. Thus, $\eta_{\rm par}$ belongs to the tangent space. Because $E_{P}\{\varpi(Y, X_{1}; \psi(P))\} = 0$ for any $P$. For any regular parametric submodel $P_{\theta}$, by taking derivative on both sides of the equation $E_{P_{\theta}}\{\varpi(Y, X_{1}; \psi(P_{\theta}))\} = 0$, we have
	\[
	   \frac{\partial}{\partial \theta^{\T}} \psi(P_{\theta}) =  \{E(\partial \varpi/\partial\psi)\}^{-1}E(\varpi S_{\theta}^{\T})
	\]
	according to the chain rule of the derivative, where $S_{\theta} = (\partial \tau / \partial \theta^{\T})^{\T}S_{\tau} + S_{X\theta}$ and $S_{X\theta}$ is the score of the marginal density of $X$ under the submodel. By straightforward calculation, we have 
	\[
	   \frac{\partial}{\partial \theta^{\T}} \psi(P_{\theta}) = E(\eta_{\rm par}S_{\theta}^{\T}),
	\]
	which proves that $\eta_{\rm par}$ is the efficient influence function of $\psi(P_{0})$. Thus, the efficient influence function of $\beta$ based on the internal data corresponds to the last $q$ components of  $\eta_{\rm par}$.
	
	Subsequently, we provide an efficient estimator for $\beta$ under the parametric density model. For any $\tau^{\dag}$, define $\varpi_{\rm int}(x; \psi^{\dag}, \tau^{\dag}) = \int \varpi(y, x_{1}; \psi^{\dag})f(y\mid x; \tau^{\dag})dy$. Let $\hat{\tau}_{\rm mle}$ be the maximum likelihood estimator for $\tau$. Then, the standard Z-estimation theory can show that the solution of the estimating equation
	\[
	    n^{-1}\sum_{i=1}^{n}\varpi_{\rm int}(X_{i}; \psi^{\dag}, \hat{\tau}_{\rm mle}) = 0
	\]
	has the influence function $\eta_{\rm par}$. Let $\hat{\psi}_{\mathrm{e}}^{\sI}$ be the solution of the above estimating equation. Then, the last $q$ components of $\hat{\psi}_{\mathrm{e}}^{\sI}$ comprise an efficient estimator for $\beta$.
	
	\begin{remark}\label{remark: integral}
		The function $\varpi_{\rm int}(x; \psi^{\dag}, \tau^{\dag})$ involves an integral which may be hard to calculate in general. However, the integral reduces to a sum and is easy to compute when $Y$ takes only finite possible values.
	\end{remark}
	
	Plugging in the efficient influence functions for $\tau$ and $\beta$ based on internal data here into the efficiency bound in Theorem \ref{efficiency theorem} in the main text, one can obtain the semiparametric efficiency bound incorporating the summary statistics. By repeatedly using the Woodbury matrix identity, it can be shown that the joint constrained maximum likelihood estimator defined in (4) in \cite{zhang2020generalized}  achieves the semiparametric efficiency bound under the parametric conditional density model when the weighting matrix $V^{-1}$ therein is set as $\Sigma_{1}^{-1}$, according to the formula (A.1) in Supplementary Material of \cite{zhang2020generalized}. Recently, \cite{zhai2024integrating} extended the method of \cite{zhang2020generalized} to account for untransportable summary statistics. The dPCML estimator proposed by \cite{zhang2020generalized} is asymptotically equivalent to a joint constrained maximum likelihood estimator that only incorporates transportable summary statistics according to Corollary 1 therein. However, it does not achieve the semiparametric efficiency bound $B^{\mathcal{A}}$ in the presence of untransportable summary statistics in general. For simplicity, consider the case with one external study. The efficiency bound can be achieved through the joint constrained maximum likelihood estimator when only transportable summary statistics are incorporated and the weighting matrix $V^{-1}$ in \cite{zhang2020generalized} is set as $\Sigma_{1,\mathcal{A}}^{-1}$ by the above discussions. According to Corollary 1 of \cite{zhai2024integrating}, it is not hard to see that the dPCML estimator is equivalent to the joint constrained maximum likelihood estimator that only incorporates transportable summary statistics with the weighting matrix $V^{-1} = (\Sigma_{1}^{-1})_{\mathcal{A}}$. Thus, the dPCML estimator does not always achieve the semiparametric efficiency bound $B^{\mathcal{A}}$ in the presence of untransportable summary statistics because $(\Sigma_{1}^{-1})_{\mathcal{A}} \neq \Sigma_{1,\mathcal{A}}^{-1}$ in general.
	
	\begin{remark}\label{remark: taylor}
		\cite{taylor2023data} considered the case where $Y\mid X$ follows generalized linear model $f(y\mid x; \tau)$ and the regression coefficient obtained from a reduced generalized linear regression between $Y$ and a $(q + 1)$-dimensional subvector $X_{1}$ of $X$ is available from an external study. They assumed the ratio between the population-level regression coefficients of $X_{1}$'s different components are (approximately) the same in the $Y \sim X$ regression in the internal study and the $Y \sim X_{1}$ regression in the external study. To simplify the discussion, we assume these ratios are identical in the two studies and the population-level regression coefficient of the first component of $X_{1}$ is nonzero in the $Y \sim X$ regression in the internal study. Let $\tilde{\vartheta} = (\tilde{\vartheta}_{1}, \dots, \tilde{\vartheta}_{q + 1})^{\T}$ be the regression coefficient in the $Y\sim X_{1}$ regression based the external data. Then, $\tilde{\beta} = (\tilde{\vartheta}_{2} / \tilde{\vartheta}_{1},\dots, \tilde{\vartheta}_{q+1} / \tilde{\vartheta}_{1})^{\T}$ is transportable.  
		Let $\tau_{j}$ be the $j$-th component of $\tau$ for $j = 1,\dots, p$. Then, $\beta = (\tau_{2} / \tau_{1}, \dots, \tau_{q+1} / \tau_{1})^{\T}$ and the efficient influence function for $\beta$ is 
		\[
		   \begin{pmatrix}
		   	-\tau_{1}^{-2}\tau_{2} & \tau_{1}^{-1} & 0 & \cdots & 0\\
		   	-\tau_{1}^{-2}\tau_{3} & 0 &  \tau_{1}^{-1} & \cdots & 0\\
		   	\vdots & \vdots & \vdots & \ddots &\vdots\\
		   	-\tau_{1}^{-2}\tau_{p} & 0 & 0 & \cdots & \tau_{1}^{-1}
		   \end{pmatrix}S_{\tau},
		\]
		according to maximum likelihood estimation theory in \cite[Chapter 3 of][]{tsiatis2006semiparametric}. Let $\hat{\tau}_{\rm mle} = (\hat{\tau}_{{\rm mle}, 1}, \dots, \hat{\tau}_{{\rm mle}, p})^{\T}$ be the maximum likelihood estimator $\hat{\tau}_{\rm mle}$ for $\tau$ based on the internal data. By the delta method, it can be shown that $(\hat{\tau}_{{\rm mle}, 2} / \hat{\tau}_{{\rm mle}, 1}, \dots, \hat{\tau}_{{\rm mle}, q+1} / \hat{\tau}_{{\rm mle}, 1})^{\T}$ is an efficient estimator for $\beta$ based on internal data.
	\end{remark}
}

\section{Discussion of the Setting When Transportablity Holds Asymptotically}\label{app: asy trans}
\subsection{Theoretical Investigations}
To mimic the situation where the internal and external studies have a minor difference, it is helpful to consider the case where the difference between some components of the internal and external distribution exists but decreases to zero as the sample size increases. 
We  give a formal description of this scenario and then investigate the asymptotic properties of our proposed estimators in this scenario. 

Suppose the internal data follow $P_0$  while the external data follow $P_{1,m}$ which may change with sample size $m$. Assume $P_{1,m}\rightarrow P_{1}$ for some $P_{1}$ in total variation metric and  $\beta(P_{1,m}) \rightarrow \beta(P_1)$ as $m\rightarrow \infty$. In this setting, we can decompose the difference $\beta(P_{1,m}) - \beta(P_0)$ into two parts:
\[
\beta(P_{1,m}) - \beta(P_0) = \{\beta(P_{1,m}) - \beta(P_1) \} + \{\beta(P_1) - \beta(P_0)\}.
\]
The first part $\beta(P_{1,m}) - \beta(P_1)$ represents the difference between $\beta(P_{1,m})$ and $\beta(P_1)$, which depends on the sample size $m$, while the second part represents the difference between $\beta(P_1)$ and $\beta(P_0)$, which is a fixed constant. In the manuscript, we assume that $P_{1,m} = P_1$, therefore the first part is zero. Under Assumption 2 in the manuscript which assumes the second part is also zero, we have $\beta(P_{1,m}) = \beta(P_0)$.  

When $P_{1,m}\neq P_1$ and Assumption 2 fails, with a slight abuse of notation, we can  partition the components of $\beta(\cdot)$ according to the order of heterogeneity level into three categories: $\mathcal{A}=\{j: \beta_{j}(P_{1,m}) - \beta_{j} = o(n^{-1/2}), j=1,\ldots,q \}$, $\mathcal{B} =\{j: n^{1/2}\{\beta_{j}(P_{1,m}) - \beta_{j}\} \rightarrow c > 0, j=1,\ldots,q \}$ and $\mathcal{C} =\{j: n^{1/2}\{\beta_{j}(P_{1,m}) - \beta_{j}\} \rightarrow \infty, j=1,\ldots,q \}$, where $\beta_{j}$ is the $j$-th component of $\beta = \beta(P_{0})$.
Let $d_{m} = \min_j\{ |\beta_j(P_{1,m}) - \beta_j|: j\in \mathcal{C} \}$, then we have $n^{1/2}d_{m} \rightarrow +\infty$. This partition has previously been adopted by  \citet{shen2020fusion} for  the integration of only summary statistics. 
Intuitively, $\mathcal{A}$, $\mathcal{B}$ and $\mathcal{C}$ are the index set of the components of the heterogeneity parameter  that are of order smaller than, the same as and larger than $n^{-1/2}$, respectively. 
We adopt the following assumption about $\tilde \beta$ for this scenario.

\begin{assumption}\label{asymptotic transportable}
	The external summary statistics $\tilde{\beta}$ is a RAL estimator of a $q$-dimensional functional $\beta(\cdot)$  of the  external data distribution  and
	$m^{1/2}\{\tilde{\beta} - \beta(P_{1,m})\} \rightarrow N(0, \Sigma_1)$, where $\Sigma_1$ only depends on $P_1$;
	a consistent covariance  estimator  $\widehat{\Sigma}_1$ for $\Sigma_1$ is available;
	$m/n\rightarrow \rho \in (0, +\infty)$.
\end{assumption}

Assumption~\ref{asymptotic transportable} states that  the limiting distribution of  $m^{1/2}\{\tilde{\beta} - \beta(P_{1,m})\}$ under $P_{1,m}$ is the same as the limiting distribution $m^{1/2}\{\tilde{\beta} - \beta(P_{1})\}$ under $P_1$. This allows us to use the information contained in $\tilde \beta$. Under Assumption \ref{asymptotic transportable}, $\tilde{\beta}_{\mathcal{A}}$, $\tilde{\beta}_{\mathcal{B}}$ and $\tilde{\beta}_{\mathcal{C}}$ can be viewed as estimates for $\beta_{\mathcal{A}}$, $\beta_{\mathcal{B}}$ and $\beta_{\mathcal{C}}$ with asymptotic biases of order smaller than, the same as and larger than $n^{-1/2}$, respectively. 

\begin{proposition}
	Under Assumption \ref{asymptotic transportable}, assuming that $\mathcal B$ and $\mathcal C$ are empty, 
	$n^{1/2}(\hat{\tau}_{\mathrm{e}}^{\sI} - \tau) \rightarrow N\{0, E(\phi_{\mathrm{e}}\phi_{\mathrm{e}}^\T)\}$ and $n^{1/2}(\hat{\beta}_{\mathrm{e}}^{\sI} - \beta) \rightarrow N\{0, E(\eta_{\mathrm{e}}\eta_{\mathrm{e}}^\T)\}$,  then $\hat{\tau}_{\mathrm{e}}$ is   consistent for $\tau$, and  the asymptotic variance of $\hat{\tau}_{\mathrm{e}}$ is equal to $B$    in Theorem  \ref{efficiency theorem}. 
\end{proposition}

Proposition S1  shows that when $\beta(P_{1,m}) - \beta(P_0) = o(n^{-1/2})$, $\hat{\tau}_{\mathrm{e}}$ can still attain the semiparametric efficiency bound $B$ in Theorem \ref{efficiency theorem}.
The proof is similar to Theorem \ref{efficient estimator}, and thus is omitted. When $\beta(P_{1,m})$ converges to $\beta(P_0)$ at a rate  slower than $n^{-1/2}$, the data-fused estimator $\hat{\tau}_{\mathrm{e}}$ is still consistent but no longer has the asymptotic distribution as in Proposition S1.

When $\mathcal{B}$ and $\mathcal{C}$ is not empty, we let $\hat{\tau}_{\mathrm{orc}} = \hat{\tau}_{\mathrm{e}}^{\mathcal{A}}$ denote an  oracle estimator  that is  obtained by incorporating only the subset $\mathcal{A}$ of external summary statistics, $\tilde{\beta}_{\mathcal{A}}$,  in the  estimation  in \eqref{efficient}. 
Note that, even if the above partition were known a priori, the components of $\tilde{\beta}$ with index in $\mathcal{B}$ and $\mathcal{C}$ are still not guaranteed to improve estimation efficiency because   the bias of these components are unknown. 
Without knowing such a partition, it's generally difficult to distinguish $\mathcal{A}$ and $\mathcal{B}$ with data, because the bias in $\mathcal{B}$ is of the same order as the variance. 
The following theorem shows that, if $\mathcal{B}$ is empty, 
the adaptive fusion estimator can still be used to  improve inference of $\tau$ when the tuning parameter $\lambda$ is suitably chosen.

\begin{theorem}\label{empty B}
	Under Assumption~\ref{asymptotic transportable} and assuming  that $\mathcal{B} = \emptyset$, $\lambda d_{m}^{\alpha} \to \infty$ and $\lambda n^{-\alpha/2} \to 0$, $n^{1/2}(\hat{\tau}_{\mathrm{e}}^{\sI} - \tau) \rightarrow N\{0, E(\phi_{\mathrm{e}}\phi_{\mathrm{e}}^\T)\}$ and $n^{1/2}(\hat{\beta}_{\mathrm{e}}^{\sI} - \beta) \rightarrow N\{0, E(\eta_{\mathrm{e}}\eta_{\mathrm{e}}^\T)\}$, then $\hat{\tau}_\mathrm{adf}$ has  the same asymptotic distribution as $\hat{\tau}_{\mathrm{orc}}$  and 
	$n^{1/2}(\hat{\tau}_{\mathrm{adf}} - \tau) \rightarrow N(0, B_{\mathcal{A}})$.
\end{theorem}

\begin{remark}
	Notice that the rate for $\lambda$ satisfies that $\lambda d_{m}^{\alpha}\rightarrow 0$, which is different from the rate in Theorem \ref{adaptive fusion}. The additional term $d_{m}^{\alpha}$ is required to distinguish the components of the summary statistics with indexes in  $\mathcal{A}$ and $\mathcal{C}$. In fact, Theorem \ref{adaptive fusion} is a special case of Theorem S1 when $d_{m}$ is a constant. The reason is that we assume $P_{1,m} = P_{1}$ in Theorem \ref{adaptive fusion}, so the difference $\beta(P_{1,m}) - \beta(P_0)$ does not depend on sample size $m$ and $d_{m}$ is a constant.   
\end{remark}

When the set $\mathcal{B}$ is not empty, one  cannot distinguish $\mathcal{A}$ from $\mathcal{B}$ with data.  Without loss of generality, assume $\mathcal{A} = \{1, \dots, q_{1}\}$ and $\mathcal{B} = \{q_{1} + 1, \dots, q_{2}\}$ for some positive integers $q_{1} < q_{2}$. Then, we have the following theorem for the adaptive fusion estimator in this scenario.
\begin{theorem}
	Under Assumption~\ref{asymptotic transportable} and assuming  that $n^{1/2}\{\beta_{\mathcal{B}}(P_{1,m}) - \beta_{\mathcal{B}}(P_0) \} \to \delta$ for some $\delta$, $\lambda d_{m}^{\alpha} \to \infty$, $\lambda n^{-\alpha/2} \to 0$, $n^{1/2}(\hat{\tau}_{\mathrm{e}}^{\sI} - \tau) \rightarrow N\{0, E(\phi_{\mathrm{e}}\phi_{\mathrm{e}}^\T)\}$, and $n^{1/2}(\hat{\beta}_{\mathrm{e}}^{\sI} - \beta) \rightarrow N\{0, E(\eta_{\mathrm{e}}\eta_{\mathrm{e}}^\T)\}$,
	then $n^{1/2}(\hat{\tau}_{\mathrm{adf}} - \tau) - \kappa \rightarrow N(0, B^{\mathcal{A} \cup \mathcal{B}})$
	where 
	\[
	\kappa = E(\phi_{\mathrm{e}}\eta_{\mathrm{e}, \mathcal{A}\cup\mathcal{B}}^\T) \left\{ \Sigma_{1, \mathcal{A} \cup \mathcal{B}}/\rho + E(\eta_{\mathrm{e}, \mathcal{A} \cup \mathcal{B}}\eta_{\mathrm{e}, \mathcal{A} \cup \mathcal{B}}^\T ) \right\}^{-1}\begin{pmatrix}
		0_{\mathcal{A}} \\ 
		\delta
	\end{pmatrix},
	\]
	and
	\[
	B^{\mathcal{A} \cup \mathcal{B}} = E(\phi_{\mathrm{e}}\phi_{\mathrm{e}}^\T ) - E(\phi_{\mathrm{e}}\eta_{\mathrm{e}, \mathcal{A} \cup \mathcal{B}}^\T) \left\{ \Sigma_{1, \mathcal{A} \cup \mathcal{B}}/\rho + E(\eta_{\mathrm{e}, \mathcal{A} \cup \mathcal{B}}\eta_{\mathrm{e}, \mathcal{A} \cup \mathcal{B}}^\T ) \right\}^{-1}E(\phi_{\mathrm{e}}\eta_{\mathrm{e}, \mathcal{A} \cup \mathcal{B}}^\T )^\T.
	\]
\end{theorem}
Proofs of Theorems S1 and S2 are  provided at the end of this section. 
Theorem S2 shows that $\hat{\tau}_{\mathrm{adf}}$ has an asymptotic bias $\kappa$, which is caused by the the use of summary statistics with  indices in $\mathcal{B}$. In summary, we show that when the bias is of order smaller than $n^{-1/2}$, the data-fused efficient estimator $\hat{\tau}_{\mathrm{e}}$ can still achieve the efficiency bound $B$ in Theorem 2.  
When bias  of order $n^{-1/2}$ or greater than $n^{-1/2}$ exists, the adaptive fusion approach can consistently select summary statistics with bias of order no greater than $n^{-1/2}$.
Therefore, when bias of order $n^{-1/2}$ does not exist, the adaptive fusion estimator has the same limiting distribution as the oracle estimator that uses only summary statistics that have bias of order smaller than $n^{-1/2}$  under suitable conditions; otherwise,  the adaptive fusion estimator will also  have asymptotic bias of order $n^{-1/2}$.


\begin{proof}[Proof of Theorem S1]
	The arguments before Theorem \ref{adaptive fusion} is still valid when $\mathcal{B} = \emptyset$. Proof of Theorem S1 follows directly from those arguments and is omitted here.
\end{proof}

\begin{proof}[Proof of Theorem S2]
	Under the assumption that $\lambda d_{m}^{\alpha} \to \infty$, $\lambda n^{-\alpha/2} \to 0$, $n^{1/2}(\hat{\tau}_{\mathrm{e}}^{\sI} - \tau) \rightarrow N\{0, E(\phi_{\mathrm{e}}\phi_{\mathrm{e}}^\T)\}$, and $n^{1/2}(\hat{\beta}_{\mathrm{e}}^{\sI} - \beta) \rightarrow N\{0, E(\eta_{\mathrm{e}}\eta_{\mathrm{e}}^\T)\}$, we have $\hat{a}_{j} \to 1$ in probability for $j \in \mathcal{A} \cup \mathcal{B}$ and $\hat{a}_{j} \to 0$ for $j \not\in \mathcal{A} \cup \mathcal{B}$. Thus, we have
	\[
	\widehat{\Sigma}_{\phi\eta}^{\mathrm{e}}\widehat{A} \to 
	\begin{pmatrix}
		E(\phi_{\mathrm{e}}\eta_{\mathrm{e}, \mathcal{A} \cup \mathcal{B}}^\T) & 0
	\end{pmatrix}
	\]
	and
	\[
	\left(I - \widehat{A} + \hat{a}\hat{a}^{\T}\right)\odot\left(\widehat{\Sigma}_1/\rho + \widehat{\Sigma}_{\eta\eta}^{\mathrm{e}}\right) \to
	\begin{pmatrix}
		\Sigma_{1, \mathcal{A} \cup \mathcal{B}} / \rho + E(\eta_{\mathrm{e}, \mathcal{A} \cup \mathcal{B}}\eta_{\mathrm{e}, \mathcal{A} \cup \mathcal{B}}^\T) & 0\\
		0 & D_{\eta, (\mathcal{A} \cup \mathcal{B})^{c}}
	\end{pmatrix}  
	\]
	in probability, where $D_{\eta, (\mathcal{A} \cup \mathcal{B})^{c}}$ is the diagonal matrix consisting of the diagonal elements of $\Sigma_{1, (\mathcal{A} \cup \mathcal{B})^{c}} / \rho + E(\eta_{\mathrm{e}, (\mathcal{A} \cup \mathcal{B})^{c}}\eta_{\mathrm{e}, (\mathcal{A} \cup \mathcal{B})^{c}}^\T)$ and $(\mathcal{A} \cup \mathcal{B})^{c} = \{j: j \not\in \mathcal{A}\cup \mathcal{B}\}$.
	This implies that
	\[
	\begin{aligned}
		\hat{\tau}_\mathrm{adf} 
		& = \hat{\tau}_{\mathrm{e}}^{\sI} - 
		\begin{pmatrix}
			E(\phi_{\mathrm{e}}\eta_{\mathrm{e}, \mathcal{A} \cup \mathcal{B}}^\T) & 0
		\end{pmatrix}
		\begin{pmatrix}
			\Sigma_{1, \mathcal{A} \cup \mathcal{B}} / \rho + E(\eta_{\mathrm{e}, \mathcal{A} \cup \mathcal{B}}\eta_{\mathrm{e}, \mathcal{A} \cup \mathcal{B}}^\T) & 0\\
			0 & D_{\eta, (\mathcal{A} \cup \mathcal{B})^{c}}
		\end{pmatrix}^{-1}
		(\hat{\beta}_{\mathrm{e}}^{\sI} - \tilde{\beta})\\
		&\quad + o_{P}\left(n^{-1/2}\right)\\
		& = \hat{\tau}_{\mathrm{e}}^{\sI} - 
		E(\phi_{\mathrm{e}}\eta_{\mathrm{e}, \mathcal{A} \cup \mathcal{B}}^\T)\left\{\Sigma_{1, \mathcal{A} \cup \mathcal{B}} / \rho + E(\eta_{\mathrm{e}, \mathcal{A} \cup \mathcal{B}}\eta_{\mathrm{e}, \mathcal{A} \cup \mathcal{B}}^\T)\right\}^{-1}
		(\hat{\beta}_{\mathrm{e}}^{\sI,\mathcal{A} \cup \mathcal{B}} - \tilde{\beta}_{\mathcal{A} \cup \mathcal{B}})
		+ o_{P}\left(n^{-1/2}\right).
	\end{aligned}
	\]
	The result of Theorem S2 then follows from the assumptions that $n^{1/2}\{\beta_{\mathcal{B}}(P_{1,m}) - \beta_{\mathcal{B}}(P_0) \} \to \delta$, $n^{1/2}(\hat{\tau}_{\mathrm{e}}^{\sI} - \tau) \rightarrow N\{0, E(\phi_{\mathrm{e}}\phi_{\mathrm{e}}^\T)\}$,  $n^{1/2}(\hat{\beta}_{\mathrm{e}}^{\sI} - \beta) \rightarrow N\{0, E(\eta_{\mathrm{e}}\eta_{\mathrm{e}}^\T)\}$, and Assumption \ref{asymptotic transportable}.
\end{proof}

\subsection{Details of the Re-bootstrap Procedure}\label{app: detail reboot}
In this section, we provide some details for the re-bootstrap procedure proposed in Section \ref{subsec: reboot} in the main text. We first describe the estimator of $q_{j}(z^{\dag}; h^{\dag})$ for any given $0 < z^{\dag} < 1$, $h^{\dag}$ and $j = 1, \dots, p$.
The error caused by normal approximation is typically of order $o(n^{-1/2})$ \citep{hall2013bootstrap} and hence negligible.
In addition, the estimation error of the variance is of order $o_{P}(n^{-1})$ which is negligible. Thus, we can approximate $q_{j}(z^{\dag}; h^{\dag})$ using the following bootstrap procedure.
\begin{enumerate}[(1)]
	\item For $r^{\dag} = 1,\dots, \bar{r}^{\dag}$, independently generate
	\[
	\begin{pmatrix}
		\hat{\tau}_{\mathrm{e}}^{\sI,(r^{\dag})}\\
		\hat{h}^{(r^{\dag})}
	\end{pmatrix}
	\sim
	N\left\{
	\begin{pmatrix}
		\hat{\tau}_{\mathrm{e}}^{\sI}\\
		h^{\dag}
	\end{pmatrix},
	\begin{pmatrix}
		\widehat{\Sigma}_{\phi\phi}^{\mathrm{e}}/n & - \widehat{\Sigma}_{\phi\eta}^{\mathrm{e}}/n\\
		- \widehat{\Sigma}_{\phi\eta}^{\mathrm{e}\T}/n & \widehat{\Sigma}_1 / m + \widehat{\Sigma}_{\eta\eta}^{\mathrm{e}}/n
	\end{pmatrix}
	\right\},
	\] 
	and obtain $\hat{\tau}_{\mathrm{adf}}^{(r^{\dag})}$ by replacing $\hat{\tau}_{\mathrm{e}}^{\sI}$ and $	\hat{\beta}_{\mathrm{e}}^{\sI} - \tilde{\beta}$ by $\hat{\tau}_{\mathrm{e}}^{\sI,(r^{\dag})}$ and
	$- \hat{h}^{(r^{\dag})}$, respectively, in the definition of $\hat{\tau}_{\mathrm{adf}}$;	
	\item Compute the $z^{\dag}$-quantile $\hat{q}_{j}(z^{\dag}; h^{\dag})$ of 
	$\left\{\hat{\tau}_{\mathrm{adf}, j}^{(r^{\dag})} - \hat{\tau}_{\mathrm{e}, j}^{\sI}\right\}_{r^{\dag} = 1}^{\bar{r}^{\dag}}$.
\end{enumerate}

The difference between $\hat{q}_{j}(z^{\dag}; h^{\dag})$ and $q_{j}(z^{\dag}; h^{\dag})$ can be neglected if $\bar{r}^{\dag}$ is sufficiently large. The value of $\bar{r}^{\dag}$ is set to be $500$ in our numerical experiments.
To obtain reasonable candidate heterogeneity parameters, we first generate $\hat{h}$ from the confidence distribution $N(\tilde{\beta} - \hat{\beta}_{\mathrm{e}}^{\sI}, \widehat{\Sigma}_1 / m + \widehat{\Sigma}_{\eta\eta}^{\mathrm{e}}/n)$ of $h$ \citep{xie2013confidence}. However, the resulting candidate heterogeneity parameter tends to overestimate the size of the true parameter. Specifically, we have
\[
\frac{E(\hat{h}_{j}^{2})}{h_{j}^{2}} = \frac{E\{(\tilde{\beta}_{j} - \hat{\beta}_{\mathrm{e}, j}^{\sI})^{2}\} + (\widehat{\Sigma}_1 / m + \widehat{\Sigma}_{\eta\eta}^{\mathrm{e}}/n)_{jj}}{E\{(\tilde{\beta}_{j} - \hat{\beta}_{\mathrm{e}, j}^{\sI})^{2}\} - \left\{\Sigma_1 / m + E(\eta_{\mathrm{e}}\eta_{\mathrm{e}}^\T)/n\right\}_{jj}} > 1.
\]
To mitigate the problem, we multiply $\hat{h}_{j}$ by the calibration factor $\hat{f}_{j} = \max\{0, (\tilde{\beta}_{j} - \hat{\beta}_{\mathrm{e}, j}^{\sI})^{2} - (\widehat{\Sigma}_1 / m + \widehat{\Sigma}_{\eta\eta}^{\mathrm{e}}/n)_{jj}\}^{1/2}/\{(\tilde{\beta}_{j} - \hat{\beta}_{\mathrm{e}, j}^{\sI})^{2} + (\widehat{\Sigma}_1 / m + \widehat{\Sigma}_{\eta\eta}^{\mathrm{e}}/n)_{jj}\}^{1/2}.$ 

Let $p_{h, j}$ be the p-value for testing $h_{j} = 0$ based on $\tilde{\beta}_{j} - \hat{\beta}_{\mathrm{e}, j}^{\sI}$ and the Wald test. An extremely small $p_{h,j}$ suggests that the heterogeneity can be consistently detected, the true $h_{j}$ is not in the $n^{-1/2}$ regime and hence the sampling for $h_{j}$ is not necessary. We propose to set the $j$-th component of the candidate heterogeneity parameter as $\tilde{\beta}_{j} - \hat{\beta}_{\mathrm{e}, j}^{\sI}$ in this case.
The resulting re-bootstrap procedure is summarized as following.
\begin{enumerate}[(1)]
	\item For $r = 1,\dots, \bar{r}$, independently generate
	$\hat{h}^{(r)} \sim N(\tilde{\beta} - \hat{\beta}_{\mathrm{e}}^{\sI}, \widehat{\Sigma}_1 / m + \widehat{\Sigma}_{\eta\eta}^{\mathrm{e}}/n)$ and calculate the calibrated heterogeneity parameter $\hat{h}_{\mathrm{cal}}^{(r)} = (\hat{h}_{\mathrm{cal}, 1}^{(r)},\dots,\hat{h}_{\mathrm{cal},q}^{(r)})^{\T}$ with $\hat{h}_{\mathrm{cal}, j}^{(r)} = \tilde{\beta}_{j} - \hat{\beta}_{\mathrm{e}, j}^{\sI}$ if $p_{h, j} 
	\leq 0.05 / \log(n)$ and $\hat{h}_{\mathrm{cal}, j}^{(r)} = \hat{f}_{j}\hat{h}_{j}^{(r)}$ otherwise for $j = 1,\dots, q$;
	\item Compute the quantiles $\hat{q}_{j}(z / 2; \hat{h}_{\mathrm{cal}}^{(r)})$  and $\hat{q}_{j}(1 - z / 2;  \hat{h}_{\mathrm{cal}}^{(r)})$ using the bootstrap procedure introduced in the last paragraph; 
	\item Construct the $(1 - z)$-confidence interval 
	\begin{equation}\label{eq: CI comp}
		\left[\hat{\tau}_{\mathrm{adf}, j} - \max_{r = 1,\dots, \bar{r}}\hat{q}_{j}(1 - z / 2; \hat{h}_{\mathrm{cal}}^{(r)}), \hat{\tau}_{\mathrm{adf}, j} - \min_{r = 1,\dots, \bar{r}}\hat{q}_{j}(z / 2; \hat{h}_{\mathrm{cal}}^{(r)})\right]
	\end{equation}
	for $\tau_{j}$.
\end{enumerate}

\newpage
\bibliographystyle{asa}
\bibliography{reference}

\end{document}